 \documentclass[final,3p,times]{elsarticle}

\usepackage{amssymb}
\usepackage{amsmath}

\usepackage{bm}

\usepackage[symbol]{footmisc}

\journal{Physica D}

\begin{document}

\begin{frontmatter}



\title{On the role of 5-wave resonances in the nonlinear dynamics of the Fermi-Pasta-Ulam-Tsingou lattice 
}


\author{Tiziana Comito${}^a$}

\author{Matteo Lotriglia${}^a$}

\author{Miguel D. Bustamante${}^{a,}$\footnote[1]{Corresponding author. E-mail: {miguel.bustamante@ucd.ie}}}

\affiliation{{School of Mathematics and Statistics, University College Dublin, Belfield, Dublin 4, Ireland}}


\begin{abstract}
  We study the dynamics of the $(\alpha+\beta)$ Fermi-Pasta-Ulam-Tsingou lattice (FPUT lattice, for short) for an arbitrary number $N$ of interacting particles, in regimes of small enough nonlinearity so that a Birkhoff-Gustavson type of normal form can be found using tools from wave-turbulence theory. Specifically, we obtain the so-called Zakharov equation for $4$-wave resonant interactions and its extension to $5$-wave resonant interactions by Krasitskii, but we introduce an important new feature: even the generic terms in these normal forms contain {resonant interactions only}, via a {unique} canonical transformation. The resulting normal forms provide an approximation to the original FPUT lattice that possesses a significant number of exact quadratic conservation laws, beyond the quadratic part of the Hamiltonian. We call the new equations "exact-resonance evolution equations" and examine their properties: (i) Heisenberg representation's slow evolution allows us to implement numerical methods with large time steps to obtain relevant dynamical information, such as Lyapunov exponents. (ii) We introduce tests, such as convergence of the normal form transformation and truncation error verification, to successfully validate our exact-resonance evolution equations. (iii) The systematic construction of new quadratic invariants (via the resonant cluster matrix) allows us to use finite-time Lyapunov exponent calculations to quantify the level of nonlinearity at which the original FPUT lattice is well approximated by the exact-resonance evolution equations.  We show numerical experiments in the case $N=9$, but the theory and numerical methods are valid for arbitrary values of $N$. We conclude that, when $3$ divides $N$, at small enough nonlinearity the FPUT lattice's dynamics and nontrivial hyperchaos are governed by $5$-wave resonant interactions.
\end{abstract}


\end{frontmatter}



\section{Introduction}

The Fermi-Pasta-Ulam-Tsingou (FPUT) lattice considers a one-dimensional chain made out of $N$ identical masses connected by nonlinear springs, where the elastic forcing is modelled as power series that include anharmonic terms up to cubic order in the displacement: the name $(\alpha+\beta)$ FPUT is commonly used for this model, to refer to quadratic terms ($\alpha$) and cubic terms ($\beta$) in the elastic forcing. Allowing interactions between first neighbours triggers complex energy exchanges that reverberate across various space and time  scales.
In the original problem \cite{fermi1955studies}, the chain has fixed boundary conditions (BCs) and displays the phenomenon of recurrence in contrast with Fermi's assumption that the presence of nonlinearity would be enough to bring the system to equipartition. This incongruity took the name of ``FPU paradox'' and triggered the interest of the scientific community, eventually leading to fundamental discoveries such as the birth of the theory of integrable systems \cite{zakharov1991integrability,zabusky1965interaction}.
Later it was shown analytically in \cite{israiljev1965statistical} that there exists a critical energy threshold above which equipartition is reachable. Further mathematical \cite{shepelyansky1997low,bambusi2006metastability,fucito1982approach}  and numerical results \cite{berchialla2004localization,paleari2007numerical} followed right after and several theories have been presented to explain why equipartition should occur eventually. One of the first ones, still controversial, is based on considerations of resonances and finite-amplitude effects. The legacy this problem carries is vast, and attempts to provide an exhaustive summary of its history is a hard task. References \cite{berman2005fermi,gallavotti2007fermi,kevrekidis2011non} can be consulted for a concise pipeline of FPUT's chronicle.

Our work focuses on the role of resonances in the dynamics of the FPUT system, from a fresh viewpoint in the context of low degrees of freedom (low number of masses $N$), where the chain has periodic boundary conditions, so that exact wave-wave resonant interactions appear: namely, particular linear combinations of wavenumbers and akin in frequencies add up to zero.
The state of the art in this subject provides a clear but incomplete picture: it has been known for a while that $3-$wave resonances do not exist,
whereas $4-$wave resonances are admitted for $N$ even only, but they turn out to be
completely integrable \cite{henrici2008results}. Thus, one must look at the next orders of interaction, namely $5$-wave resonances (admitted when $3$ divides $N$) and $6$-wave resonances, in order to get mixing across the energy spectrum. 
In the context of surface water waves with non-integrable dynamics, the question regarding the existence of quintets was undertaken in \cite{lvov1997effective} and \cite{dyachenko1995five} using a diagram scheme. More recently \cite{bustamante2019exact} tackled the same problem for the FPUT model using cyclotomic polynomials from number theory, providing an exhaustive technique to find all possible exact wave-wave interactions of the sort. As underlined in a recent report \cite{onorato2023wave}, validation of the Wave Turbulence theory to discrete systems is still an open question.
One aim of this work concerns the validation of the canonical transformation and its relation with integrability.

To remove non-resonant terms in the FPUT system, a set of quasi-identity transformations is introduced as a power series expansion with unknown coefficients to be determined. Substituting this into the evolution equations leads to the so-called Birkhoff-Gustavson normal form \cite{gustavson1966}, containing only resonant contributions up to the $5$-wave order. We follow the approach proposed by Krasitskii in \cite{krasitskii1994reduced}, validating his kernels at all necessary orders. At difference with Gustavson's and Krasitskii's works, which keep some ambiguity in the canonical transformations, we introduce new tensors to implement unique canonical transformations and normal forms so that, instead of generic terms, the equations contain exact resonant interactions only. We call the new equations ``exact-resonance evolution equations''.

In section~\ref{sec:FPUT_maths} the problem is tackled mathematically. The pivotal changes of coordinates leading to the exact-resonance evolution equations for the FPUT lattice are defined, along with the derivation of each evolution equation at every relevant order up to and including $5$-wave resonances.  Section~\ref{sec:FPUT_comp} is devoted to validation techniques. We propose two distinct methodologies to corroborate numerically the analytical results, namely the relation between the original and approximated models arising after the transformation. One approach is based on the truncation error of the mapped equations at each order. The other approach is based on the convergence of the normal form transformation. In addition, we look at time series to clearly demonstrate energy exchanges due to $5$-wave resonant interactions. Following up, the dynamical characteristics of $5$-wave interactions becomes the main topic of section~\ref{sec:FPUT_theo}: links between resonances, hyperchaos, and constants of motion are addressed from a theoretical and constructive standpoint based on the resonant cluster matrix and Lyapunov exponents, to culminate with a proposed method that demonstrates numerically how the original FPUT lattice is well approximated by the $5$-wave exact-resonance evolution equations. Finally, in section~\ref{sec:conclusion} we present conclusions, a synopsis of results, and future research venues.

\section{FPUT equations of motion, normal modes, resonances and normal-form transformation}
\label{sec:FPUT_maths}
The $(\alpha+\beta)$-FPUT model is a lattice of $N$ identical point-like particles of mass $m$ joined together by elastic springs, interacting via first neighbors, with a force derived from an anharmonic potential. Let $q_j$ be the displacement of the $j$-th particle with respect to its equilibrium position, and let $p_j$ be its linear momentum, where $j = 0,..,N-1$. The equations of motion for the $(\alpha+\beta)$-FPUT model are derived from the Hamiltonian 
\begin{equation}
    \label{eq:H}
    H = \sum_{j=0}^{N-1} \Big( \dfrac{1}{2m} p_j^2 + \dfrac{\kappa}{2}\big( q_{j+1} - q_j \big)^2 + \dfrac{\alpha}{3}\big( q_{j+1} - q_j \big)^3 + \dfrac{\beta}{4}\big( q_{j+1} - q_j \big)^4  \Big)
\end{equation}
via the canonical Hamilton equations
\begin{equation}
\begin{split}
\label{eq:q_p_dot}
  &\dot{q_j} = \dfrac{\partial H}{\partial p_j} = \dfrac{p_j}{m}\,,
\\
  &\dot{p_j} = - \dfrac{\partial H}{\partial q_j} = \kappa (q_{j-1} - 2 q_j + q_{j+ 1}) + \alpha \big( (q_{j+ 1}- q_j)^2 - (q_{j}- q_{j-1})^2  \big) + \beta \big( (q_{j+ 1}- q_j)^3 - (q_{j}- q_{j-1})^3  \big)\,, 
\end{split}
\end{equation}
where $\kappa$, $\alpha$, and $\beta$ are non-negative prefactors that weigh the contribution of the linear ($\kappa$) and nonlinear terms ($\alpha, \beta$). From  dimensional analysis we note that there are only $2$ arbitrary parameters: $\Bar{\alpha} \propto \epsilon \alpha$ and $\bar{\beta} \propto \epsilon^2 \beta$, where $\epsilon$ is the initial energy provided to the system. In other words we can set $m=\kappa=1$, but we will keep using these constants in the formulas for bookkeeping purposes.

\subsection{Periodic boundary conditions and normal modes}

Assuming periodic boundary conditions, $q_0(t) := q_{N-1}(t) \; \forall t \in \mathbb{R}$, one can apply a discrete Fourier transform in space to diagonalise the quadratic part of the Hamiltonian \eqref{eq:H} through a canonical transformation, obtaining the so-called normal modes:
\begin{eqnarray}
\label{eq:a_k_transf}
a_{k} &=& \dfrac{1}{N\sqrt{2m\omega_{k}}} \sum_{j=0}^{N-1} \left(m\omega_{k} q_j  + i p_j \right)\mathrm{e}^{- i \frac{2 \pi j k}{N}}\,,\\
\label{eq:a*_k_transf}
a_{N-k}^{*} &=& \dfrac{1}{N\sqrt{2m\omega_{k}}} \sum_{j=0}^{N-1} \left(m\omega_{k} q_j  - i p_j \right)\mathrm{e}^{- i \frac{2 \pi j k}{N}} \,,\qquad k =1, \ldots, N-1\,,
\end{eqnarray}
where we have introduced the dispersion relation 
\begin{equation}
\label{eq:FPUT_dispersion}
\omega_k = 2  \sqrt{\frac{\kappa}{m}} \sin (\pi k/N), \qquad k = 1,2,...,N-1.
\end{equation} 
Physically, these variables represent ``waves'', i.e.~oscillatory patterns with wavenumber $k$ and  natural frequency $\omega_k$. As detailed in reference \cite{bustamante2019exact}, the Hamiltonian becomes
\begin{equation}
\label{eq:H_a}
\begin{split}
    \dfrac{H}{N} &= \sum_0 \omega_0 a^*_0 a_0  +  \sum_{0,1,2} \left( \dfrac{1}{3} V_{0,1,2} a_0 a_1 a_2\delta_{0,1,2} + V_{-0,1,2} a^*_0 a_1 a_2\delta_{0}^{1,2}   + c.c.\right) \\
    &+ \sum_{0,1,2,3} \left( \dfrac{1}{4}T_{0,1,2,3} a_0a_1a_2a_3\delta_{0,1,2,3}  + \dfrac{3}{4} T_{-0,-1,2,3} a_0^*a_1^*a_2a_3 \delta_{0,1}^{2,3} +  T_{-0,1,2,3}a_0^*a_1a_2a_3\delta_{0}^{1,2,3} + c.c. \right)\,,
\end{split} 
\end{equation}
where the numerical subscript refers to wavenumber indices (e.g. $T_{0,1,2,3} := T(k_0,k_1,k_2,k_3), \,\, 1 \leq k_j \leq N-1$, and $a_0 := a_{k_0}$), whereas the Kronecker delta's sub/superscripts encode momentum conditions (e.g. $\delta_{0}^{1,2,3} = \delta(k_0 - k_1 - k_2 - k_3 \mod N)$). The tensors $V, T$ encode the interactions between normal modes in the Hamiltonian. Explicitly,

\begin{equation}
\label{eq:Vapp}
    V_{0,1,2} = -\dfrac{i \alpha}{(m \kappa)^{3/4}} \mathrm{e}^{i \pi (k_0 + k_1+k_2 )/N} \sqrt{ \sin{\dfrac{\pi k_0}{N} } \sin{\dfrac{\pi k_1}{N} } \sin{\dfrac{\pi k_2}{N} } }
\end{equation}
and
\begin{equation}
\label{eq:Tapp}
    T_{0,1,2,3} = \dfrac{\beta}{m \kappa} \mathrm{e}^{i \pi (k_0 + k_1+k_2+k_3 )/N} \sqrt{ \sin{\dfrac{\pi k_0}{N} } \sin{\dfrac{\pi k_1}{N} } \sin{\dfrac{\pi k_2}{N} } \sin{\dfrac{\pi k_3}{N} } }  \,,
\end{equation}
whereas negative indices indicate that the corresponding wavenumber $k_i$ is replaced with $N-k_i$. For example, $V_{-0,1,2}$ is equal to the RHS of equation \eqref{eq:Vapp} where $k_0$ is replaced with $N-k_0$.
The Hamilton equations are simply 
\vspace{-0.2cm}
$$i \dot{a}_k = \frac{1}{N} \dfrac{\partial H}{\partial a^*_k}\,, \qquad  i \dot{a}^*_k = -\frac{1}{N} \dfrac{\partial H}{\partial a_k}\,, \qquad k=1, \ldots, N-1 \,,
$$ 
and explicitly, using the short-hand notation  $a_0:=a_{k_0}, a_1:=a_{k_1}$, etc., we have
\begin{equation}
\begin{split}
\label{eq:a_k}
i \dot{a}_0 &=  \omega_0 a_0 + \sum_{1, 2} \big( V_{-0,1,2} a_1 a_2 \delta_{0}^{1,2} - 2 V_{0,1,-2} a^*_1 a_2 \delta_{0,1}^{2} - V_{0,1,2} a^*_1 a^*_2 \delta_{0,1,2} \big) + 
\\& +  \sum_{1,2,3} \big(  T_{-0,1,2,3} a_1a_2a_3 \delta_{0}^{1,2,3} + 3 T_{-0,-1,2,3} a_1^*a_2a_3 \delta_{0,1}^{2,3} + 3 T_{0,1,2,-3} a_1^*a^*_2a_3 \delta_{0,1,2}^{3} + T_{0,1,2,3} a_1^*a^*_2a^*_3 \delta_{0,1,2,3} \big)\,.
\end{split}
\end{equation}
This equation carries the same dynamical information as Eq.~\eqref{eq:q_p_dot}. It shows three- and four-wave interactions (respectively coming from the terms involving the $V$ and the $T$ tensors), inherited from the quadratic and cubic nonlinearity in the original equations \eqref{eq:q_p_dot}. In fact, $V_{0,1,2}\propto \alpha$  and $T_{0,1,2,3} \propto \beta$ and these tensors must respect specific symmetries under index exchanges, implied in the multiple summations in equation \eqref{eq:H_a}.  

The presence of Kronecker delta terms in equation \eqref{eq:a_k} means that not every arrangement of wavenumbers is allowed to interact and, therefore, some terms contribute more meaningfully than others to the dynamics. A heuristic motivation of our rigorous analytical treatment in the next section is as follows: irreversible (i.e., chaotic) energy exchanges can occur only if those interactions allowed by the Kronecker deltas in wavenumber space satisfy extra conditions based on the resonance of their frequencies. In such a case, an exact resonance is present, involving $S (\geq 0)$ incoming waves and $T(\geq 0)$ outgoing waves:
\begin{equation}
\label{eq:resonant_condition}
    \begin{cases}
   & k_1 + \ldots +k_S = k_{S+1}+ \ldots +k_{S+T} \pmod N\,,\\
   & \omega(k_1) + \ldots + \omega(k_S)  = \omega(k_{S+1}) + \ldots + \omega(k_{S+T})\,, \quad  1\leq k_i\leq N-1, \quad i = 1,\ldots,S+T\,,
    \end{cases}
\end{equation}
where, as before, the dispersion relation is $\omega(k) = 2 \sqrt{\kappa/m} \sin( \pi k/N)$. System \eqref{eq:resonant_condition} is a Diophantine system of equations, namely it is a set of equations for a total of $S+T$ integer unknowns. No solutions exist when one of $(S,T)$ is greater than one and the other is equal to $1$ or $0$. This includes the case $S+T=3$, which means that there are no triad exact resonances. Trivial solutions exist when $S=T$. Remarkably, nontrivial solutions exist when $S=T=2$ and $N$ is even (resonant quartets), when $S=2, T=3$ (or $S=3, T=2$) and $N$ is divisible by $3$   (resonant quintets), when $S+T=6$ (resonant sextuplets), and so on. See \cite{bustamante2019exact} for details on explicit parameterisations of these solutions.

\subsection{Resonances and near-identity canonical transformations to normal-form variables}
One objective of this paper is to arrive, via canonical transformations, to equivalent equations driven by exact resonances only, and show that the lowest relevant order to obtain chaos in FPUT necessitates five-wave exact resonances.
We achieve this by using classical tools from turbulence theory: via successive near-identity transformations, it is possible to remove non-resonant terms order by order and recover a reduced equation driven by exact resonances. The procedure was successfully applied by Zakharov \cite{zakharov1974hamiltonian} and afterwards revised by Krasitskii ~\cite{krasitskii1990canonical,krasitskii1994reduced} in the Hamiltonian theory of weakly nonlinear surface waves where they postulated a canonical transformation in the form of an integral power series. This change of variables aims to remove unimportant terms from the evolution equation by setting the tensors of the transformation appropriately. The transformations are carried out iteratively, allowing us to retrieve the evolution equations truncated at each relevant order. In our context, we will introduce three such transformations, corresponding respectively to the elimination of non-resonant triads, quartets and quintets:
        \begin{equation}
        \label{eq:b_to_a}
        a_0 = b_0 + \sum_{1, 2} \big( A^{(1)}_{0,1,2}b_1b_2 \delta_0^{1,2}+ A^{(2)}_{0,1,2}b_1^*b_2 \delta_{0,1}^{2} + A^{(3)}_{0,1,2}b_1^*b_2^* \delta_{0,1,2} \big)\,,
        \end{equation} 
        \begin{equation}
        \begin{split}
        \label{eq:c_to_a}
        a_0 = c_0 &+ \sum_{1, 2} \big( A^{(1)}_{0,1,2}c_1c_2 \delta_0^{1,2}+ A^{(2)}_{0,1,2}c_1^*c_2 \delta_{0,1}^{2} + A^{(3)}_{0,1,2}c_1^*c_2^* \delta_{0,1,2} \big)\\
        &+  \sum_{1,2,3} \big( B^{(1)}_{0,1,2,3}c_1c_2c_3 \delta_0^{1,2,3} +B^{(2)}_{0,1,2,3}c^*_1c_2c_3 \delta_{0,1}^{2,3} + {B^{(3)}_{0,1,2,3}c_1^*c_2^*c_3 \delta_{0,1,2}^{3}} +  B^{(4)}_{0,1,2,3}c_1^*c_2^*c_3^* \delta_{0,1,2,3} \big)\,,
        \end{split}
        \end{equation}
        \begin{equation}
        \begin{split}
        \label{eq:d_to_a}
        a_0 = d_0 &+ \sum_{1, 2} \big( A^{(1)}_{0,1,2}d_1d_2 \delta_0^{1,2}+ A^{(2)}_{0,1,2}d_1^*d_2 \delta_{0,1}^{2} + A^{(3)}_{0,1,2}d_1^*d_2^* \delta_{0,1,2} \big)\\
        &+  \sum_{1,2,3} \big( B^{(1)}_{0,1,2,3}d_1d_2d_3 \delta_0^{1,2,3} +B^{(2)}_{0,1,2,3}d^*_1d_2d_3 \delta_{0,1}^{2,3} + {B^{(3)}_{0,1,2,3}d_1^*d_2^*d_3 \delta_{0,1,2}^{3}} +  B^{(4)}_{0,1,2,3}d_1^*d_2^*d_3^* \delta_{0,1,2,3} \big)\\
        &+ \sum_{1,2,3,4} \big( C^{(1)}_{0,1,2,3,4}d_1d_2d_3d_4 \delta_0^{1,2,3,4} + C^{(2)}_{0,1,2,3,4}d_1^*d_2d_3d_4 \delta_{0,1}^{2,3,4} +  C^{(3)}_{0,1,2,3,4}d_1^*d_2^*d_3d_4 \delta_{0,1,2}^{,3,4} 
        \\
        & \qquad \qquad + C^{(4)}_{0,1,2,3,4}d_1^*d_2^*d_3^*d_4 \delta_{0,1,2,3}^{4} +  C^{(5)}_{0,1,2,3,4}d_1^*d_2^*d_3^*d_4^* \delta_{0,1,2,3,4} 
        \big)\,,
        \end{split}
        \end{equation}
        
where $A^{(i)}_{0,1,2}$, $B^{(i)}_{0,1,2,3}$, and $C^{(i)}_{0,1,2,3,4}$ are unknown tensor coefficients to be determined as follows. Substituting Eqs.~\eqref{eq:b_to_a}--\eqref{eq:d_to_a} into Eq.~\eqref{eq:a_k} makes higher powers of the new variables appear, via combinations of the lower orders ones in the original equations.
After the substitutions, common terms are collected together and suitably symmetrised. Remarkably, in this process, factors of the form
\begin{equation}
    \Delta_{1+2-0} := \omega_1 + \omega_2 - \omega_0\,,\qquad 
\Delta_{1+2+3-0} := \omega_1 + \omega_2 +\omega_3  - \omega_0\,,\qquad 
\Delta_{2+3-0-1} :=  \omega_2 +\omega_3  - \omega_0 - \omega_1
\end{equation}
appear, as well as other similar factors such as $\Delta_{2 + 3 + 4 - 0 - 1} := \omega_2 +\omega_3 +\omega_4  - \omega_0 - \omega_1$ and so on. Notice that a simple comparison with the resonant conditions \eqref{eq:resonant_condition} shows that, for a given $(S+T)$-wave resonance, the corresponding  coefficient $\Delta_{k_1 \ldots k_S}^{k_{S+1}\ldots k_{S+T}}$ is equal to zero.

Next, within these equations of motion, non-resonant terms are eliminated, order by order, by setting the coefficients $A^{(i)}_{0,1,2}$, $B^{(i)}_{0,1,2,3}$, and $C^{(i)}_{0,1,2,3,4}$ appropriately. The detailed procedure is well explained in the above-cited papers \cite{zakharov1974hamiltonian,krasitskii1990canonical,krasitskii1994reduced}, so in this exposition we will be brief, leaving some of the details involving long expressions to \ref{appendix:supplementary_material}, while commenting here on the new ideas we introduce.

\subsubsection{Eliminating non-resonant triad interactions}

Formulas for the coefficients $A^{(i)}_{0,1,2}$ are easy to obtain, because they arise at the lowest order of interaction: triads, which are never resonant in the FPUT case. We have:
\begin{equation}
\label{eq:A_transform}
         A^{(1)}_{0,1,2} =  \big( \Delta_{1+2-0} \big)^{-1} V_{-0,1,2} \,,  \qquad 
         A^{(2)}_{0,1,2} = 2  \big(\Delta_{0+1-2} \big)^{-1} V_{0,1,-2}\,, \qquad
        A^{(3)}_{0,1,2} = \big( \Delta_{0+1+2} \big)^{-1} V_{0,1,2} \,,       
\end{equation}
and we stress that these formulas are always valid, as the factors $\Delta_{1+2-0}$ and so on are never equal to zero in FPUT because the dispersion relation is subadditive (see \cite{bustamante2019exact} for details).  As a result of the canonical transformation \eqref{eq:b_to_a}, the variables $b$ satisfy the approximate equations of motion
\begin{equation}
\label{eq:b}
i \dot{b_0}  =  \omega_0 b_0 +  \mathcal{O}(b^3)\,,
\end{equation}
with a truncation error $\mathcal{O}(b^3)$ representing a homogeneous polynomial of degree $3$ in the $b$- and $b^*$-variables.

\subsubsection{Eliminating non-resonant quartet interactions}
Formulas for the coefficients $B^{(i)}_{0,1,2,3}$ are classified into two types: ``always non-resonant'' type, producing formulas for $B^{(1)}_{0,1,2,3}$, $B^{(3)}_{0,1,2,3}$ and $B^{(4)}_{0,1,2,3}$, and ``potentially resonant'' type, producing a formula for $B^{(2)}_{0,1,2,3}$. This distinction is made because the only potential $4$-wave resonance occurs when $S=T=2$, namely only $\Delta_{0+1-2-3}$ can possibly be equal to zero at a resonance. This follows again from the subadditivity of the FPUT dispersion relation and we refer to \cite{bustamante2019exact} for details. The ``always non-resonant'' case gives:
\begin{equation}
\label{eq:B_transform}
\begin{split}
         B^{(1)}_{0,1,2,3} = -\big(\Delta_{0-1-2-3} \big)^{-1} \Big(T_{-0,1,2,3} + Z^{(1)}_{0,1,2,3}\Big) \,,\\
         B^{(3)}_{0,1,2,3} =  -\big(\Delta_{0+1+2-3} \big)^{-1} \Big(3T_{0,1,2,-3} +  Z^{(3)}_{0,1,2,3}\Big) \,,\\
         B^{(4)}_{0,1,2,3} =  -\big(\Delta_{0+1+2+3} \big)^{-1} \Big(T_{0,1,2,3} + Z^{(4)}_{0,1,2,3} \Big)\,,
    \end{split}
\end{equation}
where $T$ is given in equation \eqref{eq:Tapp}, while $Z^{(1)}, Z^{(3)}$ and $Z^{(4)}$ are given in equations \eqref{eq:Z1app}--\eqref{eq:Z4app}  in \ref{appendix:supplementary_material}. In contrast, the ``potentially resonant'' case distinguishes two subcases: the resonant subcase (when $\Delta_{0+1-2-3} = 0$) and the non-resonant subcase (when $\Delta_{0+1-2-3} \neq 0$), as follows:
\begin{equation}
\label{eq:B2_transform_full}
        B^{(2)}_{0,1,2,3} =  \Lambda_{0,1,2,3} +  \lambda_{0,1,2,3}\,,
\end{equation}
with
\begin{equation}
\label{eq:Lambda_0123}
     \Lambda_{0,1,2,3} := 
        - A^{(3)}_{0,1,-0-1}A^{(3)}_{-2-3,2,3} + A^{(1)}_{0,2,0-2}A^{(1)}_{3,1,3-1} + A^{(1)}_{0,3,0-3}A^{(1)}_{2,1,2-1}             + A^{(1)}_{0+1,0,1}A^{(1)}_{2+3,2,3} - A^{(1)}_{1,2,1-2}A^{(1)}_{3,3-0,0} - A^{(1)}_{1,3,1-3}A^{(1)}_{2,2-0,0} 
\end{equation}
and
\begin{equation}
\label{eq:lambda_0123}
        \lambda_{0,1,2,3} := \begin{cases}
            0 & \text{if} \quad \Delta_{0+1-2-3} = 0\,,\\
       ~\\
        -\big(\Delta_{0+1-2-3}  \big)^{-1} \left(3T_{-0,-1,2,3} + \frac{1}{4}\left(Z^{(2)}_{0,1,2,3}+Z^{(2)}_{1,0,2,3}+Z^{(2)}_{2,3,0,1}+Z^{(2)}_{3,2,0,1}\right)\right) & \text{if} \quad \Delta_{0+1-2-3} \neq 0\,,
        \end{cases}
\end{equation}
where $A^{(1)}, A^{(3)}$ are given in equation \eqref{eq:A_transform}, $T$ is given in equation \eqref{eq:Tapp}, and  $Z^{(2)}$ is given in equation \eqref{eq:Z2app} in \ref{appendix:supplementary_material}.
It is important to remark here that, at difference with Krasitskii \cite{krasitskii1994reduced} (where the tensor $\lambda$ is set to be identically zero), in our approach the tensor takes non-zero values precisely at places where the quartets are not resonant. Our choice of $\lambda$ in equation \eqref{eq:lambda_0123} satisfies the required symmetries under the permutations $0 \leftrightarrow 1$ and  $2 \leftrightarrow 3$, and antisymmetry under the permutation $\{0,1\}\leftrightarrow \{2,3\}$, and is chosen so that the equations of motion for the $c$-variables contain only non-resonant terms. 
Explicitly, as a result of the canonical transformation  \eqref{eq:c_to_a}, the variables $c$ satisfy the approximate equations of motion
\begin{equation}
\label{eq:c}
i \dot{c_0}  =  \omega_0 c_0 + \sum_{1, 2, 3} \widetilde{T}_{0,1,2,3} c_1^*c_2c_3 \delta_{0,1}^{2,3} +  \mathcal{O}(c^4)\,,
\end{equation}
this time with a truncation error $\mathcal{O}(c^4)$ representing a homogeneous polynomial of degree $4$ in the $c$-variables, and with
\begin{equation}
\label{eq:T_tilde}
    \widetilde{T}_{0,1,2,3} = 
    \begin{cases}
        3T_{-0,-1,2,3} + \dfrac{1}{4} \big( Z^{(2)}_{0,1,2,3} + Z^{(2)}_{1,0,2,3} + Z^{(2)}_{2,3,0,1}+ Z^{(2)}_{3,2,0,1} \big) & \text{if} \quad \Delta_{0+1-2-3} = 0\,,\\
        0 & \text{if} \quad \Delta_{0+1-2-3} \neq 0\,,
    \end{cases}
\end{equation}
where again $Z^{(2)}$ is given in equation \eqref{eq:Z2app} in \ref{appendix:supplementary_material}.
At this level of approximation, the Hamiltonian \eqref{eq:H}  is
\begin{equation}
\label{eq:H_c}
    \frac{H}{N}  = \sum_0 \omega_0 c^*_0 c_0 + \sum_{0,1,2,3} \widetilde{T}_{0,1,2,3} c_0^*c_1^*c_2 c_3 \delta_{0,1}^{2,3}  + \mathcal{O}(c^5)\,
\end{equation}
and the equations of motion \eqref{eq:c} are canonical:
\vspace{-0.2cm}
$$i \dot{c}_k = \frac{1}{N} \dfrac{\partial H}{\partial c^*_k}+  \mathcal{O}(c^4)\,, \qquad  i \dot{c}^*_k = -\frac{1}{N} \dfrac{\partial H}{\partial c_k} +  \mathcal{O}(c^4)\,, \qquad k=1, \ldots, N-1 \,.
$$ 

\subsubsection{Eliminating non-resonant quintet interactions}

Finally, formulas for the coefficients $C^{(i)}_{0,1,2,3,4}$ are obtained in a similar way. Again there is a distinction between ``always non-resonant'' and ``potentially resonant'' types, because the only possible FPUT $5$-wave resonances occur when $\{S,T\} = \{2,3\}$, namely only $\Delta_{0+1+2-3-4}$ or $\Delta_{0+1-2-3-4}$ can possibly equal zero at a resonance. The ``always non-resonant'' case gives
\begin{equation}
\label{eq:C_transform}
    \begin{split}
         C^{(1)}_{0,1,2,3,4} = \big(  \Delta_{0-1-2-3-4} \big)^{-1} \left(-i \,X^{(1)}_{0,1,2,3,4} \right)\,, \\
         C^{(4)}_{0,1,2,3,4} = \big( \Delta_{4-0-1-2-3} \big)^{-1}\left(- i\,  X^{(4)}_{0,1,2,3,4}\right)\,, \\
         C^{(5)}_{0,1,2,3,4} = \big( \Delta_{0+1+2+3+4} \big)^{-1}  \left(-i\,X^{(5)}_{0,1,2,3,4}\right) \,,
    \end{split}
\end{equation}
where the tensors $X^{(1)}, X^{(4)}, X^{(5)}$ are given in equations \eqref{kernel:X1}--\eqref{kernel:X5} in \ref{appendix:supplementary_material}. In contrast, the ``potentially resonant'' case is separated in two subcases: the resonant subcases (when $\Delta_{0+1+2-3-4} = 0$ or $\Delta_{0+1-2-3-4} = 0$) and the non-resonant subcases (when $\Delta_{0+1+2-3-4} \neq 0$ or $\Delta_{0+1-2-3-4} \neq 0$), as follows: 
\begin{equation}
\label{eq:C2}
        {C}^{(2)}_{0,1,2,3,4} = \begin{cases}
            {C}^{(2,\text{Krasitskii})}_{0,1,2,3,4} +{C}^{(2,\lambda)}_{0,1,2,3,4} & \text{if} \quad \Delta_{0+1-2-3-4} = 0\,,\\
       ~\\
      \big(\Delta_{0+1-2-3-4} \big)^{-1} \left( i\,X^{(2)}_{0,1,2,3,4} \right)& \text{if} \quad \Delta_{0+1-2-3-4} \neq 0\,,
        \end{cases}
\end{equation}
\begin{equation}
\label{eq:C3}
        {C}^{(3)}_{0,1,2,3,4} = \begin{cases}
            {C}^{(3,\text{Krasitskii})}_{0,1,2,3,4}  + C^{(3,\lambda)}_{0,1,2,3,4} & \text{if} \quad \Delta_{0+1+2-3-4} = 0\,,\\
       ~\\
      \big(\Delta_{0+1+2-3-4} \big)^{-1} \left( -i\,X^{(3)}_{0,1,2,3,4} \right)& \text{if} \quad \Delta_{0+1+2-3-4} \neq 0\,,
        \end{cases}
\end{equation}
where: $X^{(2)}, X^{(3)}$ are given in equations \eqref{kernel:X2}--\eqref{kernel:X3} in \ref{appendix:supplementary_material}; ${C}^{(2,\text{Krasitskii})}, {C}^{(3,\text{Krasitskii})}$ are the usual resonant-case Krasitskii tensors \cite[equations (3.30) and (3.35)]{krasitskii1994reduced}, given by 
\begin{equation}
    \begin{split}
        \label{eq:C2_C3_Krasitskii}
             {C}^{(2,\text{Krasitskii})}_{0,1,2,3,4}  &= -i\,\left(p_{0,1,2,3,4} +  p_{0,1,3,2,4} +  p_{0,1,4,2,3} - p_{1,0,2,3,4} -  p_{1,0,3,2,4} -  p_{1,0,4,2,3}\right)\,,\\
     {C}^{(3,\text{Krasitskii})}_{0,1,2,3,4}  &= \frac{i}{2} \left(Q_{0,1,2,3,4}+Q_{0,2,1,3,4}+Q_{0,1,2,4,3}+Q_{0,2,1,4,3}\right)\,,
    \end{split}
\end{equation}
with $p_{0,1,2,3,4}$ and $Q_{0,1,2,3,4}$ given in equations \eqref{eq:P,p,Q} in \ref{appendix:supplementary_material};  finally, $C^{(2,\lambda)}, C^{(3,\lambda)}$ are new tensors that we obtain from the contribution of the tensor $\lambda$ to these Krasitskii resonant tensors, which we have calculated explicitly, and present here for the first time, because there was no such contribution in Krasitskii \cite{krasitskii1994reduced}  due to the fact that $\lambda$ was taken to be identically zero there. We have:
\begin{equation}
    \begin{split}
    \label{eq:C2_C3_lambda}
         {C}^{(2, \lambda)}_{0,1,2,3,4}  &= -\frac{1}{3} \left(A^{(1)}_{0,2,0-2} \lambda_{0-2, 1,3,4} 
         + A^{(1)}_{0,3,0-3} \lambda_{0-3, 1,2,4}
         + A^{(1)}_{0,4,0-4} \lambda_{0-4, 1,2,3}\right.\\
         & \qquad\qquad \left.
         - A^{(1)}_{1,2,1-2} \lambda_{1-2, 0,3,4} 
         - A^{(1)}_{1,3,1-3} \lambda_{1-3, 0,2,4}
         - A^{(1)}_{1,4,1-4} \lambda_{1-4, 0,2,3}
         \right)\,,
         \\
          {C}^{(3,\lambda)}_{4,3,2,1,0}  &= -\frac{1}{2}\left(A^{(1)}_{0,2,0-2} \lambda_{0-2, 1,3,4}
          +A^{(1)}_{1,2,1-2} \lambda_{1-2, 0,3,4}
          +A^{(1)}_{0,3,0-3} \lambda_{0-3, 1,2,4}
          +A^{(1)}_{1,3,1-3} \lambda_{1-3, 0,2,4}\right.\\
          &\qquad\qquad \left.
          -A^{(1)}_{0,4,0-4} \lambda_{0-4, 1,2,3}
          -A^{(1)}_{1,4,1-4} \lambda_{1-4, 0,2,3}
          -2 A^{(1)}_{2+4, 4, 2} \lambda_{2+4, 3, 0, 1}
          -2 A^{(1)}_{3+4, 4, 3} \lambda_{3+4, 2, 0, 1}
          \right)\,.\\
    \end{split}
\end{equation}
With our choice \eqref{eq:C2}--\eqref{eq:C3} of tensors $C^{(2)}$ and $C^{(3)}$, the equations of motion for the $d$-variables contain only non-resonant terms. 
Explicitly, as a result of the canonical transformation  \eqref{eq:d_to_a}, the $d$-variables satisfy the approximate equations of motion
\begin{equation}
\label{eq:d}
i \dot{d_0}  =  \omega_0 d_0 + \sum_{1, 2, 3} \widetilde{T}_{0,1,2,3} d_1^*d_2d_3 \delta_{0,1}^{2,3} +  \sum_{1,2,3,4} \big( {\widetilde{W}_{0,1,2,3,4}} d_1^*d_2d_3d_4 \delta_{0,1}^{2,3,4} - \frac{3}{2} {\widetilde{W}_{4,3,2,1,0}} d_1^*d_2^*d_3d_4 \delta_{0,1,2}^{3,4} \big) + \mathcal{O}(d^5)\,,
\end{equation}
this time with a truncation error $\mathcal{O}(d^5)$ representing a homogeneous polynomial of degree $5$ in the $d$-variables,  with $\widetilde{T}$ given in equation \eqref{eq:T_tilde} and with $\widetilde{W}$ given by
\begin{equation}
\label{eq:W2_tilde}
    \widetilde{W}_{0,1,2,3,4} = 
    \begin{cases}
        -i\,X^{(2)}_{0,1,2,3,4} & \text{if} \quad \Delta_{0+1-2-3-4} = 0\,,\\
        0 & \text{if} \quad \Delta_{0+1-2-3-4} \neq 0\,,
    \end{cases}
\end{equation}
where $X^{(2)}$ is given in equation \eqref{kernel:X2} in \ref{appendix:supplementary_material}. The Hamiltonian in terms of the $d$-variables becomes
\begin{equation}
\label{eq:H_d}
    \frac{H}{N}  = \sum_0 \omega_0 d^*_0 d_0 + \sum_{0,1,2,3} \widetilde{T}_{0,1,2,3} d_0^*d_1^*d_2d_3 \delta_{0,1}^{2,3} +  \sum_{0,1,2,3,4} \dfrac{1}{2} \widetilde{W}_{0,1,2,3,4} \big( d_0^*d_1^*d_2d_3d_4  - d_0d_1d^*_2d_3^*d_4^* \big) \delta_{0,1}^{2,3,4} + \mathcal{O}(d^6)\,
\end{equation}
and the equations of motion \eqref{eq:d} are canonical:
\vspace{-0.2cm}
$$i \dot{d}_k = \frac{1}{N} \dfrac{\partial H}{\partial d^*_k}+  \mathcal{O}(d^5)\,, \qquad  i \dot{d}^*_k = -\frac{1}{N} \dfrac{\partial H}{\partial d_k} +  \mathcal{O}(d^5)\,, \qquad k=1, \ldots, N-1 \,.
$$ 

Notice that, because of the identities
\begin{equation}
    \begin{split}
    \label{eq:W2_W3_tilde}
         \widetilde{W}_{0,1,2,3,4}  &= -i\,X^{(2)}_{0,1,2,3,4} + \Delta_{0+1-2-3-4} C^{(2)}_{0,1,2,3,4}\,,
         \\
          -\frac{2}{3}\widetilde{W}_{4,3,2,1,0}  &= i\,X^{(3)}_{0,1,2,3,4}+ \Delta_{0+1+2-3-4} C^{(3)}_{0,1,2,3,4}\,,
    \end{split}
\end{equation}
when $\Delta_{0+1-2-3-4} = 0$ we have $X^{(2)}_{0,1,2,3,4} = \frac{2}{3} X^{(3)}_{4,3,2,1,0}$ and all the symmetries of $\widetilde{W}$ hold, namely: $X^{(2)}_{0,1,2,3,4} = X^{(2)}_{1,0,2,3,4} = X^{(2)}_{0,1,3,2,4} = X^{(2)}_{0,1,2,4,3}$.

\subsubsection{Exact-resonance evolution equations}
Note how non-resonant triads, quartets and quintets do not completely disappear, as their presence rests within the canonical transformation \eqref{eq:d_to_a}. In contrast, the equations of motion \eqref{eq:d} are  ``exact'' from a resonance point of view: the right-hand side of these equations contains exact-resonance contributions only, namely each nonzero term is parameterised by wavenumbers that satisfy equations \eqref{eq:resonant_condition}. Because of this, we will call these equations ``exact-resonance evolution equations''. The associated weights of the nonlinear terms are collected in the compact tensors $\widetilde{T}, \widetilde{W}$ defined in equations \eqref{eq:T_tilde}, \eqref{eq:W2_tilde}, which are sparse in the sense that most of these coefficients are zero except when the wavenumbers are in exact resonance. We reserve the ``tilde'' notation for such tensors. Eq.\eqref{eq:c} is known as the Zakharov equation \cite{zakharov1968stability} and is extensively used in several works, including FPUT \cite{pistone2019}. Eq.~\eqref{eq:d} instead is somewhat new (due to the fact that we perform extra elimination of non-resonant terms both for $4$-wave and $5$-wave interactions) and has never been explored before in the context of FPUT. 

Another very important aspect, from both conceptual and numerical points of view, is that the ``exact-resonance evolution equations'' \eqref{eq:d} can be mapped to the so-called Heisenberg representation, via the linear (but non-autonomous) transformation $d_0(t) = D_0(t) \exp(-i \omega_0 t)$, which gives the equations of motion
\begin{equation}
\label{eq:d_Heisenberg}
i \dot{D_0}  =  \sum_{1, 2, 3} \widetilde{T}_{0,1,2,3} D_1^*D_2D_3 \delta_{0,1}^{2,3} +  \sum_{1,2,3,4} \big( {\widetilde{W}_{0,1,2,3,4}} D_1^*D_2D_3D_4 \delta_{0,1}^{2,3,4} - \frac{3}{2} {\widetilde{W}_{4,3,2,1,0}} D_1^*D_2^*D_3D_4 \delta_{0,1,2}^{3,4} \big) + \mathcal{O}(D^5)\,,
\end{equation}
namely the $D$-variables satisfy an autonomous system, containing only resonances, but the leading term in the equations of motion is cubic (as opposed to linear, as in the case of the $d$-variables), so the $D$-variables are very slow when compared with the $d$-variables. Numerically, this allows us to apply numerical forward-integration methods with very large time steps, obtaining very accurate solutions while reducing the wall time required for long-time studies such as equipartition/thermalisation and hyperchaos characterisation. Conceptually, the fact that these new equations are still autonomous follows directly from the fact that each term in these equations correspond to an exact resonance. For example, terms like $\exp(-i (\omega_2+\omega_3+\omega_4-\omega_1) t)\delta_{0,1}^{2,3,4}$ reduce to $\exp(i\omega_0  t)\delta_{0,1}^{2,3,4}$, which allows for this result. Moreover, while the transformation from the $d$-variables to the $D$-variables is not canonical, one can verify that system \eqref{eq:d_Heisenberg} is indeed canonical:
\vspace{-0.2cm}
$$i \dot{D}_k = \frac{1}{N} \dfrac{\partial H_D}{\partial D^*_k}+  \mathcal{O}(D^5)\,, \qquad  i \dot{D}^*_k = -\frac{1}{N} \dfrac{\partial H_D}{\partial D_k} +  \mathcal{O}(D^5)\,, \qquad k=1, \ldots, N-1 \,,
$$ 
with Hamiltonian 
$$\frac{H_D}{N}  = \sum_{0,1,2,3} \widetilde{T}_{0,1,2,3} D_0^*D_1^*D_2D_3 \delta_{0,1}^{2,3} +  \sum_{0,1,2,3,4} \dfrac{1}{2} \widetilde{W}_{0,1,2,3,4} \big( D_0^*D_1^*D_2D_3D_4  - D_0D_1D^*_2D_3^*D_4^* \big) \delta_{0,1}^{2,3,4} + \mathcal{O}(D^6)\,.$$

In the next section we will verify the validity of the approximate equations \eqref{eq:b}, \eqref{eq:c} and \eqref{eq:d} via direct comparisons with the original system \eqref{eq:a_k}. 

\section{Validation of the exact-resonance evolution equations for $b, c$ and $d$}
\label{sec:FPUT_comp}

In order to validate the exact-resonance evolution equations \eqref{eq:b}--\eqref{eq:d}, we will consider the following approach. By truncating these equations, namely by discarding their respective truncating errors, we obtain a family of models.

Numerical simulations of these models for the $b, c$ and $d$ variables are evolved using a mixed Runge-Kutta time scheme \cite{grant2022perturbed}. Tensor multiplication is employed to update the expressions by building multi-dimensional array-like objects so that element-wise multiplication is systematically applied without the need for nested loops. 
As for the evolution equations \eqref{eq:a_k} for the $a$ variables, we use the same Runge-Kutta scheme, with the same parameters and time steps, as for the $b,c$ and $d$ variables, so the comparison described below depends only on the ``size'' of the solution and not on extra parameters such as the time step chosen.

\subsection{Preliminary definitions: energy norm and generic initial condition}
In the validation studies below we will use the following definitions. First, given a state in the $a$-variables, namely the vector $\textbf{a} := (a_1, \ldots, a_{N-1})^T \in \mathbb{C}^{N-1}$, we define the energy norm of $\textbf{a}$ by
\begin{equation}
\label{eq:norm_def}
    \lVert \textbf{a} \rVert := \sqrt{\sum_{k=1}^{N-1}\omega_k a_k^* a_k }\,.
\end{equation}
Second, we define a type of ``generic'' initial condition that we will be using for the scaling method and for the study of Lyapunov exponents:
\begin{equation}
    \label{eq:IC_f}
    f_k =  \frac{N}{(N-1)  \left(N-1+k\right)\sqrt{\omega_k}} \,, \qquad k=1,\ldots, N-1\,,
\end{equation}
The idea is that the ``energy'' at wavenumber $k$, $\omega_k \left|f_k\right|^2$, is not symmetric under $k\to N-k$, which makes the system's dynamics more interesting. We have $\omega_k \left|f_k\right|^2 \propto (1+(k-1)/N)^{-2}$, so the energy is preferentially distributed at low wavenumbers: at $k=1$ the energy is about 4 times larger than at $k=N-1$.

\subsection{The scaling method to check the exact-resonance evolution equation's truncation error}

\begin{figure}[!htbp]
    \centering
    \includegraphics[width=0.49 \textwidth]{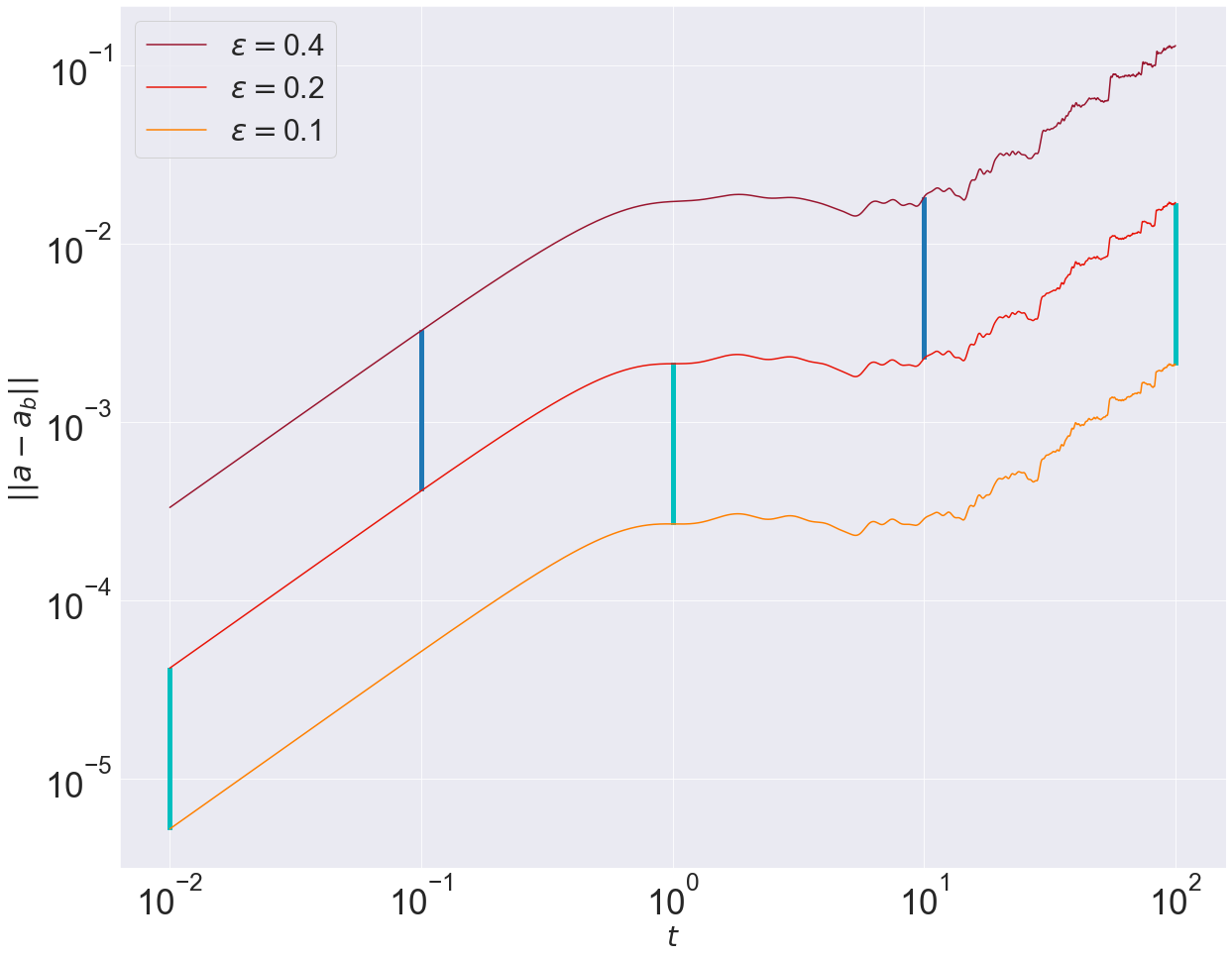}
     \includegraphics[width=0.49 \textwidth]{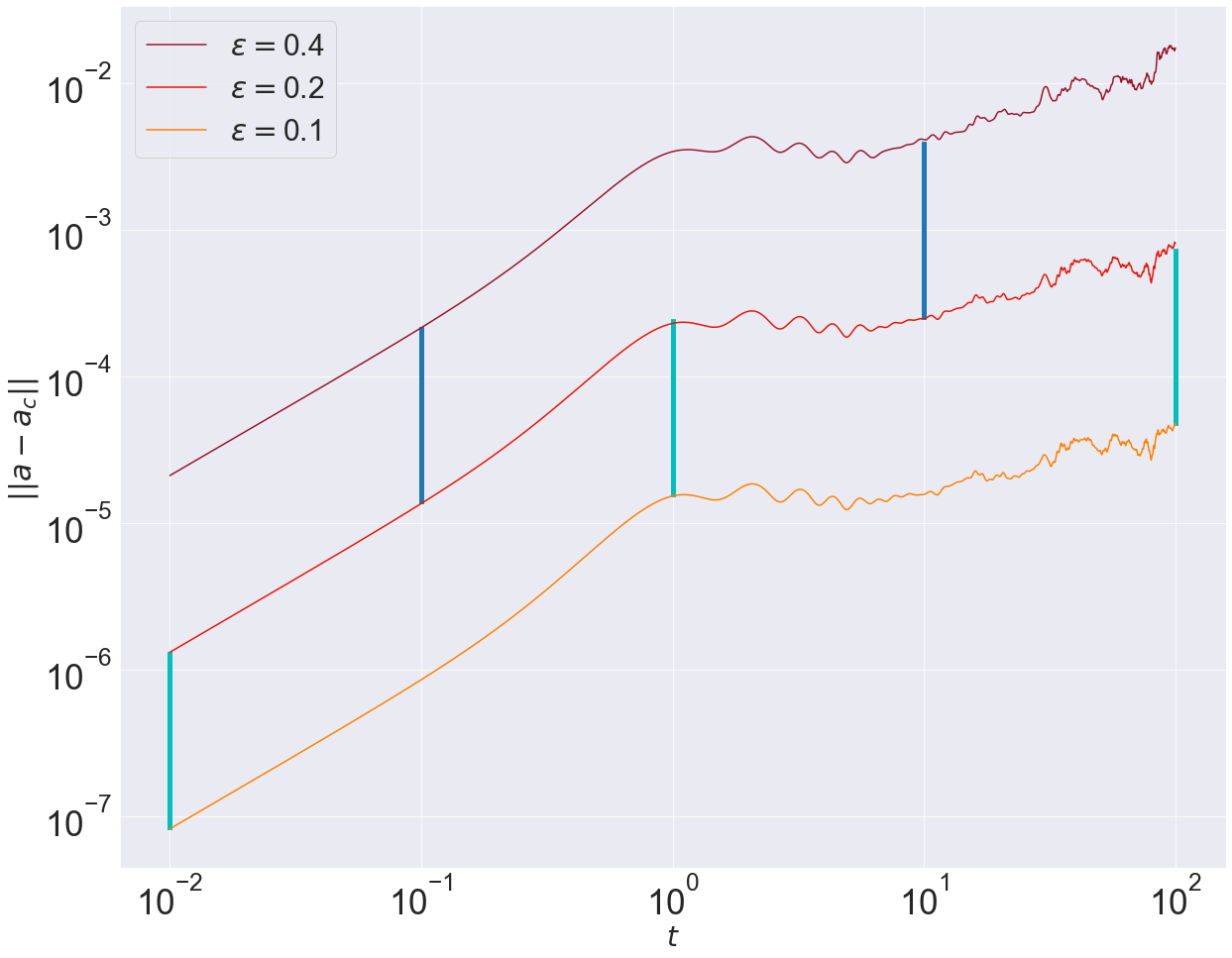}  
     \includegraphics[width=0.49 \textwidth]{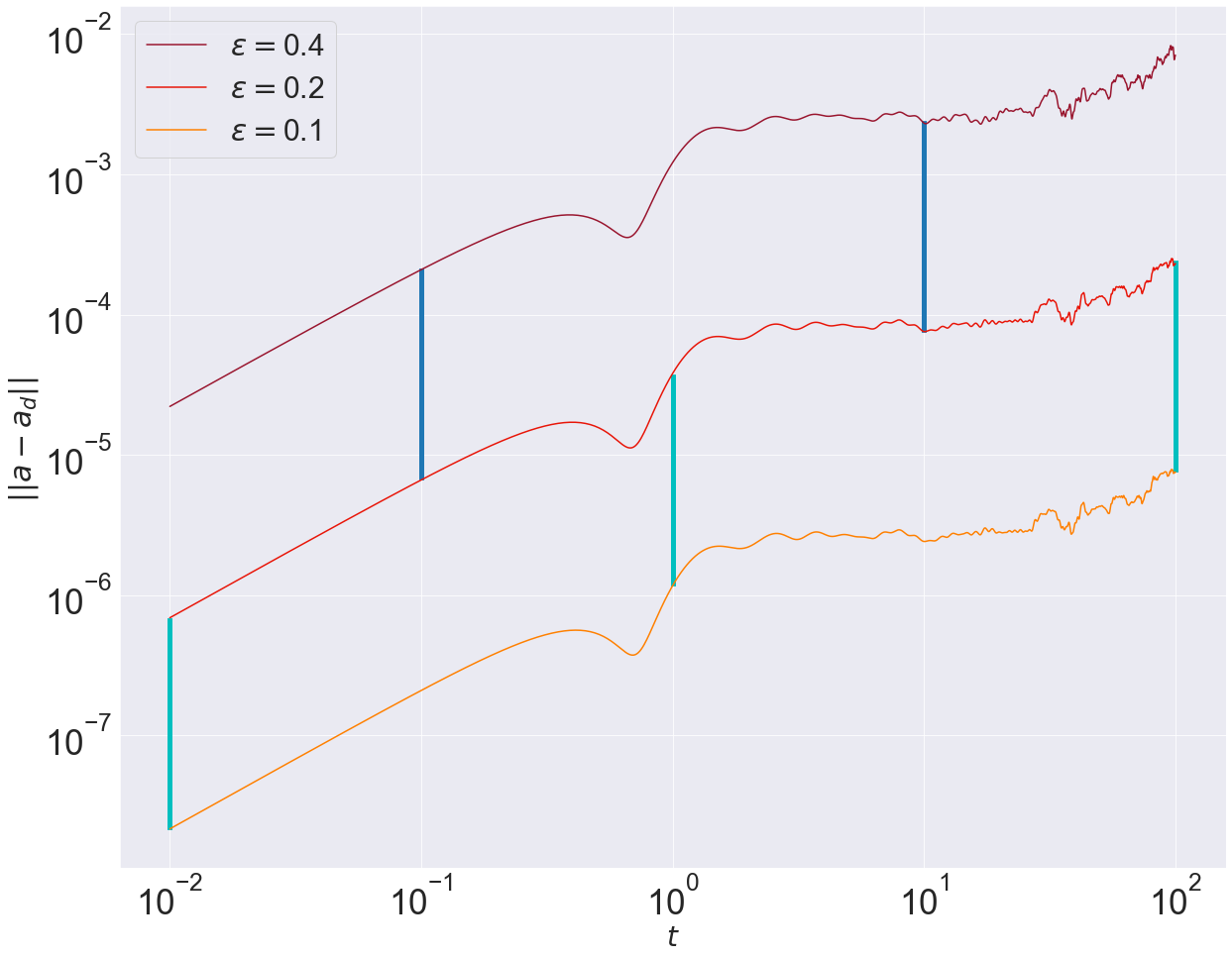}    \caption{Numerical implementation of the scaling method, case $N=9$. (a) Time series of the energy norm of the difference between the mapped solution $a_b$ (based on truncated equations \eqref{eq:b} and mapping \eqref{eq:b_to_a}) and the solution $a$ of equations \eqref{eq:a_k}, for initial conditions \eqref{eq:IC_b}, \eqref{eq:IC_f}, with $\epsilon=0.1$ (lighter plot), $\epsilon=0.2$ (dark plot), $\epsilon=0.4$ (darker plot). The vertical bars start at their low end from one of these curves, and go upwards to a value $2^3$ times the value of the abscissa at their low end. (b) Similar to panel (a), but comparing $a$ with $a_c$ (based on truncated equations \eqref{eq:c} and mapping \eqref{eq:c_to_a}), with initial conditions \eqref{eq:IC_c}, \eqref{eq:IC_f}, and the vertical lines going up a factor of $2^4$ times the value of the abscissa at their low end. (c) Similar to panel (a), but comparing $a$ with $a_d$ (based on truncated equations \eqref{eq:d} and mapping \eqref{eq:d_to_a}), with initial conditions \eqref{eq:IC_d}, \eqref{eq:IC_f}, and the vertical lines going up a factor of $2^5$ times the value of the abscissa at their low end.  \label{fig:scaling}}
\end{figure}

We introduce the ``scaling method'' to verify numerically the truncation error present in the exact-resonance evolution equations \eqref{eq:b}--\eqref{eq:d}. We illustrate the method via an example. After discarding the truncation error in equation \eqref{eq:b} we get the simple model $i \dot{b}_0 = \omega_0 b_0$, which has simple solutions (notice that, because exact solutions are generally not available in the case of the $c$- and $d$-variables, we will always refer to accurate numerical solutions rather than exact solutions). Then, applying the transformation \eqref{eq:b_to_a} will produce, by definition, a solution to equation \eqref{eq:a_k}, but with a RHS consisting of just the linear and quadratic terms in it, plus a cubic truncation error $\mathcal{O}(a^3)$, inherited from the $\mathcal{O}(b^3)$ truncation error in equation \eqref{eq:b} that we had discarded. Thus, we can compare the ``reference'' numerical solution $\textbf{a}(t)$ of the original system \eqref{eq:a_k} against the ``mapped'' solution $\textbf{a}_b(t)$ obtained by mapping, via the transformation \eqref{eq:b_to_a}, the solution to the truncated version of equation \eqref{eq:b}. These two time series should differ by a quantity of $\mathcal{O}(a^3)$ as per the above analysis. In terms of the norm defined in equation \eqref{eq:norm_def}, this is
\begin{equation}
    \label{eq:scaling_b}
    \lVert \textbf{a}(t) - \textbf{a}_b(t)\rVert \propto \lVert \textbf{a}(t)\rVert^3 \qquad \text{for all}\qquad t \in [0,T)\,,  
\end{equation}
where $T$ is reasonably small (but finite), such that $\lVert \textbf{a}(t)\rVert$ remains small enough so that the transformation \eqref{eq:b_to_a} remains being a near-identity transformation. Now, to test quantitatively the scaling \eqref{eq:scaling_b}, we impose the following initial condition for the $b$-variables:
\begin{equation}
    \label{eq:IC_b}
    b_k(0) = \epsilon f_k\,,\quad k=1, \ldots, N-1\,,
\end{equation}
where $\epsilon$ is a positive parameter and $f_k$ is defined in \eqref{eq:IC_f}. Then, the scaling \eqref{eq:scaling_b} becomes

\begin{equation}
    \label{eq:scaling_b_test}
    \lVert \textbf{a}(t) - \textbf{a}_b(t)\rVert \propto \epsilon^3 g_b(t)\qquad \text{for all}\qquad t \in [0,T)\,,
\end{equation}
where $g_b(t)$ is an unspecified positive function that does not depend on the parameter $\epsilon$. We perform three numerical tests of relation \eqref{eq:scaling_b_test}, taking $N=9$ (so there will be non-trivial resonant quintets in the $d$-variables), using the initial conditions \eqref{eq:IC_b} with three respective values $\epsilon = 0.1, 0.2, 0.4$, clearly related by factors of $2$. The resulting distances $\lVert \textbf{a}(t) - \textbf{a}_b(t)\rVert$ are plotted in 
figure \ref{fig:scaling}(a), in a log-log plot for ease of visualisation. These clearly verify the relation \eqref{eq:scaling_b_test}, as graphically supported by the vertical lines in the plot corresponding to a scaling factor of $8$, namely corresponding to $2^3$, which is the expected scaling as per the RHS of the relation \eqref{eq:scaling_b_test}. The apparent linear growth in time of these distances is basically due to the fact that this distance is obtained from the time integral of the $\mathcal{O}(a^3)$ terms that were discarded.

We repeat this study for the more complicated equation  \eqref{eq:c}, which includes the resonant quartet interactions but discards higher orders. The scaling method in this case compares the numerical solution $\textbf{a}(t)$ of the original system \eqref{eq:a_k} against the mapped solution $\textbf{a}_c(t)$, obtained by mapping via the transformation \eqref{eq:c_to_a} the solution $\textbf{c}(t)$ to the truncated version of equation \eqref{eq:c}. These two time series should differ by a quantity of $\mathcal{O}(a^4)$ as per the above analysis, namely if we take the initial conditions
\begin{equation}
    \label{eq:IC_c}
    c_k(0) = \epsilon f_k\,,\quad k=1, \ldots, N-1\,,
\end{equation}
then we expect the following scaling:
\begin{equation}
    \label{eq:scaling_c_test}
    \lVert \textbf{a}(t) - \textbf{a}_c(t)\rVert \propto \epsilon^4 g_c(t)\qquad \text{for all}\qquad t \in [0,T)\,,
\end{equation}
where $g_c(t)$ is a positive function that does not depend on  $\epsilon$. We perform three numerical tests of relation \eqref{eq:scaling_c_test}, taking $N=9$ again, using the initial conditions \eqref{eq:IC_c} with three respective values $\epsilon = 0.1, 0.2, 0.4$. The resulting distances $\lVert \textbf{a}(t) - \textbf{a}_c(t)\rVert$ are plotted in 
figure \ref{fig:scaling}(b), clearly verifying the relation \eqref{eq:scaling_c_test}, as graphically supported by the vertical lines in the plot corresponding to a scaling factor of $16$, namely corresponding to $2^4$, as predicted by the  RHS of  \eqref{eq:scaling_c_test}.

Finally, we repeat the study for the even more complicated equation \eqref{eq:d}, which includes all resonant quintets and quartets. The scaling method in this case compares the numerical solution $\textbf{a}(t)$ of the original system \eqref{eq:a_k} against the mapped solution $\textbf{a}_d(t)$, obtained by mapping via the transformation \eqref{eq:d_to_a} the solution $\textbf{d}(t)$ to the truncated version of equation \eqref{eq:d}. These two time series should differ by a quantity of $\mathcal{O}(a^5)$ as per the above analysis, so if we take the initial conditions
\begin{equation}
    \label{eq:IC_d}
    d_k(0) = \epsilon f_k\,,\quad k=1, \ldots, N-1\,,
\end{equation}
then we expect the following scaling:
\begin{equation}
    \label{eq:scaling_d_test}
    \lVert \textbf{a}(t) - \textbf{a}_d(t)\rVert \propto \epsilon^5 g_d(t)\qquad \text{for all}\qquad t \in [0,T)\,,
\end{equation}
where $g_d(t)$ is a positive function that does not depend on  $\epsilon$. We perform three numerical tests of relation \eqref{eq:scaling_d_test}, taking $N=9$ again, using the initial conditions \eqref{eq:IC_d} with three respective values $\epsilon = 0.1, 0.2, 0.4$. The resulting distances $\lVert \textbf{a}(t) - \textbf{a}_d(t)\rVert$ are plotted in 
figure \ref{fig:scaling}(c), clearly verifying the relation \eqref{eq:scaling_d_test}, as graphically supported by the vertical lines in the plot corresponding to a scaling factor of $32$, namely corresponding to $2^5$, as predicted by the  RHS of  \eqref{eq:scaling_d_test}.

With these studies, the evolution equations \eqref{eq:b}, \eqref{eq:c} and \eqref{eq:d}, along with the transformations \eqref{eq:b_to_a}, \eqref{eq:c_to_a} and \eqref{eq:d_to_a}, can be considered validated. This means that both the analytical formulae for the tensors involved and their numerical implementation are validated.

\subsection{Convergence of the normal-form transformation}
\label{subsec:convergence}

Focusing now on the deepest model, namely the evolution equation \eqref{eq:d} for the $d$-variables, one of the pivotal goals of our work is to show that this model, via its $4$-wave and $5$-wave resonant interactions, is able to capture the dynamics of the original FPUT system \eqref{eq:a_k}. In order to do this, the first question we need to answer is whether the normal-form transformation \eqref{eq:d_to_a} keeps the order between the terms involved, at the amplitudes that the system explores throughout its dynamics. We will say that the normal-form transformation \eqref{eq:d_to_a} \emph{converges} if the sizes of the subsequent terms are ordered. Namely, let us write \eqref{eq:d_to_a} in the form
$$\textbf{a} = \textbf{d} + \textbf{Y}^{(2)}(\textbf{d}) + \textbf{Y}^{(3)}(\textbf{d}) + \textbf{Y}^{(4)}(\textbf{d})$$
where, for each $j=2,3,4$, $\textbf{Y}^{(j)}(\textbf{d})$ is a vector of homogeneous polynomials of degree $j$ in the $d$-variables, explicitly given by
\begin{eqnarray}    
\label{eq:Y2}
{Y}_0^{(2)}(\textbf{d}) \!\!\!\!&=&\!\!\!\! \sum_{1, 2} \big( A^{(1)}_{0,1,2}d_1d_2 \delta_0^{1,2}+ A^{(2)}_{0,1,2}d_1^*d_2 \delta_{0,1}^{2} + A^{(3)}_{0,1,2}d_1^*d_2^* \delta_{0,1,2} \big)\,,\\
\label{eq:Y3}
{Y}_0^{(3)}(\textbf{d})\!\!\!\! &=&\!\!\!\!
        \sum_{1,2,3} \big( B^{(1)}_{0,1,2,3}d_1d_2d_3 \delta_0^{1,2,3} +B^{(2)}_{0,1,2,3}d^*_1d_2d_3 \delta_{0,1}^{2,3} + {B^{(3)}_{0,1,2,3}d_1^*d_2^*d_3 \delta_{0,1,2}^{3}} +  B^{(4)}_{0,1,2,3}d_1^*d_2^*d_3^* \delta_{0,1,2,3} \big)\,, \qquad \\
\nonumber
{Y}_0^{(4)}(\textbf{d}) \!\!\!\!&=& \!\!\!\!
 \sum_{1,2,3,4} \big( C^{(1)}_{0,1,2,3,4}d_1d_2d_3d_4 \delta_0^{1,2,3,4} + C^{(2)}_{0,1,2,3,4}d_1^*d_2d_3d_4 \delta_{0,1}^{2,3,4} +  C^{(3)}_{0,1,2,3,4}d_1^*d_2^*d_3d_4 \delta_{0,1,2}^{,3,4} 
        \\
\label{eq:Y4}        
        & &\qquad \qquad + C^{(4)}_{0,1,2,3,4}d_1^*d_2^*d_3^*d_4 \delta_{0,1,2,3}^{4} +  C^{(5)}_{0,1,2,3,4}d_1^*d_2^*d_3^*d_4^* \delta_{0,1,2,3,4} 
        \big)\,.
\end{eqnarray}
Then, we say that \eqref{eq:d_to_a} converges if the order relations $\lVert\textbf{d}\rVert >  \lVert\textbf{Y}^{(2)}(\textbf{d})\rVert > \lVert\textbf{Y}^{(3)}(\textbf{d}) \rVert > \lVert\textbf{Y}^{(4)}(\textbf{d})\rVert$ hold, where $\lVert \cdot\rVert$ is the energy norm defined in \eqref{eq:norm_def}.

\begin{figure}[h!]
    \centering
    \includegraphics[width=0.47\textwidth]{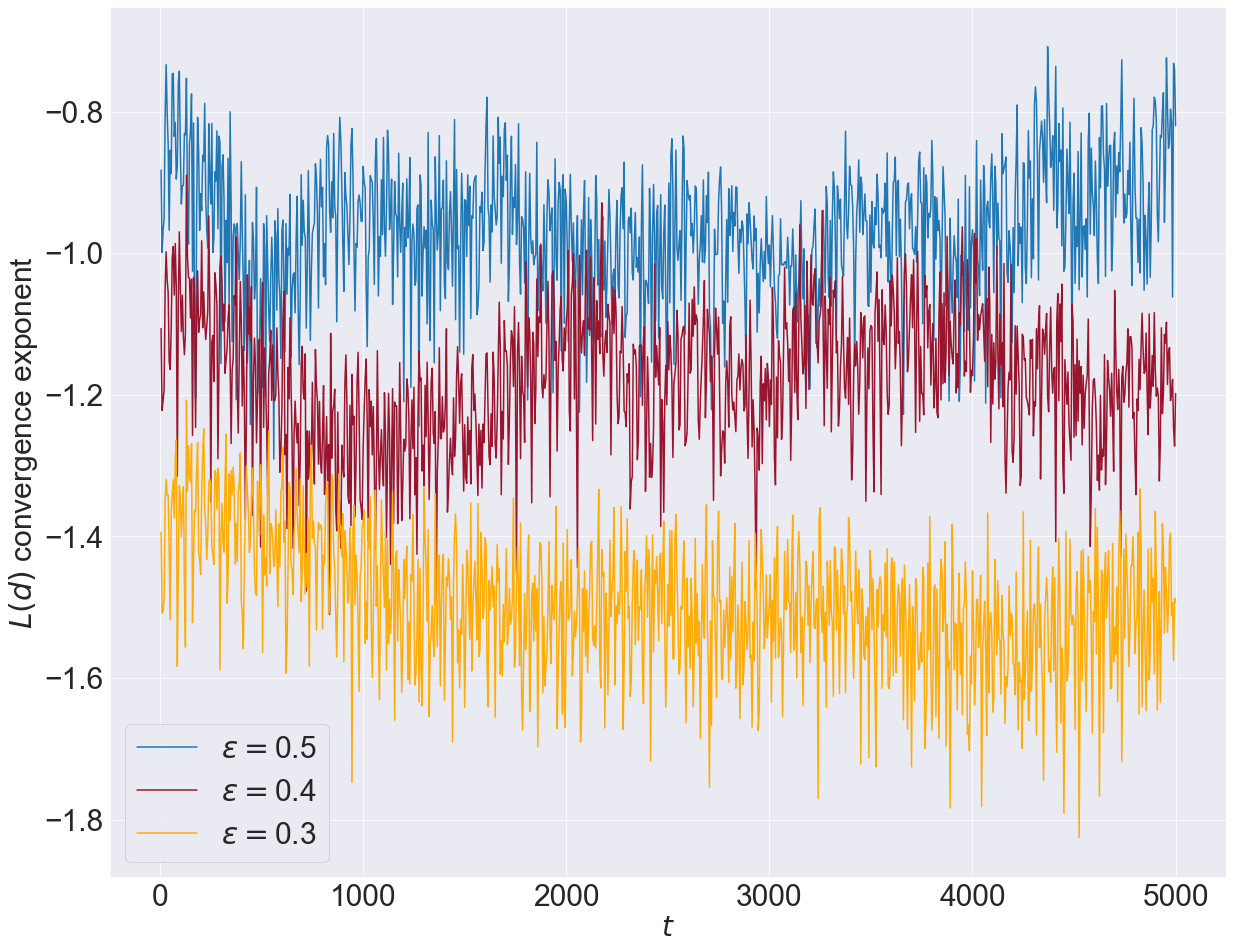}
    \hfill
        \includegraphics[width=0.47\textwidth]{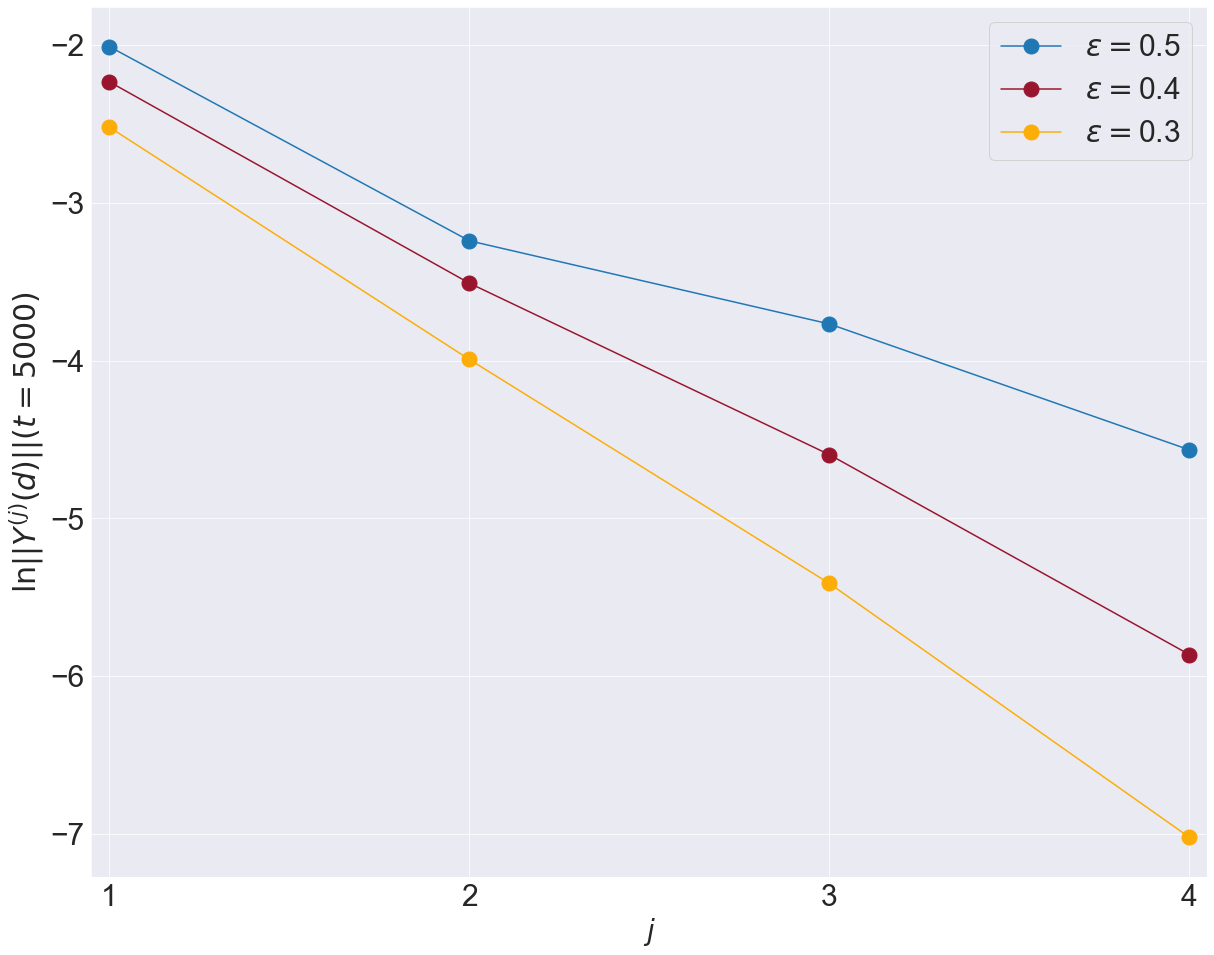}
    \caption{For the case $N=9$, for initial conditions \eqref{eq:IC_d}, \eqref{eq:IC_f} with $\epsilon=0.3$ (lower curve), $\epsilon=0.4$ (middle curve) and $\epsilon=0.5$ (upper curve). (a) Time series of the convergence exponent $L(\textbf{d})$, equation \eqref{eq:L_fit}. (b) At $t=5000$, values of the logarithms of the energy norms of the successive terms \eqref{eq:Y2}--\eqref{eq:Y4} in the normal-form transformation from $d$- to $a$-variables. \label{fig:L_conv}}
    
\end{figure}

It goes without saying that if the normal-form transformation \eqref{eq:d_to_a} does not converge under this definition then the models \eqref{eq:a_k} and \eqref{eq:d} are expected to provide divergent results.
Thus, it is important to quantify the relative sizes involved in the above order relations. Clearly, in the $2(N-1)$-dimensional space where the $d$-variables evolve, this convergence will be worse far from the origin and better close to the origin. We introduce a practical method to numerically assess the quality of this convergence, based on ideas from \cite{walsh2020convergence}. Denoting $\textbf{Y}^{(1)}(\textbf{d}) := \textbf{d}$, it is natural to attempt a fit of the form 
\begin{equation}
\label{eq:fit_LM}
    \lVert \textbf{Y}^{(j)}(\textbf{d})\rVert_{\text{fit}} = \exp(L j + M)\,,\qquad j=1,2,3,4,  
\end{equation}
and the relevant fit parameter is $L$: convergence, in the sense defined above, requires $L<0$. Notice that, by looking at the above fit, the quantity $\exp(L)$ can be thought of as the ``level of nonlinearity'', namely a dimensionless measure of how large the amplitudes are. In practice, using least squares on the logarithm of the above fit, we obtain a running estimate of $L$ as a function of the state $\textbf{d}$:
\begin{equation}
    \label{eq:L_fit}
    L(\textbf{d}) := \sum_{j=1}^4 (j/5-1/2)\ln \lVert \textbf{Y}^{(j)}(\textbf{d})\rVert\,.
\end{equation}
A value of $L(\textbf{d}) = -1$ indicates, via the fit \eqref{eq:fit_LM}, that the size of the last term $\lVert \textbf{Y}^{(4)}(\textbf{d})\rVert$ is about $5\%$ of the size of the first term $\lVert \textbf{d}\rVert$.  
In figure \ref{fig:L_conv}(a), we show running estimates of this convergence exponent $L(\textbf{d})$, using again $N=9$ and the initial condition \eqref{eq:IC_d} but for other three values of the scaling parameter: $\epsilon = 0.3, 0.4, 0.5$. This time we run the simulations until $t=5000$, to show how the exponent varies over time. It is evident that the case $\epsilon = 0.3$ has a very good convergence exponent, well below $-1.2$, which implies, via the fit \eqref{eq:fit_LM},  a relative-error relation $\lVert \textbf{Y}^{(4)}(\textbf{d})\rVert < 3\% \lVert \textbf{d}\rVert$ throughout the simulation. As the initial condition increases in size, the convergence exponent worsens. For $\epsilon = 0.4$ it reaches maximum values of about $-0.89$, which corresponds to $\lVert \textbf{Y}^{(4)}(\textbf{d})\rVert < 7\% \lVert \textbf{d}\rVert$. This relative error rises to $12\%$ in the case $\epsilon=0.5$. In figure \ref{fig:L_conv}(b) we show, at the final time $t=5000$, the actual values of $\ln \lVert \textbf{Y}^{(j)}(\textbf{d})\rVert$ used to get the slopes $L(\textbf{d})$. It is evident from the plot that the case $\epsilon = 0.5$ (and also the case $\epsilon=0.4$ to some extent) shows a more wild behaviour, not a clean linear decay like the $\epsilon=0.3$ case. 

Although technically a negative exponent above $-1$ could still indicate convergence if we had higher-order terms to verify it, at this level of approximation an error of $7\%$ obviously matters, and more so when it compounds over time. However, the question about the amplitudes (represented by the prefactor $\epsilon$) at which the behaviour of the $d$-system begins to be different in nature from the behaviour of the original $a$-system will be tackled when we study the Lyapunov spectrum of the respective systems in section \ref{sec:FPUT_theo}.

\subsection{Time series analysis: time scales of resonant quartets and quintets}

In this subsection we want to show in more detail the dynamical signature of $4$-wave and $5$-wave resonances. The focus of this article is on $5$-wave resonances, as $4$-wave resonances alone are known to be integrable, so we look at the case $N=9$, where only resonant quintets are able to exchange energy. 

\begin{figure}[h!]
    \centering
\includegraphics[width=0.47\textwidth]{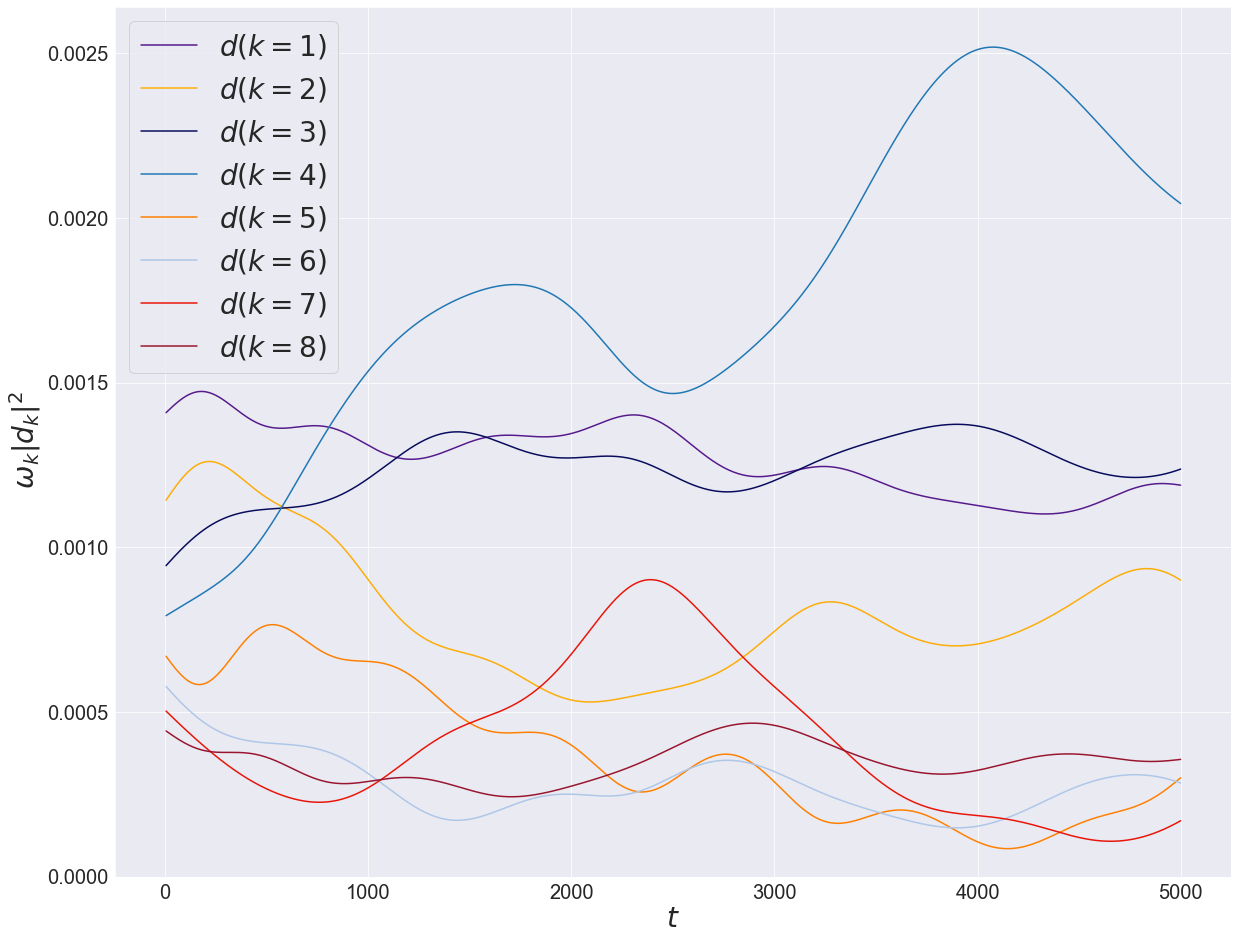}
    \hfill
        \includegraphics[width=0.47\textwidth]{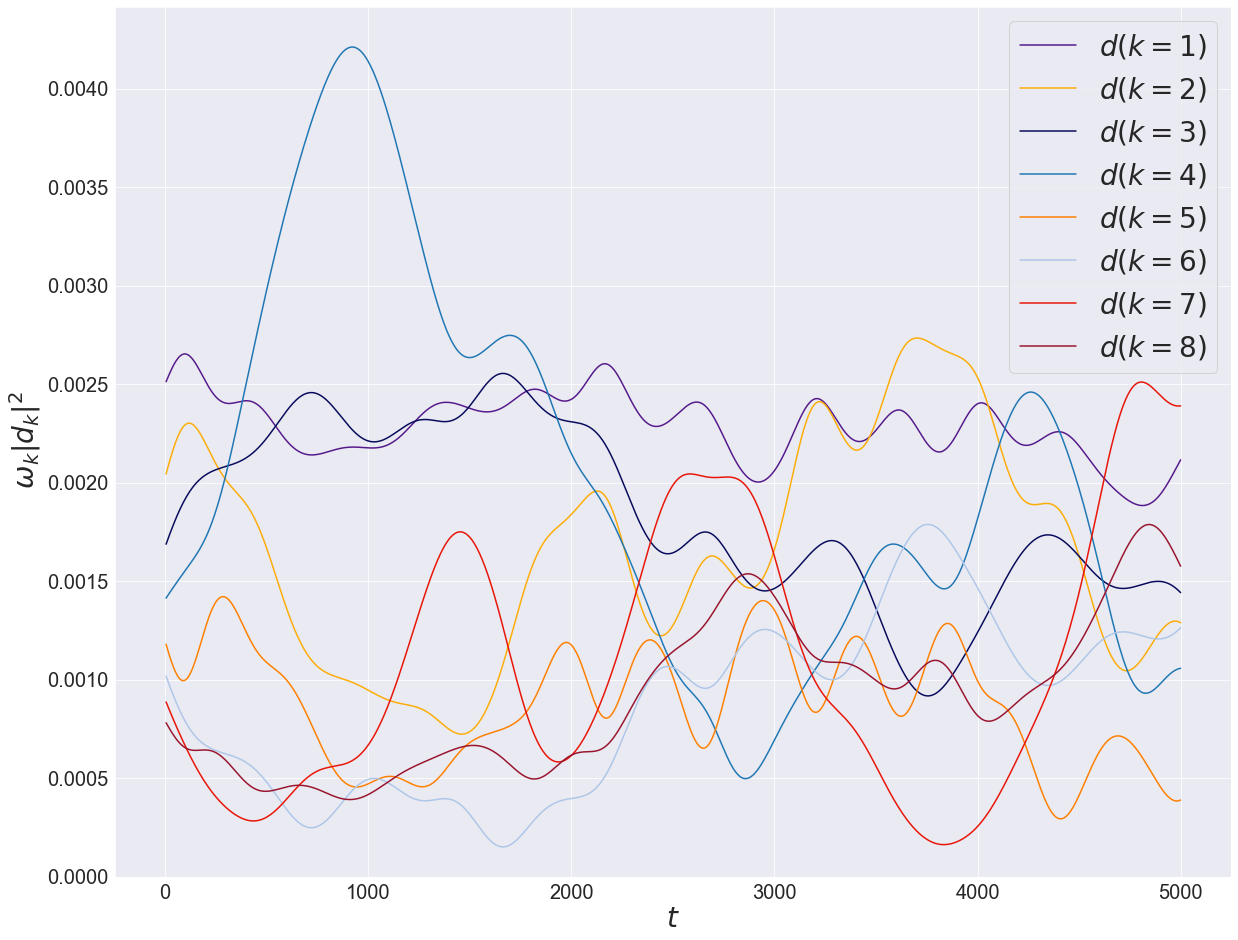}
    \caption{Time evolution of individual energies in the $d$-variables for initial conditions \eqref{eq:IC_d}, \eqref{eq:IC_f} with $\epsilon = 0.3$ (a) and $\epsilon = 0.4$ (b).  \label{fig:time_series_chaos}}    
\end{figure}

To begin with, we present the time evolution of the solution to equation \eqref{eq:d}, focusing on the individual mode energies $\omega_k |d_k(t)|^2, \,\,\, k=1, \ldots, 8$, in terms of two of the  initial conditions considered in the last subsection: 
initial condition \eqref{eq:IC_d}, \eqref{eq:IC_f} with $\epsilon = 0.3$ (figure \ref{fig:time_series_chaos}(a)) and $\epsilon = 0.4$ (figure \ref{fig:time_series_chaos}(b)). Qualitatively, it is apparent from these plots that there is chaos in the system (to be quantified thoroughly in section \ref{sec:FPUT_theo}) and some modes' energies can have high-amplitude variations. All these features are enhanced when the initial condition is larger ($\epsilon=0.4$) (i.e. when the level of nonlinearity, or simply  ``nonlinearity'', is larger).

\begin{figure}[h!]
    \centering
    \includegraphics[width=0.47\textwidth]{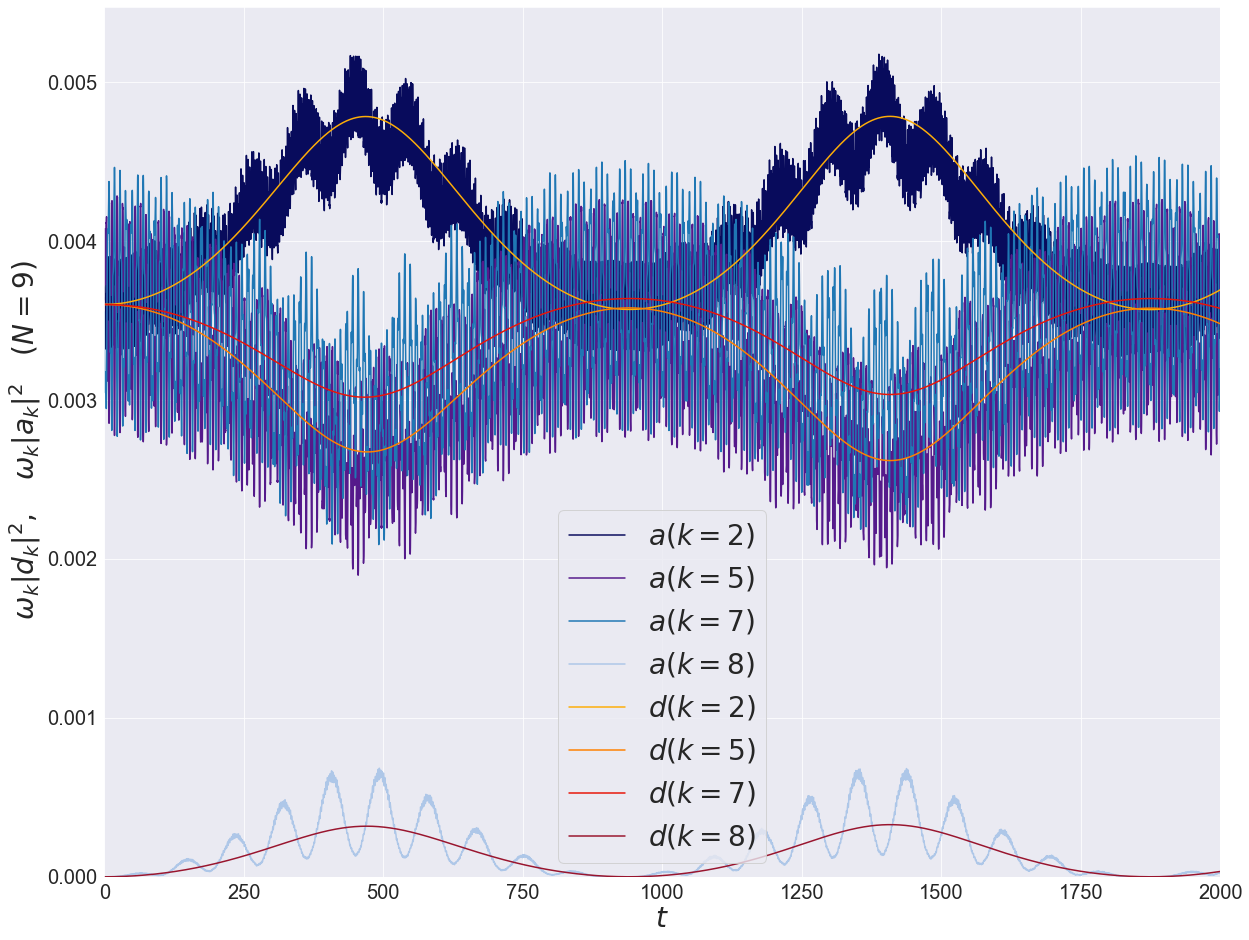}
    \includegraphics[width=0.47\textwidth]{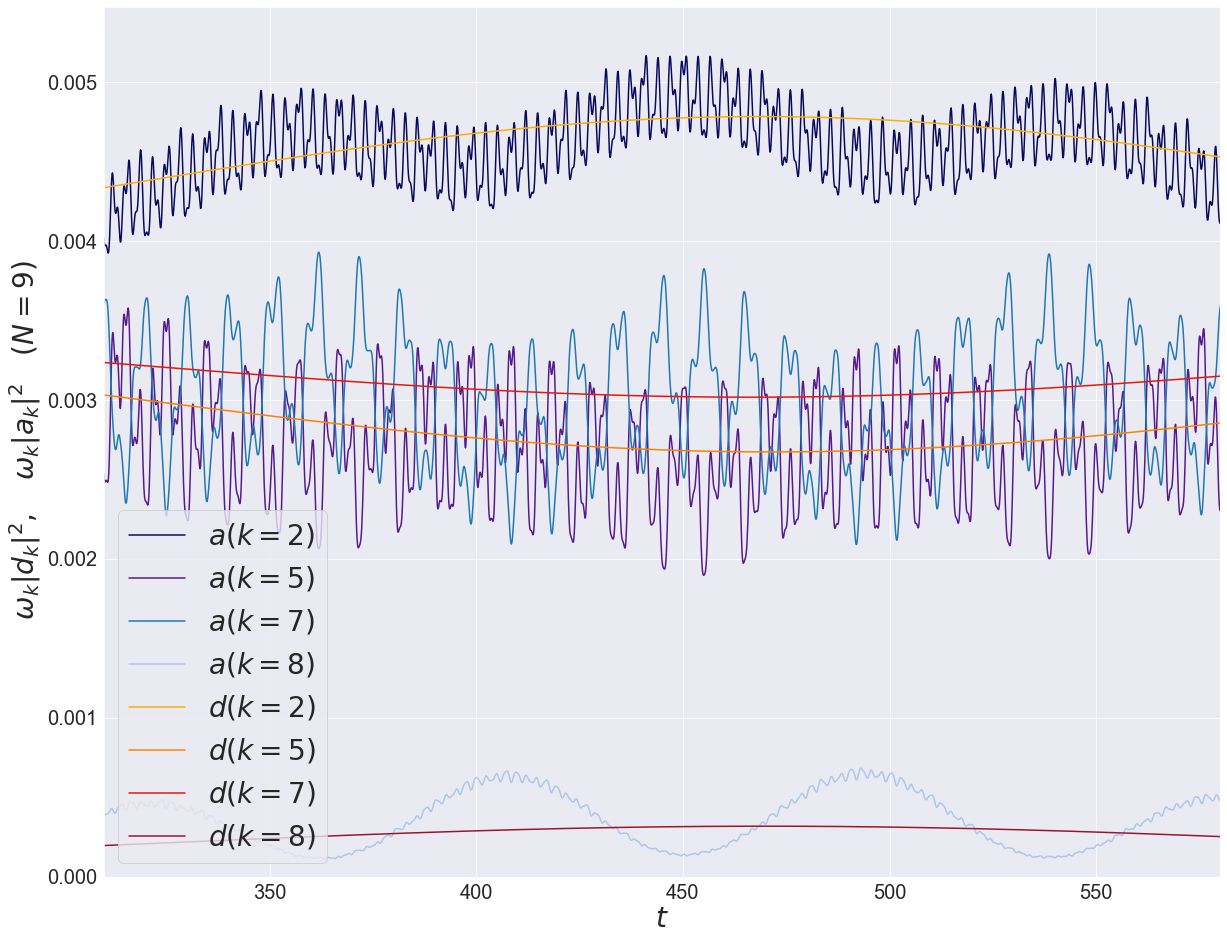}
    \caption{(a) Time evolution of individual energies in a resonant quintet (case $N=9$) stemming from initial conditions \eqref{eq:IC_quintet}. The simulations used time steps $dt=0.1$ for the $a$-variables and $dt=1$ for the $d$-variables. (b) Zoom-in of (a) over the time range $[310, 580]$.
    \label{fig:time_series_superimposed}}
\end{figure}

The above analysis is based on too generic initial conditions. To provide a more heuristic proof that $5$-wave resonances have an effect on the dynamics of this system, we resort to more specific initial conditions (still in the case $N=9$), based on a single resonant quintet, and initialise the system so only modes in that quintet have energy initially. In this way the energy transfer will be  clearly due to the resonant interaction. In this case, we focus on the resonant quintet $\{2,2,8;5,7\}$ (one of the eight accessible quintets in the Single Octahedron cluster \cite{bustamante2019exact}) and explore the situation when energy is initially distributed uniformly, but only to modes with $k=2, 5$ and $7$:
\begin{equation}
\label{eq:IC_quintet}
\omega_2 |d_2(0)|^2 = \omega_5 |d_5(0)|^2 = \omega_7 |d_7(0)|^2 = 0.06\,, 
\end{equation}
while $d_k(0)=0$ for $k =1,3,4,6,8$. 
In this way, the $5$-wave resonance should produce a sustained energy growth in the mode $k=8$. That is exactly what we observe in figure \ref{fig:time_series_superimposed}. That figure shows the energies $\omega_k |d_k|^2$ of the modes $k=2,5,7,8$ (involved in the selected quintet resonance), as a result of the simulation of system \eqref{eq:d}. It is evident that a quasi-periodic behaviour is observed, characterised by energy exchanges amongst the modes in the quintet, and as predicted the mode with $k=8$ receives a significant share of the energy in any of the other modes. To this plot are superimposed on the same plot the corresponding energies $\omega_k |a_k|^2$ of the same modes, but as a result of simulating the system \eqref{eq:a_k} with the mapped initial conditions via \eqref{eq:d_to_a}. This illustrates two things: first,  a hierarchy of time scales is evident in the form of envelopes, going from the slowest one (quintets, of period $T_5 \approx 900$) to the quartet-interaction time scale (of period $T_4 \approx 90$), to the triad-interaction time scale (period $T_3 \approx 9$). Second, the fact that the $a$-variables follow the $d$-variables indicates that the nonlinearity is relatively small: in fact, the convergence exponent $L$ achieves a maximum value of $-1.34$ throughout the simulation.

\begin{figure}[h!]
    \centering
    \includegraphics[width=0.47\textwidth]{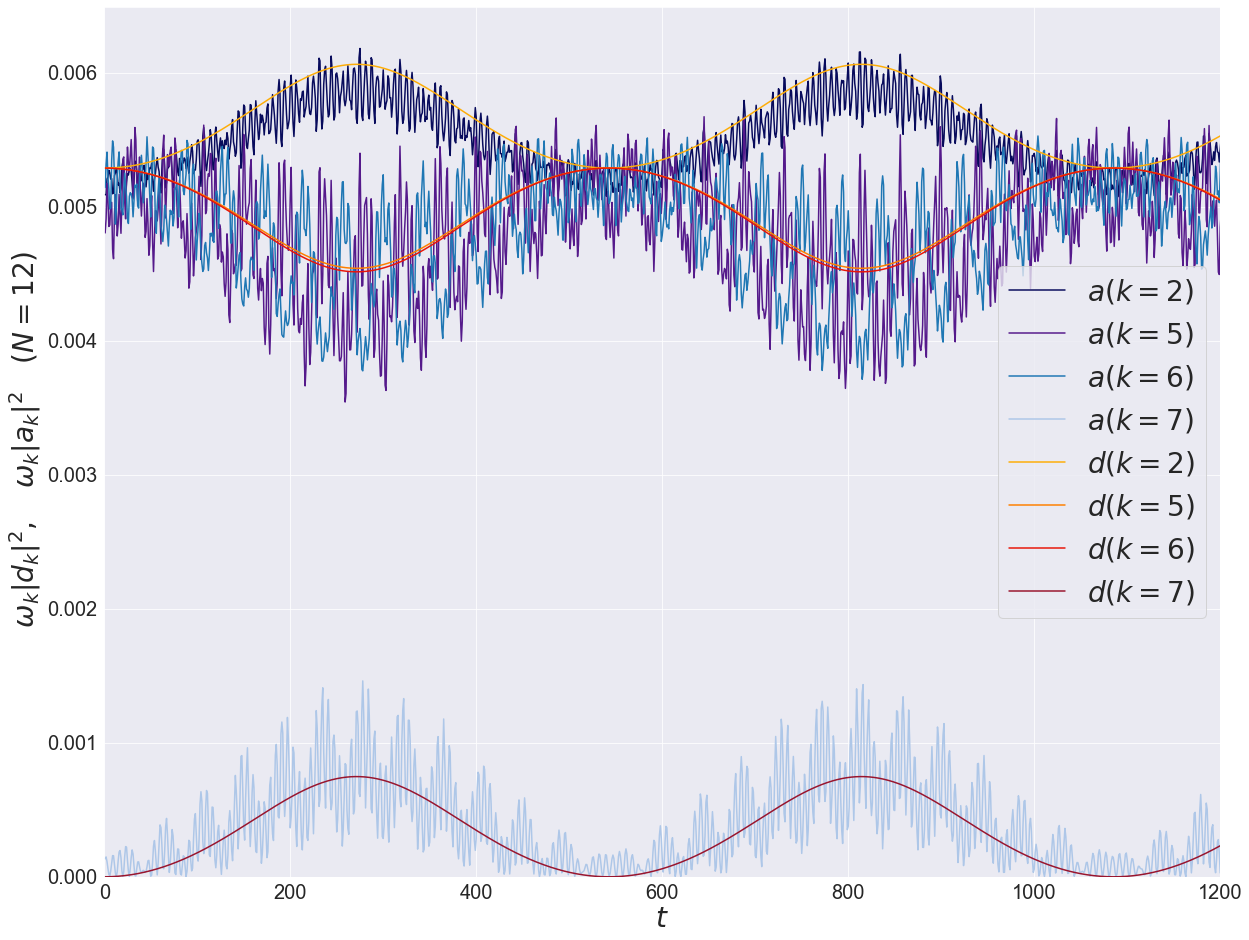}
    \includegraphics[width=0.47\textwidth]{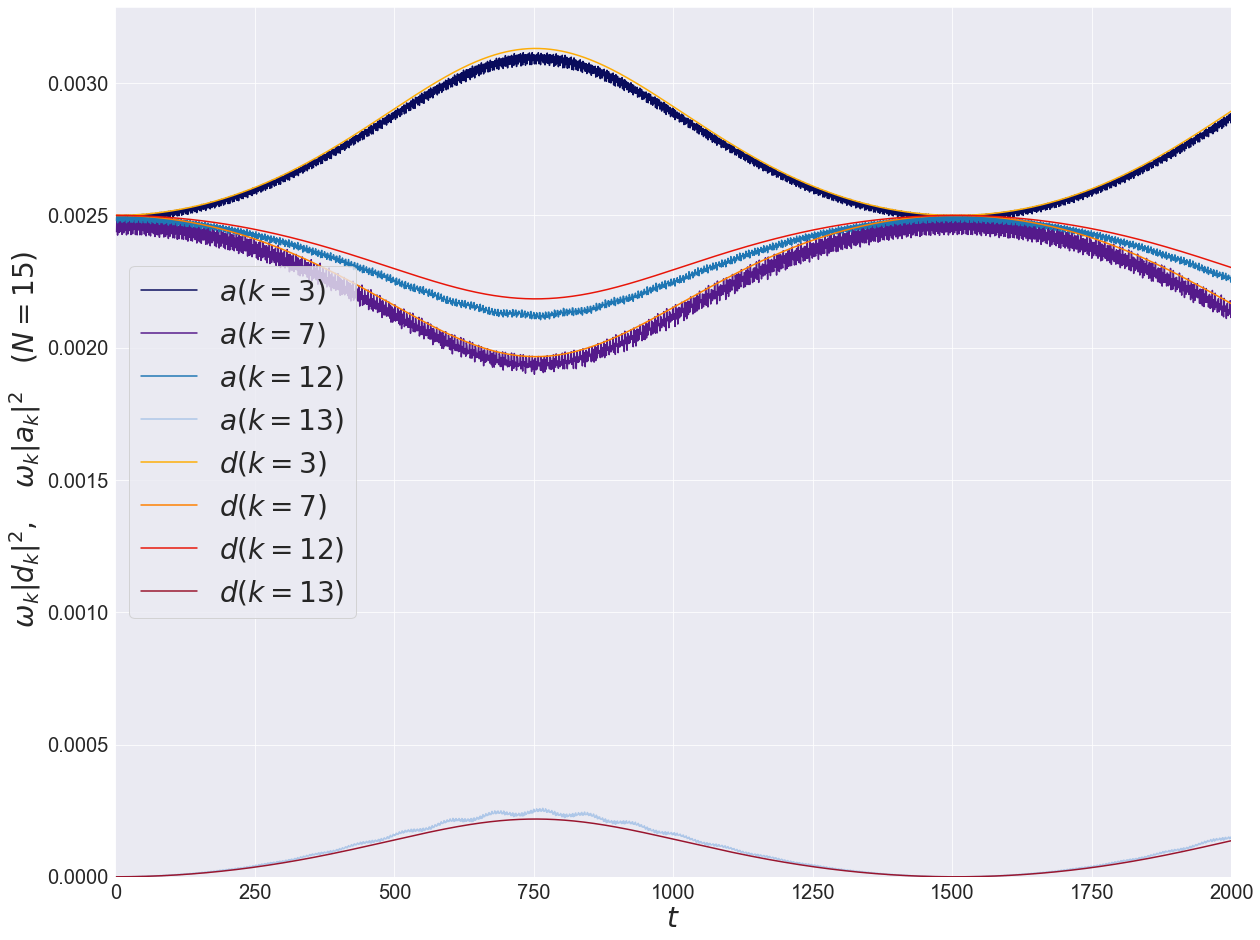}
    \caption{Time evolution of individual energies in a resonant quintet: (a) $N=12$ resonant quintet, initial conditions \eqref{eq:IC_quintet_12}. (b) $N=15$ resonant quintet, initial conditions \eqref{eq:IC_quintet_15}. 
    \label{fig:time_series_superimposed_12_15}}
\end{figure}

To illustrate that the above ideas are independent of the number of particles (as long as it is divisible by $3$), we now show results for $N=12$ and $N=15$. Although these systems have quintets with non-repeated modes, for simplicity of presentation we prefer to consider quintets that have a repeated mode, just like the quintet we considered in the case $N=9$. For $N=12$ we choose the resonant quintet  $\{2,2,7;5,6\}$ with ``target'' mode $k=7$ so the initial conditions are
\begin{equation}
\label{eq:IC_quintet_12}
\omega_2 |d_2(0)|^2 = \omega_5 |d_5(0)|^2 = \omega_6 |d_6(0)|^2 = 0.0053\,,
\end{equation}
all other modes being initially zero, leading to the plot in figure \ref{fig:time_series_superimposed_12_15}(a) showing again the main quintet interaction in a low-nonlinearity scenario (maximum convergence exponent $-1.17$). For $N=15$ the chosen quintet is $\{3,3,13;7,12\}$ with ``target'' mode $k=13$ so the initial conditions are
\begin{equation}
\label{eq:IC_quintet_15}
\omega_3 |d_3(0)|^2 = \omega_7 |d_7(0)|^2 = \omega_{12} |d_{12}(0)|^2 = 0.0025\,,
\end{equation}
again in a low-nonlinearity scenario (maximum convergence exponent $-1.34$).

\section{Chaos characterization: Lyapunov exponents, resonances and constant of motion}
\label{sec:FPUT_theo}

We now focus on the relation between the original variables $a$ and the mapped variables $d$, having validated the exact-resonance evolution equations \eqref{eq:d} for the variables $d$ (or equations \eqref{eq:d_Heisenberg} for the slow variables $D$) in the previous section, where we showed that these exact-resonance evolution equations will produce meaningful solutions with regards to the original variables $a$ if and only if the transformation from $a$ to $d$ converges. Unfortunately, a complete and definitive test of convergence would require us to go beyond the variables $d$, which contain up to $5$-wave resonances: one would have to formally continue the process up to $M$-wave interactions, and check whether the transformation converges as $M$ goes to infinity. Such a calculation is beyond the scope of this paper and we do not claim that it would converge (see \cite{bogomolnyi1984higher,wood1987limitations, kaluza1992improved, robnik1993pade} for a discussion and further references on the asymptotic character of the normal form transformation).  

However, what we do know from the previous section is that the \emph{truncated} system \eqref{eq:d} for the variables $d$ (or \eqref{eq:d_Heisenberg} for the slow variables $D$) indeed represents the original system \eqref{eq:a_k}, with an arbitrary accuracy that can be controlled by reducing the amplitudes of the modes. This will be true for any subsequent truncation up to $M$-wave interactions, where $M$ is fixed. This is the same idea as in the theory of asymptotic expansions: a function's  asymptotic expansion truncated at any given order $M$ will be as close as desired to the original function, if the domain is appropriately restricted.

It is in this sense of restricted domains, namely of ``small amplitudes'' or ``weak nonlinearity'', that we can draw quite strong conclusions when comparing the original system of equations \eqref{eq:a_k} for the variables $a$ against the truncated systems \eqref{eq:d} (resp. \eqref{eq:d_Heisenberg}) for the mapped variables $d$ (resp. $D$).

Motivated by the above discussion, a pivotal idea that we want to introduce in this section is to quantify numerically, in terms of Lyapunov exponents, the accuracy at which the truncated models describe the original ones. We believe Lyapunov exponents can be a strong tool to compare 
different dynamical systems, such as system \eqref{eq:a_k} for the variables $a$ and system \eqref{eq:d_Heisenberg} for the mapped ``slow'' variables $D$: if these systems are related by a one-to-one mapping that is accurate numerically, then their Lyapunov spectra must coincide. Thus, comparing these Lyapunov spectra will allow us to establish unequivocally the level of nonlinearity at which the two systems behave in a truly different manner.   

A second powerful idea, to be combined with the previous one, is that system \eqref{eq:d} for the mapped  variables $d$ is amenable to analytical calculations: because it is based exclusively on resonant interactions, one can apply the ``resonant cluster matrix'' method introduced by Harper et al. \cite{harper2013quadratic} to construct explicitly a number of independent constants of motion that are quadratic in the amplitudes $|d_k|, \,k=1, \ldots, N-1$. Thus, the effective number of degrees of freedom can be explicitly reduced, and thus the maximum number of positive Lyapunov exponents can be calculated. This result can then be used to validate a numerical calculation of Lyapunov exponents: we will take the case study $N=9$, which according to this method has a maximum of $4$ positive Lyapunov exponents. On the one hand, this number is verified numerically for system \eqref{eq:d} in the variables $d$. On the other hand, this number can be used to establish at which level of nonlinearity the original system \eqref{eq:a_k} for the variables $a$ departs from the $d$ system, by looking at the fifth largest Lyapunov exponent as a function of the system's nonlinearity: while it should be equal to zero at all levels of nonlinearity for the $d$ system, in the $a$ system it suffers a transition from zero (at weak nonlinearity) to non-zero at larger nonlinearity. The threshold occurs at a level of nonlinearity that is related to the loss of convergence of the normal form transformation.

\subsection{Integrability and chaos of the exact-resonance evolution equations for the $d$ variables}

\subsubsection{Canonical Poisson bracket (commutator)}
The exact-resonance evolution equations \eqref{eq:d} are Hamiltonian and canonical. It is useful to introduce the Poisson bracket (a.k.a. commutator), a bi-linear operation that takes two scalar functions $F$ and $G$ of the $d$-variables and gives another scalar function, via
\begin{equation}
\label{eq:Poisson_Bracket}
\{F,G\}:= -i \sum_{k=1}^{N-1} \frac{\partial F}{\partial d_k} \frac{\partial G}{\partial d^*_k} - \frac{\partial F}{\partial d^*_k} \frac{\partial G}{\partial d_k}\,.
\end{equation}
It is useful here to recall the basic commutation relations
\begin{equation}
    \{d_k, d_{k'}\}=\{d_k^*, d_{k'}^*\} = 0\,,\qquad  \{d_k, d_{k'}^*\}  = -i \delta_{k,k'},
\end{equation}
where $\delta$ denotes the Kronecker delta symbol.
In this way, a given scalar function $F$ of the $d$-variables satisfies
$$\frac{\mathrm{d} F}{\mathrm{d}t} = \frac{1}{N} \{F,H\}\,,$$
where as usual $\frac{\mathrm{d} F}{\mathrm{d}t} $ is calculated using the chain rule and the equations of motion \eqref{eq:d}.

\subsubsection{Finding quadratic constants of motion} 
Following the method introduced by Harper et al. \cite{harper2013quadratic}, we look for constants of motion that are quadratic in the $d$-variables, namely, functions of the form
\begin{equation}
    \label{eq:CoM}
    I = \sum_{k=1}^{N-1} \Omega_k d_k d^*_k\,,
\end{equation}
where $\Omega_k\,,\,\,k=1, \ldots, N-1$, are constants to be found
so that ${\mathrm{d} I}/{\mathrm{d}t} = 0$ under the evolution  \eqref{eq:d}, that is, so that $\{I,H\} = 0$ (in words, $I$ commutes with $H$).
While it is somewhat evident that the quadratic part of the Hamiltonian, $\sum_{k=1}^{N-1} \omega_k d_k d^*_k$, is a constant of motion, where $\omega_k$ is given by the FPUT dispersion relation \eqref{eq:FPUT_dispersion}, there are usually more than one independent quadratic invariants for a given $N$. 

Because of the bi-linearity of the Poisson bracket, the condition $\{I,H\} = 0$ can be separated into smaller problems. In fact, regarding the Hamiltonian $H$ defined in equation \eqref{eq:H_d}, we will now show that, in order for $I$ to commute with $H$, $I$ must commute with every term in every sum in the definition of $H$.  First, notice that the quadratic part of $H$ trivially commutes with $I$, because $\{d_k d_k^*, d_{k'}d_{k'}^*\} = 0$ for all $k,k'$. Next, the remaining terms in the Hamiltonian \eqref{eq:H_d} are of two types: (i) monomial of degree $4$ in the amplitudes, corresponding to a resonant quartet, and (ii) monomial of degree $5$ in the amplitudes, corresponding to a resonant quintet. Clearly, these monomials are independent when considered as functions of the $2(N-1)$ variables $\{d_k\}_{k=1}^{N-1}$ and $\{d_k^*\}_{k=1}^{N-1}$. Now, let us study the commutator between $I$ and a given monomial representing a resonant term. A useful identity is easily derived from equation \eqref{eq:Poisson_Bracket}:
\begin{equation}
    \label{eq:basic_commut}
    \{d_{k}^*d_{k},G\} \equiv i \left(d_{k}\frac{\partial G}{\partial d_k}-d_{k}^*\frac{\partial G}{\partial d^*_k}\right)\,.
\end{equation} 
From the above identity, it follows that the operator $\{d_{k}^*d_{k},\cdot\}$ preserves the degree of a monomial, and moreover, monomials are eigenfuctions of the operator, for any $k$. Thus, from bi-linearity of the commutator, monomials are eigenfunctions of the operator $\{I,\cdot\}$.  Therefore, $\{I,H\}$ is a linear combination  of the independent monomials appearing in equation \eqref{eq:H_d}, and hence, $\{I,H\} = 0$ if and only if $\{I,G\} =0$ for every single monomial $G$ in equation \eqref{eq:H_d}.

Starting with the resonant quartets in \eqref{eq:H_d}, let us use the more explicit notation $d_{k_0}^*d_{k_1}^*d_{k_2}d_{k_3}$ for the monomial, where $k_0+k_1 = k_2+k_3 \pmod N$ and $\omega_0+\omega_1 = \omega_2+\omega_3$. We have, using \eqref{eq:CoM} and \eqref{eq:basic_commut},
$$\{I,d_{k_0}^*d_{k_1}^*d_{k_2}d_{k_3}\} =  i \sum_{k=1}^{N-1} \Omega_k \left( -\delta_{k_0}^k - \delta_{k_1}^k + \delta_{k_2}^k + \delta_{k_3}^k\right)d_{k_0}^*d_{k_1}^*d_{k_2}d_{k_3}\,.$$
We conclude that for each resonant quartet the above commutator must be equal to zero. In equations,
\begin{equation}
\label{eq:cluster_quartet}
    \Omega_{k_0}+\Omega_{k_1} = \Omega_{k_2}+\Omega_{k_3} \quad\text{when} \quad \begin{cases}
   & k_0+k_1 = k_2+k_3 \pmod N\,,\\
   & \omega_0+\omega_1 = \omega_2+\omega_3\,.
    \end{cases} 
\end{equation}
As for resonant quintets, we use the explicit notation $d_{k_0}^*d_{k_1}^*d_{k_2}d_{k_3}d_{k_4}$ for the monomial, where $k_0+k_1 = k_2+k_3 + k_4 \pmod N$ and $\omega_0+\omega_1 = \omega_2+\omega_3 + \omega_4$. We obtain, similarly,
$$\{I,d_{k_0}^*d_{k_1}^*d_{k_2}d_{k_3}d_{k_4}\} = i \sum_{k=1}^{N-1} \Omega_k \left(-\delta_{k_0}^k - \delta_{k_1}^k + \delta_{k_2}^k + \delta_{k_3}^k + \delta_{k_4}^k\right)d_{k_0}^*d_{k_1}^*d_{k_2}d_{k_3}d_{k_4}\,,$$
so we conclude
\begin{equation}
\label{eq:cluster_quintet}    \Omega_{k_0}+\Omega_{k_1} = \Omega_{k_2}+\Omega_{k_3}+\Omega_{k_4}\quad\text{when}     \quad\begin{cases}
   & k_0+k_1 = k_2+k_3+k_4 \pmod N\,,\\
   & \omega_0+\omega_1 = \omega_2+\omega_3+\omega_4\,.
    \end{cases}
\end{equation}

\subsubsection{Resonant cluster matrix}
In summary, $I$ as given by equation \eqref{eq:CoM} is a constant of motion for our system \eqref{eq:d} if and only if, for every single exact resonance (quartet or quintet), equations \eqref{eq:cluster_quartet} and \eqref{eq:cluster_quintet} are satisfied.
These equations are linear and homogeneous in the set of unknowns $\{\Omega_k\}_{k=1}^{N-1}$, which can be understood as null vectors of the so-called resonant cluster matrix $\mathcal{A}$, constructed algorithmically  as follows: each resonance will constitute a row. The order of the rows does not matter. For a quartet resonance of the form \eqref{eq:cluster_quartet}, assign an arbitrary row label $r$. The components of this row are $\mathcal{A}_{r,k} :=  -\delta_{k_0}^k - \delta_{k_1}^k + \delta_{k_2}^k + \delta_{k_3}^k\,, \,\,k=1, \ldots N-1\,.$ 
For a quintet resonance of the form \eqref{eq:cluster_quintet}, assign an arbitrary row label $r'$. The components of this row are $\mathcal{A}_{r',k} :=  -\delta_{k_0}^k - \delta_{k_1}^k + \delta_{k_2}^k + \delta_{k_3}^k + \delta_{k_4}^k\,, \,\,k=1, \ldots N-1\,.$
The resulting matrix $\mathcal{A}$ will thus have $N-1$ columns and $R_4+R_5$ rows, where $R_4$ is the total number of quartet resonances, namely the total number of independent terms in the quartic sum in equation \eqref{eq:H_d}, and $R_5$ is the total number of quintet resonances, namely the total number of independent terms in the quintic sum in equation \eqref{eq:H_d}. In this way, equations \eqref{eq:cluster_quartet}--\eqref{eq:cluster_quintet} for the variables $\{\Omega_k\}_{k=1}^{N-1}$ become the null-space equations
$$\mathcal{A}{\bm \Omega} = 0\,,\quad {\bm \Omega} := (\Omega_1,\ldots, \Omega_{N-1})^T\,.$$

\subsubsection{Independent quadratic constants of motion}
Thus, the number of independent quadratic constants of motion of system \eqref{eq:d} is given by the dimension of the  null space of the resonant cluster matrix $\mathcal{A}$. We denote this dimension by ${\mathcal J}_N$, as it depends on the number of particles $N$. To get an idea of how ${\mathcal J}_N$ depends on $N$, notice that when $N$ is odd all quartet resonances are trivial (namely, $k_0=k_2$ and $k_1=k_3$), so equations \eqref{eq:cluster_quartet}
 are satisfied identically. Also, when $N$ is not divisible by $3$ there are no quintet resonances. Thus, when $N$ is odd and not divisible by $3$, the cluster matrix vanishes (this means that one should go to higher orders of wave-wave interactions in order to find nontrivial resonances) and therefore ${\mathcal J}_N = N-1$: all the individual energies $|d_1|^2, \ldots , |d_{N-1}|^2$ are constants of motion. A less trivial case is when $N$ is even (but not divisible by $3$, so it has no quintets). This case has nontrivial resonant quartets, which are all of pair-off form: a basis for the resonant cluster matrix rows is given by the rows $(0,\ldots,0, k,0,\ldots,0,N-k,0,\ldots,0)$, with $k=1,\ldots, N/2$. Thus, ${\mathcal J}_N = N/2$. In this case, all $\lfloor{N/4}\rfloor$  monomials commute with each other so the system is integrable (see \cite{rink2006proof, rink2008integrable}).

The only truly nontrivial case is when $N$ is divisible by $3$ and $N>6$: resonant quintets exist. In addition, if $N$ is even then nontrivial resonant quartets exist. Table \ref{tab:invariants_J_N} shows the number of quadratic invariants as a function of $N$ for selected cases. The apparent wild behaviour of ${\mathcal J}_N$ is due to the nontrivial dependence of the number of resonant quartets and quintets  on the divisors of $N$ (see \cite{bustamante2019exact} for details).  

\begin{table}[h]
    \centering
    \begin{tabular}{@{}ccccccccc@{}}
$N$ &9&12&15&18&21&24&27&30\\
${\mathcal J}_N$ &3&6&5&8&9&11&9&13 \\
$\mathcal{P}_N = N-2-{\mathcal J}_N$ & 4&4&8&8&10&11&16&15
    \end{tabular}
\caption{\label{tab:invariants_J_N} Number of independent quadratic constants of motion ${\mathcal J}_N$ and maximum number of positive Lyapunov exponents ${\mathcal P}_N$ of system \eqref{eq:d} for selected values of $N$ divisible by $3$.}
\end{table}

\subsubsection{Effective number of degrees of freedom}
\label{subsubsec:eff_dof}
A priori, system \eqref{eq:d} has $2(N-1)$ degrees of freedom, as there are $N-1$ complex variables $\{d_k\}_{k=1}^{N-1}$. It is instructive to introduce the so-called amplitude-phase representation, as follows: $d_k := |d_k| \exp(i \phi_k), \,\, d_k^* := |d_k| \exp(-i \phi_k)$, where now $|d_k|$ (amplitudes) and $\phi_k$ (phases) are two independent real variables, so the number of degrees of freedom is still $2(N-1)$. However, in the Hamiltonian \eqref{eq:H_d} we see that only certain combinations of these variables appear. In fact, we can make use of the resonant cluster matrix to construct precisely all the  combinations of phases that appear in the Hamiltonian. By inspection, these combinations are simply the rows in
${\mathcal A} \Phi$, where $\Phi := (\phi_1,\ldots, \phi_{N-1})^T$. While in general there will be many rows (as many rows as the number of nontrivial quartet and quintet resonances), most of these combinations are dependent. To calculate the number of independent rows of $\mathcal A$, namely the number of independent phase combinations that appear in the Hamiltonian, we make use of the rank-nullity theorem, which states 
$$\# \text{ independent rows of }{\mathcal A} = (\#\text{ columns of }{\mathcal A}) - (\text{dimension of null-space of }{\mathcal A})\,,$$
which is equal to $N-1-{\mathcal J}_N$. So, the number of independent phase combinations appearing in the Hamiltonian is equal to $N-1-{\mathcal J}_N$. This is to be interpreted as the effective number of dynamical phases. As for the real amplitudes $|d_k|$, there are a total of $N-1$, but the fact that there are ${\mathcal J}_N$ independent constants of motion that depend on these amplitudes allows us to fix these constants and thus reduce the effective number of dynamical amplitudes to $N-1 - {\mathcal J}_N$. Finally, if we now recall that the Hamiltonian is a constant of motion and along with it we have the time translation invariance, this gives us a reduction of $2$ degrees of freedom. In summary, we have a total of $2(N-2 - {\mathcal J}_N)$ effective degrees of freedom.

\subsection{Lyapunov exponents}
\label{subsec:LE}
Lyapunov exponents constitute the natural quantifiers of chaos or hyperchaos in any system, because they define the perturbation growth (or decay) rates in a system. For the FPUT system of $N$ particles (be it the original system \eqref{eq:a_k} in the $a$-variables, or the truncated system of exact-resonance evolution equations \eqref{eq:d} in the $d$-variables), the number of degrees of freedom is $2(N-1)$ and therefore there are $2(N-1)$ Lyapunov exponents. We order them from largest to smallest: $\lambda_1 \geq \lambda _2 \geq \ldots \geq \lambda_{2(N-1)}$. The set of Lyapunov exponents is called the Lyapunov spectrum. Because the FPUT system is Hamiltonian, the Lyapunov spectrum is symmetric around zero: $\lambda_{2(N-1)-(n-1)} = -\lambda_n$, for $n=1, \ldots, N-1$. Thus, the first $(N-1)$ Lyapunov exponents are non-negative and we have
$\lambda_1 \geq \ldots \geq \lambda_{N-1}\geq 0$. Furthermore, as the Hamiltonian is a constant of motion, we necessarily have $\lambda_{N-1}=0$. In fact, any independent constant of motion of the system will have an  associated Lyapunov exponent that is equal to zero. 

\subsubsection{Maximum number of positive Lyapunov exponents of the exact-resonance evolution equations \eqref{eq:d} in the $d$-variables} 
In the $d$-variables, in section \ref{subsubsec:eff_dof} we found that, after considering the ${\mathcal J}_N$ independent quadratic invariants and the Hamiltonian $H$, the number of effective degrees of freedom reduced to $2(N-2-{\mathcal J}_N)$. In terms of the Lyapunov spectrum, this is equivalent to  $\lambda_{N-1}=\lambda_{N-2}=\ldots=\lambda_{N-1-{\mathcal J}_N}=0$. Therefore, calling ${\mathcal P}_N$ the maximum number of positive Lyapunov exponents, we have ${\mathcal P}_N = N-2 - {\mathcal J}_N$. 
Table \ref{tab:invariants_J_N} shows this quantity as a function of $N$, for selected values of $N$ divisible by $3$. The lowest value ${\mathcal P}_N$ takes is $4$ (for $N=9$ or $N=12$), suggesting the presence of hyperchaos. Whether this number of positive Lyapunov exponents is attained in reality can be checked by a numerical calculation of Lyapunov exponents, as we will do in section \ref{subsubsec:LE_N=9} for the case $N=9$.

\subsubsection{Using Lyapunov exponents to compare the original FPUT system \eqref{eq:a_k} ($a$-variables) with the exact-resonance evolution equations \eqref{eq:d} ($d$-variables)}

In the past, analysis of the largest positive Lyapunov exponent ($\lambda_1$ in our notation) has been extensively used to address the FPUT paradox. For example, in \cite{benettin2018fermi} Benettin et al. investigate the asymptotic behaviour of $\lambda_1$ for a set of FPUT-like models and recover its dependence on the energy and number of particles. 

Our plan is to use  numerical computations of Lyapunov spectra to provide evidence that $5$-wave resonances are responsible for the hyperchaos observed in the original FPUT system, at least for small nonlinearity. To this end, we propose a method to compare the original system \eqref{eq:a_k} with the mapped equations \eqref{eq:d}, truncated up to and including quintet resonances. The basic idea is that, if two given systems of evolution equations $S_1$ and $S_2$ are related via a one-to-one smooth mapping, and no truncation is performed in either system, then the calculation of their respective Lyapunov spectra should give the same results. One way to see this is by noting that a Lyapunov exponent is calculated as the limit as time goes to infinity of the ratio (logarithm of the distance between two nearby solution trajectories) / (time). When a smooth map relates two systems, then the distance in one system is equal to the distance in the other system times a sub-determinant of the Jacobian of the mapping. The latter is finite, so in the limit the logarithms of the respective distances converge asymptotically and the respective ratios have the same limit. 

Now, we are trying to compare the original FPUT system \eqref{eq:a_k} in the $a$-variables with the exact-resonance evolution equations \eqref{eq:d} in the $d$-variables. Although the mapping from $d$ to $a$ is one-to-one and smooth, system \eqref{eq:d} is truncated up to and including resonant quintets, so, at high nonlinearity, the solutions to the respective evolution equations are not necessarily related by a one-to-one mapping, and we expect the Lyapunov spectra in the $a$-variables to be different from the Lyapunov spectra in the $d$-variables. Here we recall the idea of convergence of the normal-form transformation, which we discussed extensively in section \ref{subsec:convergence}. There, we demonstrated that when the amplitudes are small enough (weak nonlinearity regime) the mapping from $d$ to $a$ should converge and the solutions of the respective equations should be mapped in a one-to-one fashion. Based on this convergence, in that section we found a reasonable (but admittedly ad-hoc) criterion to decide at which level of nonlinearity we should expect the original system and the exact-resonance evolution equations to truly depart, in terms of the nature of their respective solutions. 

Here we propose a more quantitative method, devoid of ad-hoc assumptions, to decide on this very same question: At which level of nonlinearity do the original system and the exact-resonance evolution equations truly depart? 
Let $\{\lambda_n^{(d)}\}_{n=1}^{N-1}$ be the non-negative Lyapunov exponents in the $d$-system \eqref{eq:d} and let $\{\lambda_n^{(a)}\}_{n=1}^{N-1}$ be the non-negative Lyapunov exponents in the $a$-system  \eqref{eq:a_k}. Our comparison method makes use of our result that the exact-resonance evolution equations possess a number of constants of motion apart from the Hamiltonian, so  $\lambda_{N-1-{\mathcal J}_N}^{(d)} = 0$ must hold (see table \ref{tab:invariants_J_N} for representative values of ${\mathcal J}_N$). Thus, if $\lambda_{N-1-{\mathcal J}_N}^{(a)} > 0$ then we can assert that the $a$-system  \eqref{eq:a_k} and the $d$-system \eqref{eq:d}   truly depart.

How do we determine numerically whether $\lambda_{N-1-{\mathcal J}_N}^{(a)} > 0$ with enough confidence? In a numerical calculation of Lyapunov exponents in the $d$-system, knowing a priori that $\lambda_{N-1-{\mathcal J}_N}^{(d)} = 0$ is a great advantage, as it allows us to reduce the computational time to get these exponents, focusing on the first $N-1-{\mathcal J}_N$ exponents. We will use the method of cloned trajectories \cite{soriano2012method}, which is particularly well suited to this situation. Another advantage of knowing that $\lambda_{N-1-{\mathcal J}_N}^{(d)} = 0$ is that  the numerically computed exponent $\lambda_{N-1-{\mathcal J}_N}^{(d),num}$ will always have a non-zero value (because the computation is done using a finite-time signal), and thus this value can be used as the ``error'' in the estimation of this and any other exponent, being somehow a measure of the inherent error of the method and the system of equations. Usually, this error is larger than the standard deviation estimate of the error based on an average over the Lyapunov exponent time signal. Thus, if we find numerically that $\lambda_{N-1-{\mathcal J}_N}^{(a)} > 0$ even after allowing for error bars based on these errors, then we can assert with confidence that the two systems truly differ. In section  
\ref{subsubsec:LE_N=9} we implement such a study as a function of the nonlinearity in the case $N=9$, and obtain a sharp, quantitative transition in terms of the nonlinearity level: the systems are close at small nonlinearity and truly different at large nonlinearity.

\subsection{A case study: The FPUT lattice with $N=9,\,\alpha=1, \,\beta=0.05$}
\label{subsec:case_study_N9}
To showcase the potential of these combined methods, let us work with the FPUT system carrying $N=9$ masses, with interaction coefficients $\alpha=1, \,\beta=0.05$. We first analyse the system in the $d$-variables. Here, the relevant degrees of freedom are $d_1,\ldots,d_8$ and their complex conjugates. The Hamiltonian \eqref{eq:H_d} reduces to $H^{(d)} = H_2^{(d)} + H_4^{(d)} + H_5^{(d)}$, with:
\begin{equation} \label{eq:H5_N9}
    \begin{split} 
        H_5^{(d)} &= 3 \widetilde{W}_{5,7,2,2,8}  (d_7^*d_5^*d_2d_8d_2 - d_2^*d_4^*d_7d_1d_7) + 3 \widetilde{W}_{5,8,1,1,2} (d_8^*d_5^*d_2d_1d_1 - d_1^*d_4^*d_8d_8d_7 ) \\
        &+ 6 \widetilde{W}_{3,4,2,6,8} ( d_3^*d_4^*d_6d_8d_2 - d_6^*d_5^*d_3d_1d_7) 
        + 3 \widetilde{W}_{4,4,1,2,5} (d_4^*d_4^*d_5d_1d_2  -  d_5^*d_5^*d_4d_8d_7 )+ c.c.
    \end{split}
\end{equation}
where c.c. indicates complex conjugates, and the interaction coefficients are given numerically by 
$$\widetilde{W}_{5,7,2,2,8} = - i \,2.8909506487978,\qquad \widetilde{W}_{5,8,1,1,2} = - i\, 1.7415459095483\,,$$
$$\widetilde{W}_{3,4,2,6,8} = i \,2.4446465948986, \qquad \widetilde{W}_{4,4,1,2,5} = - i \, 2.1433819261863\,.$$ 
Because $H_4^{(d)}$ contains so-called trivial resonant terms only, which depend on the amplitude squares $|d_k|^2$ only, its explicit form   is not needed to analyse the constants of motion. The form of $H_2^{(d)}$ is the usual one, $H_2^{(d)}=\sum_{k=1}^{N-1}\omega_k |d_k|^2$. 

\subsubsection{Resonant cluster matrix for the $N=9$ case ($d$-variables)}
The resonant cluster matrix $\mathcal A$ is constructed by looking at the $8$ monomials in equation \eqref{eq:H5_N9}: for each monomial, a row of $\mathcal A$ of length $N-1$ is formed by recording, from $k=1$ to $k=N-1$, the multiplicities of the factors $d_k$ (positive multiplicities) and $d_k^*$ (negative multiplicities). For example, the first monomial $d_7^*d_5^*d_2d_8d_2$ indicates that the first row of $\mathcal A$ is $(0, 2, 0, 0,  -1, 0, -1, 1)$. The resulting matrix and its null space are: 

\begin{equation}
\label{eq:res_quintet_cluster}
\mathcal{A}  =
\left(\begin{smallmatrix}
0 & 2 & 0 & 0 & -1 & 0 & -1 & 1 \\
2 & 1 & 0 & 0 & -1 & 0 & 0 & -1 \\
0 & 1 & -1 & -1 & 0 & 1 & 0 & 1 \\
1 & 1 & 0 & -2 & 1 & 0 & 0 & 0 \\
0 & 0 & 0 & 1 & -2 & 0 & 1 & 1 \\
1 & 0 & 1 & 0 & -1 & -1 & 1 & 0 \\
-1 & 0 & 0 & -1 & 0 & 0 & 1 & 2 \\
1 & -1 & 0 & -1 & 0 & 0 & 2 & 0 
\end{smallmatrix}\right) \,, \quad \mathrm{Nul}(\mathcal{A})  = \mathrm{Span}\left(
\left( \begin{smallmatrix}
1\\
0\\
0\\
1\\
1\\
0\\
0\\
1
\end{smallmatrix}\right)
,
\left( \begin{smallmatrix}
0\\
1\\
0\\
1\\
1\\
0\\
1\\
0
\end{smallmatrix}\right)
,
\left( \begin{smallmatrix}
0\\
0\\
1\\
0\\
0\\
1\\
0\\
0
\end{smallmatrix}\right)
\right)\,.
\end{equation}

Note that $\mathcal{A}$ is not necessarily a square matrix (it is a coincidence here). From the null-space in \eqref{eq:res_quintet_cluster} we infer that ${\mathcal J}_9=3$ is the number of independent quadratic constants of motion. These can be read off directly from the generating vectors given:
\begin{equation}
\label{eq:invariants_N9}
    \begin{split}
        I_1 &= |d_1|^2 + |d_4|^2 + |d_5|^2 + |d_8|^2 \,,\\
        I_2 &= |d_2|^2 + |d_4|^2 + |d_5|^2 + |d_7|^2 \,,\\
        I_3 &= |d_3|^2 + |d_6|^2\,.
    \end{split}
\end{equation}
It is instructive to remark that the quadratic part of the Hamiltonian is given by $H_2^{(d)} = \omega_1 I_1 + \omega_2 I_2 + \omega_3 I_3\,.$ This is easy to show, by recalling the trigonometric identity $\omega_1+\omega_2 = \omega_4$, namely $\sin(\pi/9)+\sin(2\pi/9)=\sin(4\pi/9)$.

\subsubsection{Lyapunov spectra for the $N=9$ case: establishing the nonlinearity level below which $5$-wave resonances accurately describe the hyperchaos of the original FPUT system}
\label{subsubsec:LE_N=9}

Having just found that the $d$-system has ${\mathcal J}_9=3$  independent quadratic constants of motion, we conclude that the $d$-system has a maximum of $9-2-{\mathcal J}_9 = 4$ positive Lyapunov exponents. We thus turn to the numerical calculation of (finite-time) Lyapunov exponents, to compare the dynamics of the $d$-system versus that of the $a$-system. For this study we will use a generic type of initial conditions $d_k^{(0)} = \epsilon f_k$ with $f_k$ given in equation \eqref{eq:IC_f}, and $a_k^{(0)}$ is calculated from $d_k^{(0)}$ via the mapping in equation \eqref{eq:d_to_a}. The parameter $\epsilon$ is the same scaling parameter used in the numerical experiments in section \ref{subsec:convergence} and is a proxy for the level of nonlinearity. Admittedly, the study we propose of a system's behaviour along a fixed ray in the state space parameterised by $\epsilon$ does not provide the full picture of bifurcations, islands of order and stochastic webs, that we would expect to find in a thorough study. However, our study allows us to quantify the idea of convergence of the normal form transformation by looking at the behaviour of the Lyapunov exponents in the $d$-system versus the $a$-system. In fact, in section \ref{subsec:convergence} it was apparent that the normal form transformation converges (according to our ad-hoc criterion) when $\epsilon < 0.4$, at least for small times. We want to quantify this by measuring the Lyapunov exponents.

To calculate Lyapunov exponents we use the method of cloned trajectories \cite{soriano2012method}. The basic idea of the method is as follows. We consider the given initial condition and let the system evolve over long times, producing the so-called `fiducial' (or reference) trajectory. In parallel, we consider a slightly separated initial condition (by a distance $\delta := 10^{-5}$, this value being arbitrarily chosen) and let it evolve as well, producing a `cloned' trajectory. Due to the existence of positive Lyapunov exponents, the trajectories will separate exponentially. Thus, after a certain time (called $T_{\mathrm{cycle}}$, whose value must not be too large but it must be large enough so that the exponential separations between trajectories can be measured) one has to `reset' the position of the cloned trajectory, so the vector joining the two trajectories is rescaled to have length $\delta$. Then, the trajectories continue to be calculated numerically for an extra time $T_{\mathrm{cycle}}$, to be reset again at the end. Each such step is called an `iteration'. In this process, the exponential growth factors and the directions are recorded at each iteration and used to compute the Lyapunov exponents after several iterations. In fact, combining this with the use of multiple independent cloned trajectories and a Gram-Schmidt orthonormalisation procedure it is possible to construct recursively the whole spectrum of Lyapunov exponents.

\begin{figure}[]
    \centering
    \includegraphics[width=0.45\textwidth]{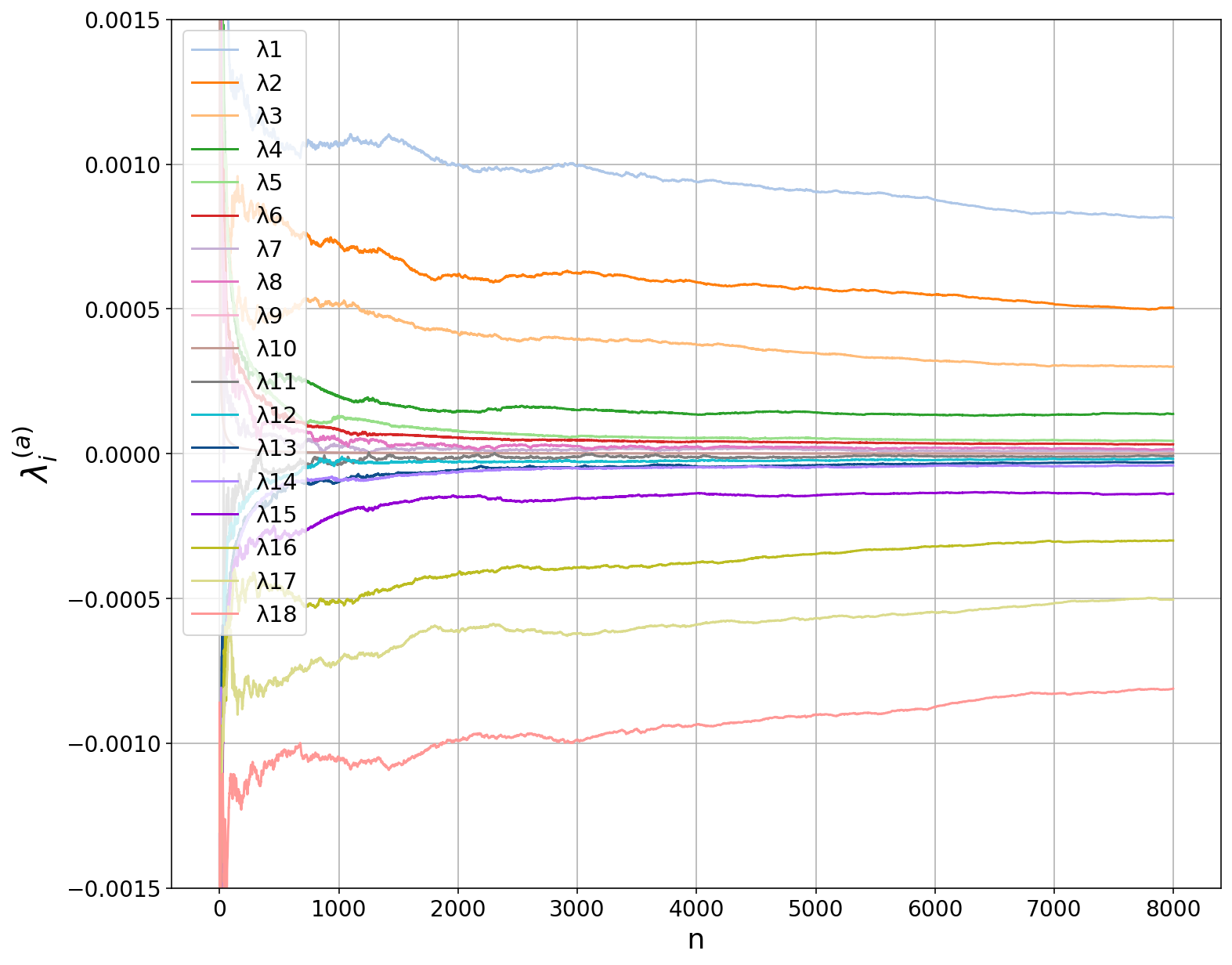}
    \includegraphics[width=0.45\textwidth]{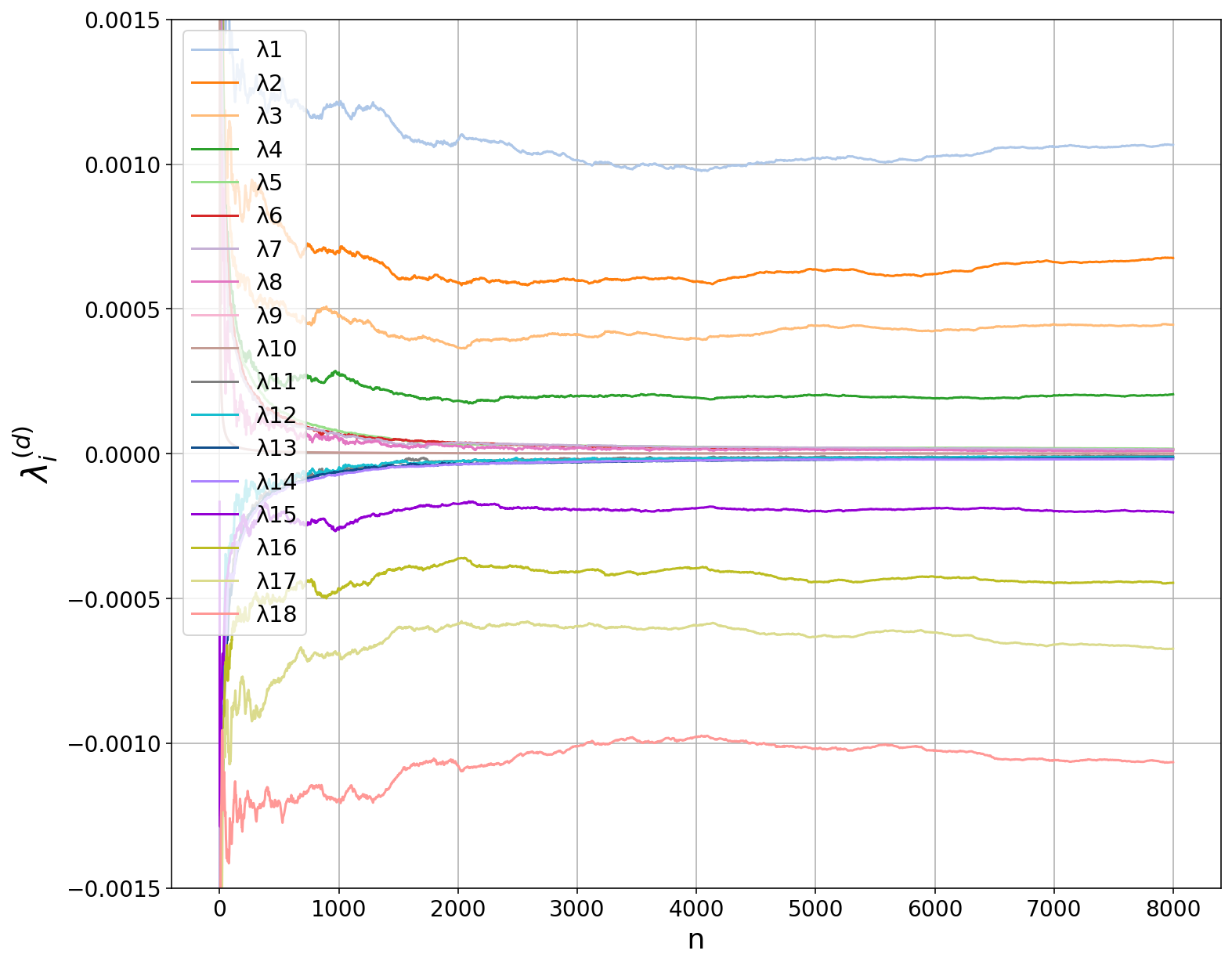}
    \includegraphics[width=0.45\textwidth]{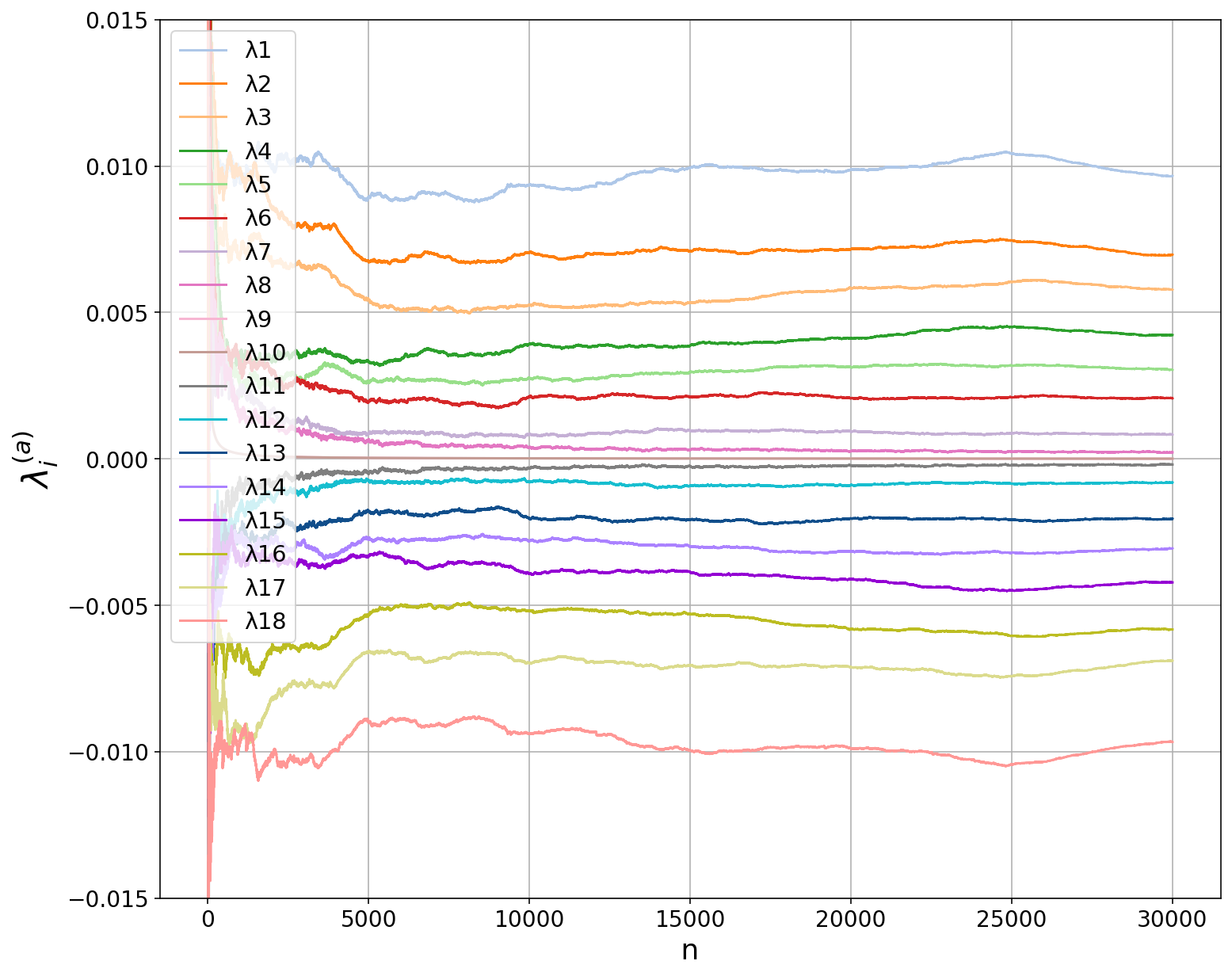}
    \includegraphics[width=0.45\textwidth]{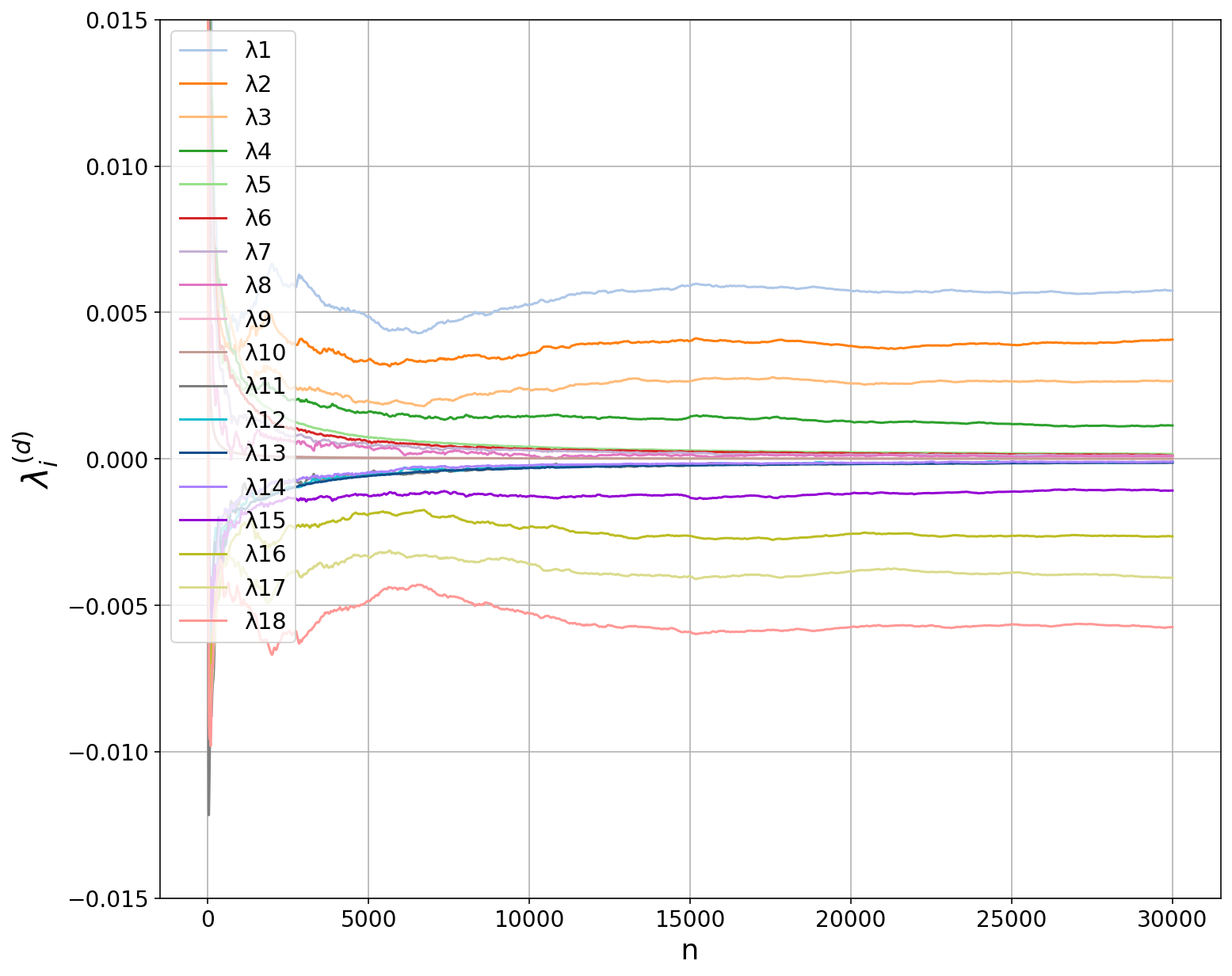}
    \caption{For FPUT in the case $N=9, \alpha=1, \beta=0.05$, Lyapunov spectra comparison between the original $a$-system (left panels) and exact-resonance evolution equations, $d$-system (right panels), for two selected values of the scaling parameter $\epsilon$: $\epsilon = 0.4$ (top panels) and $\epsilon=0.75$  (bottom panels). Initial conditions are 
    $d_k^{(0)} = \epsilon f_k$ with $f_k$ given in equation \eqref{eq:IC_f}, and $a_k^{(0)}$ is calculated from $d_k^{(0)}$ via the mapping in equation \eqref{eq:d_to_a}. In all these panels, the abscissa $n$ denotes the `resetting' or `iteration' number. At the beginning of each iteration, the distance from the clone trajectories to the fiducial trajectory is reset to $10^{-5}$, and the trajectories are evolved during a time $T_{\mathrm{cycle}}$, at which the information on the separation between trajectories is recorded to update the Lyapunov exponent calculation, and a new resetting or iteration begins. $T_{\mathrm{cycle}}=100$ for $\epsilon=0.4$ and $T_{\mathrm{cycle}}=2$ for $\epsilon=0.75$. Notice that for the $d$-system only the first $4$ and the last $4$ Lyapunov exponents are supposed to be different from zero, and thus the measured value of $\lambda_5^{(d)}$ serves as a proxy for the error bars in all calculations of Lyapunov exponents (cf. figure \ref{fig:chi_delta}). To this error, the standard deviation from the running Lyapunov exponent data from the last quarter of the iterations is added, but this extra error is quite small in comparison.}
    \label{fig:spectrum_ad_new}
\end{figure}

\begin{figure}[]
    \centering
    \includegraphics[width=0.4\textwidth]{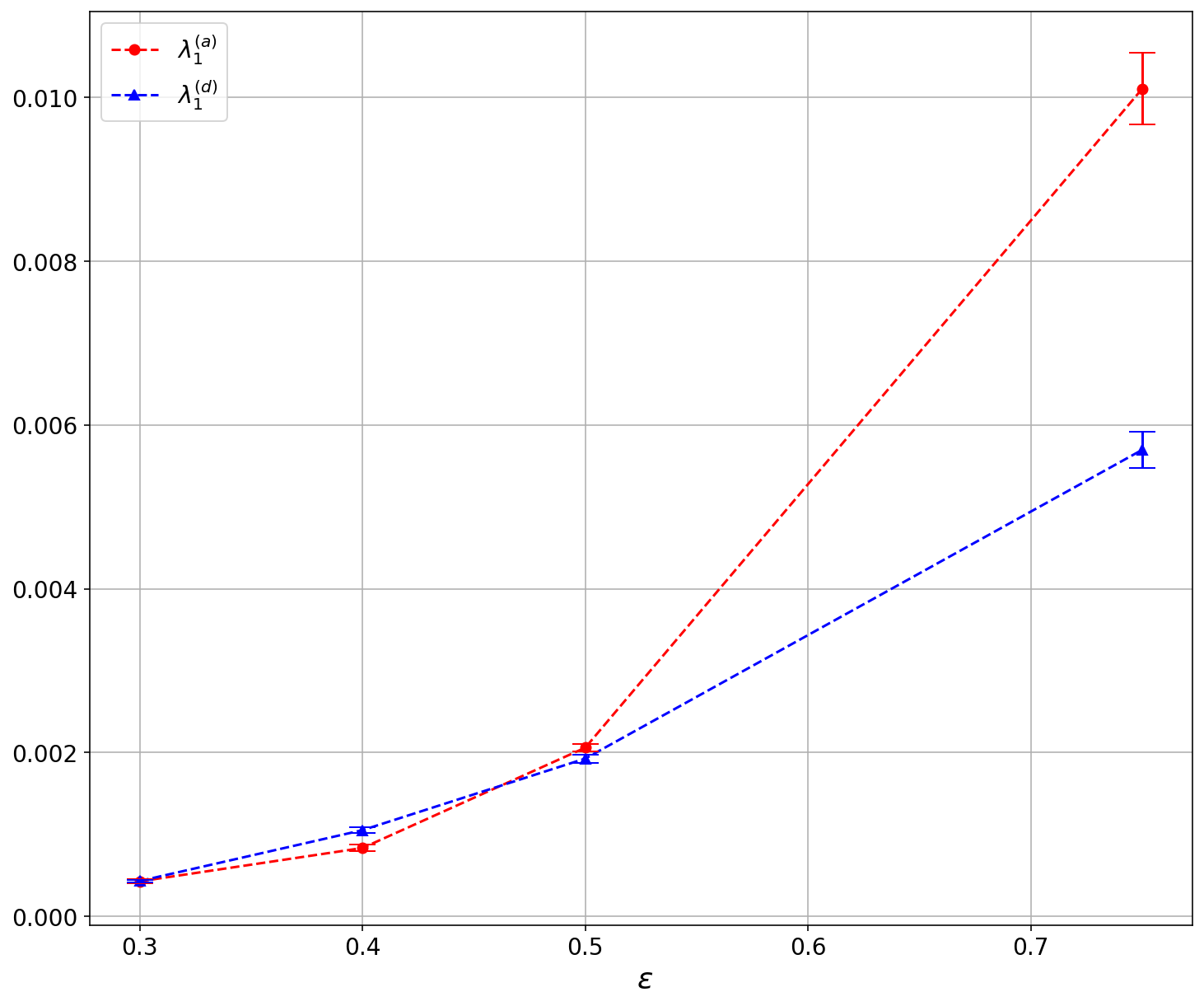}    
    \includegraphics[width=0.4\textwidth]{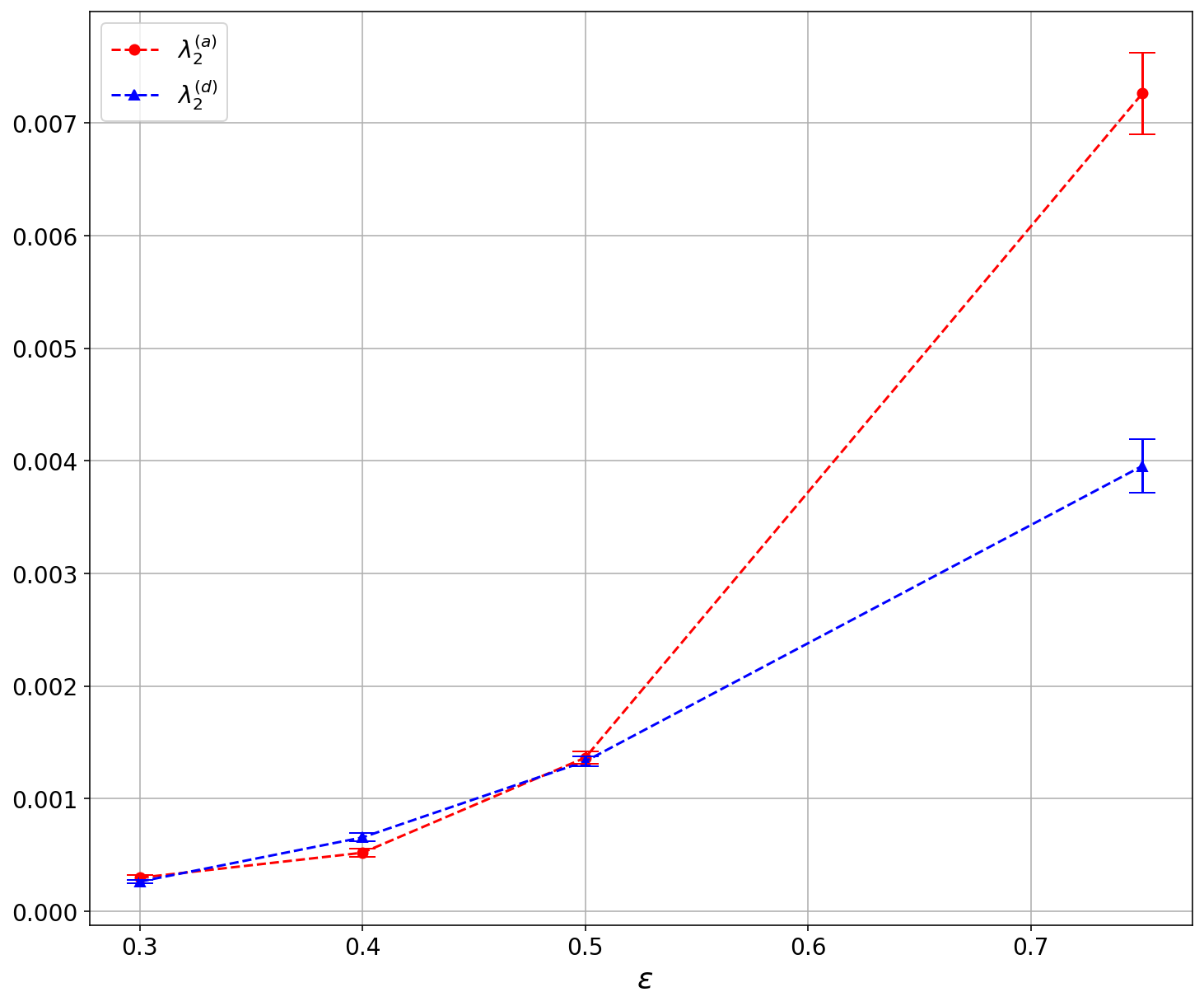}
    \includegraphics[width=0.4\textwidth]{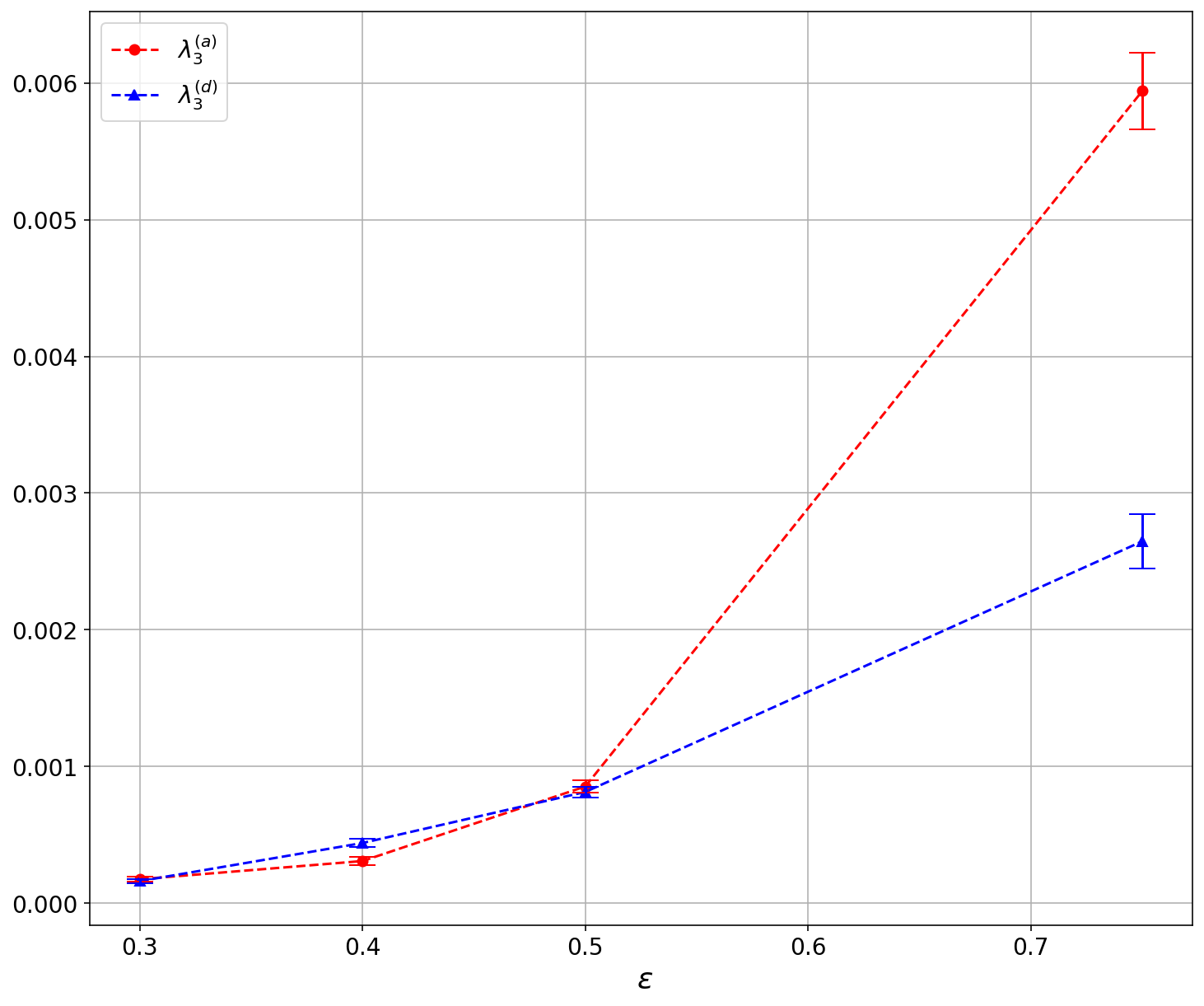}
    \includegraphics[width=0.4\textwidth]{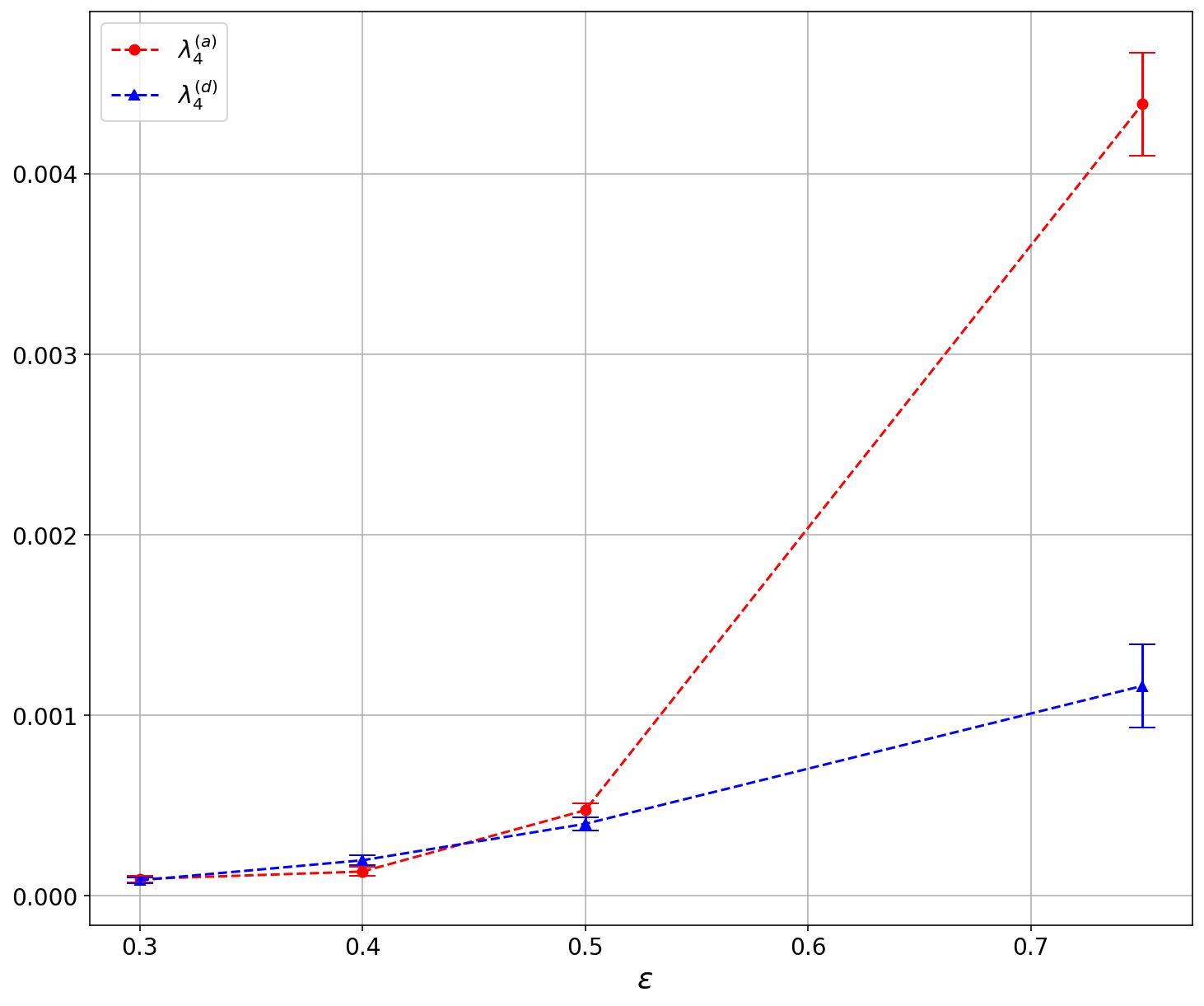}
    \includegraphics[width=0.45\textwidth]{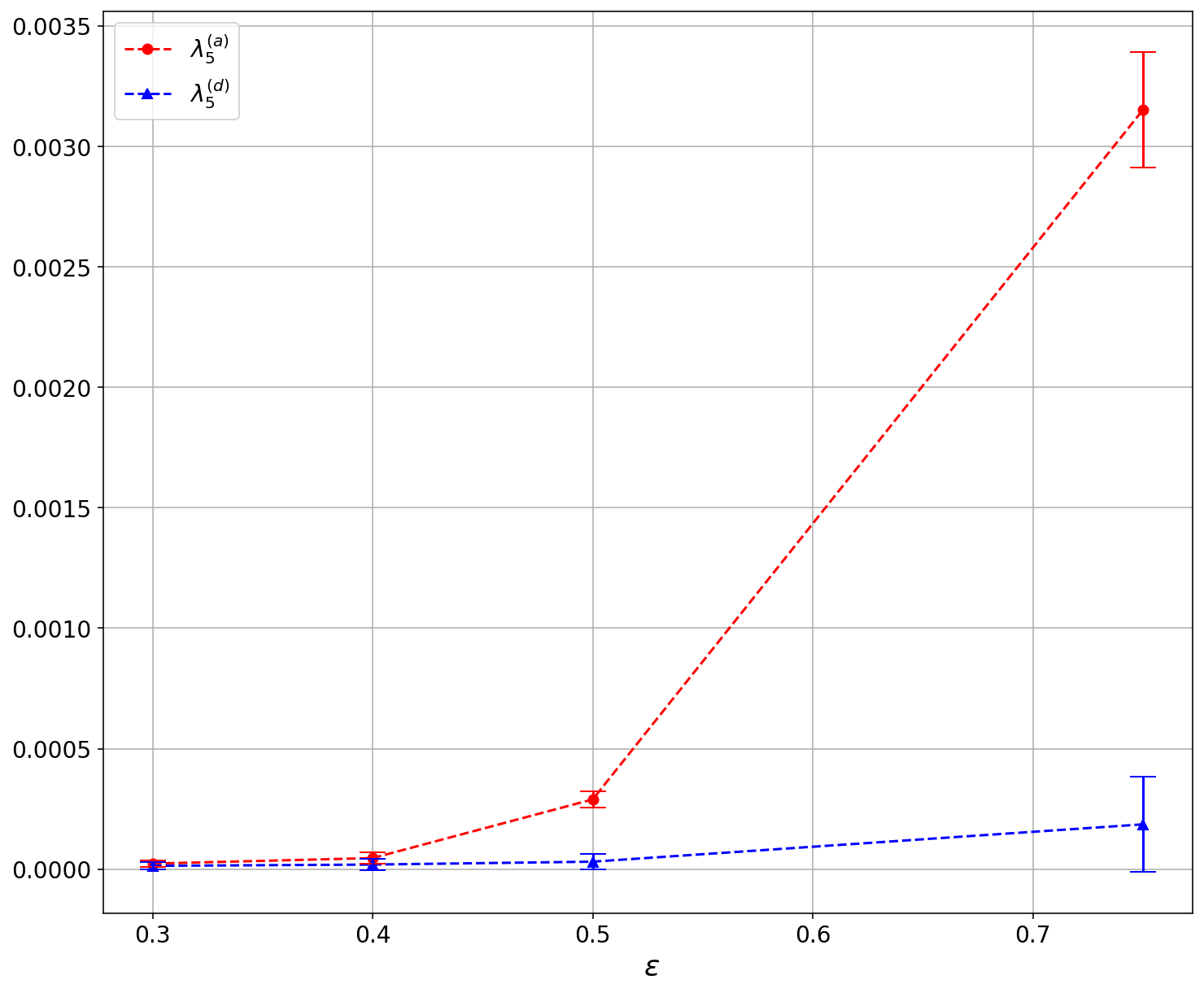}
    \caption{For FPUT in the case $N=9, \alpha=1, \beta=0.05$, study of convergence, towards a regime of small amplitude, of the five largest Lyapunov exponents calculated separately for the $a$-system (circle markers) and the $d$-system (triangle markers), as a function of the scaling parameter $\epsilon$ (same definition and same initial conditions as in the caption of Fig. \ref{fig:chi_delta}). The values of the exponents plotted are calculated as the average over the last quarter of the iterations of long-time running exponents as in figure \ref{fig:spectrum_ad_new}. The error bars are defined as the numerically calculated value of $\lambda_5^{(d)}$, which in theory should be equal to zero, plus a small correction from the standard deviation of the last quarter of the running data of Lyapunov exponents (cf. figure \ref{fig:spectrum_ad_new}).}
    \label{fig:chi_delta}
\end{figure}

Figure \ref{fig:spectrum_ad_new} shows the running values of Lyapunov spectra obtained for both the $a$-system (left panels) and $d$-system (right panels), and for two representative values of the scaling parameter: $\epsilon=0.4$ (top panels; $T_{\mathrm{cycle}}=100$) and $\epsilon=0.75$ (bottom panels; $T_{\mathrm{cycle}}=2$). First, we notice that the spectra are symmetric around zero, which is expected as the systems are Hamiltonian. Second, looking at the $d$-systems (right panels) we see that, as the theory predicts, of the non-negative Lyapunov exponents only $4$ of them are clearly positive, while the others seem to tend to zero. Due to the fact that these exponents are calculated in a finite time, the numerical result for the fifth exponent $\lambda_5^{(d)}$ is never exactly zero (while it is zero theoretically), and this allows us to use $\lambda_5^{(d)}$ as an uncertainty (error bars) for all other numerically calculated exponents (cf. figure \ref{fig:chi_delta}). Third, the case $\epsilon=0.4$ (top panels) seems to indicate that the spectra for the $a$-system is quite close to the spectra for the $d$-system: in particular, it appears that only $4$ positive Lyapunov exponents exist, whereas the case $\epsilon = 0.75$ shows at least $7$ positive Lyapunov exponents, demonstrating a clear departure from the $d$-system, which always has $4$ positive Lyapunov exponents.

More quantitatively, in either $a$-system or $d$-system with initial-condition scaling parameter values of $\epsilon=0.3, 0.4, 0.5, 0.75$, we perform  simulations over several iterations and calculate numerically the average of the running values of a given Lyapunov exponent over the last quarter of the iterations. We plot these Lyapunov exponent averages in figure \ref{fig:chi_delta}. The successive panels from left to right and from top to bottom  correspond to the Lyapunov exponents $\lambda_1, \ldots, \lambda_5$. We see that for the first $4$ Lyapunov exponents it seems that at nonlinearities below or at $\epsilon=0.5$ the two systems $a$ and $d$ coincide, within the error bars. However, this would lead to the wrong conclusion, as the decider is the fifth exponent (last panel), which shows that the systems diverge at $\epsilon=0.5$ already. So, from the perspective of the Lyapunov exponents, one can conclude that at around $\epsilon=0.3-0.4$ the original system develops a sharp transition, when going from small to large nonlinearity: at small nonlinearity it is a system dominated by $5$-wave resonances and with a controlled number of positive Lyapunov exponents, obtained from the theory of the exact-resonance equations of motion, and the normal form transformation from $d$- to $a$-variables appears to converge; at large nonlinearities, the system truly departs from the simple picture based on $5$-wave resonances, and develops hyperchaos at higher dimensions. In particular, the normal form transformation from $d$- to $a$-variables definitely does not converge anymore at this level of nonlinearity, and the concept of $M$-wave resonances loses meaning.

\section{Conclusions}
\label{sec:conclusion}

In this paper we have presented an all-round assessment of the approximate models arising from a weakly-nonlinear mathematical treatment of the FPUT lattice, using tools from wave-turbulence theory, starting from the Zakharov equation and Krasitskii's extension to $5$-wave resonances, and further improving the theory to eliminate all non-resonant terms, producing a unique transformation of coordinates that `distils' the equations to exact resonant terms only (section \ref{sec:FPUT_maths}). The main tools presented in section \ref{sec:FPUT_comp} to support this construction are: (i) We validated the hypermatrices containing the nonlinear coefficients appearing in the normal-form transformation (obtained when eliminating all non-resonant terms while keeping up to and including $5$-wave resonances), via a scaling approach based on the error introduced by the truncation of the mapped equations in the normal-form variables. (ii) We demonstrated the concept of convergence of the normal-form transformation and introduced an ad-hoc method of assessment of such a convergence. (iii) We generalised our previous results \cite{harper2013quadratic} to construct a number of quadratic constants of motion based on the resonant interactions in the FPUT system. Our results resemble the general results of Gustavson \cite{gustavson1966}, but in our case there is a clear advantage in using and extending Zakharov's method: our normal form transformations are unique (as we eliminate all non-resonant terms) and conserve momentum, and all resonances are clearly classified in terms of cyclotomic polynomials \cite{bustamante2019exact} so the quadratic constants of motion can be constructed a priori.

In order to obtain nontrivial $5$-wave resonances, the number of particles $N$ must be divisible by $3$. From our results obtained in Sec.~\ref{sec:FPUT_theo} we can draw some important considerations: the number of quadratic constants of motion, while depending on the number of divisors of $N$, grows linearly overall with the number of particles $N$ in the system (as shown in Table~\ref{tab:invariants_J_N}). While in future work we will show how this number of invariants (and the condition on divisibility of $N$) changes as $6$-wave resonances, $7$-wave resonances, and so on, are included in the normal form transformation, we can state the following results: 
\begin{itemize}
    \item The theory gives us an upper limit to the chaotic directions expected in the truncated system.
    \item If the number of maximum chaotic directions is equal to the number of positive finite-time Lyapunov exponents, then we have found all possible constants of motion.
    \item We can understand up to which order of the normal-form transformation the system is almost completely represented by its approximated representation.
\end{itemize}
Finally, in the same section we proposed a method based on the numerical calculation of finite-time Lyapunov exponents to estimate the level of nonlinearity at which the original FPUT system is well approximated by our $5$-wave exact-resonance equations (which generalise Zakharov equations). While the theory is valid for any $N$ divisible by $3$ and greater than $6$, we focused on the case study $N=9$, and compared our numerical calculations of the Lyapunov exponents to establish that at high enough nonlinearity the dynamics of the original FPUT system is clearly different from that of the exact-resonance equations, while at small enough nonlinearity the dynamics of these systems are indistinguishable (see Fig.~\ref{fig:spectrum_ad_new} and Fig.~\ref{fig:chi_delta}). 
We found that the exact-resonance equations are a good approximation of the original problem when the time scales are separable. 

The main conclusion of this study applies to any $N$ divisible by $3$ and greater than $6$ and is very optimistic: for small enough nonlinearity the $5$-wave exact-resonance equations of motion constitute an excellent approximation to the original FPUT system, and this can be quantified by looking at a particular Lyapunov exponent (specifically, the one at position $N-1-{\mathcal J}_N$, where ${\mathcal J}_N$ is the number of independent quadratic constants of motion), which must be equal to zero if these two systems are well approximated. 

In future work we will study the implications of our results on the question of equipartition in FPUT, in particular in the case of fixed boundary conditions.

\section{Acknowledgements}
This publication has emanated from research conducted with the financial support of Taighde \'Eireann – Research Ireland under Grant number 18/CRT/6049. For the purpose of Open Access, the authors have applied a CC BY public copyright licence to any Author Accepted Manuscript version arising from this submission.

\appendix

\section{Supplementary material}
\label{appendix:supplementary_material}
This paper extends  Krasitskii's results \cite{krasitskii1990canonical} and shows their applicability in the FPUT context. The mathematical treatment has been adapted to the specific problem at hand. In this appendix, we provide  backup material to the analytical tensor calculations presented in the main text of the paper. \\



Always non-resonant case of $4$-wave interactions:

\begin{equation}
\label{eq:Z1app}
\begin{split}
        Z^{(1)}_{0,1,2,3} = \dfrac{2}{3} \big( &V_{-0,1,0-1} A^{(1)}_{2+3,2,3} + V_{-0,2,0-2} A^{(1)}_{1+3,1,3} + V_{-0,3,0-3} A^{(1)}_{1+2,1,2} \\
        & - V_{-1,1-0,0} A^{*(3)}_{-2-3,2,3} - V_{-2,2-0,0} A^{*(3)}_{-1-3,1,3} - V_{-3,3-0,0} A^{*(3)}_{-1-2,1,2} \big)\,.
    \end{split}
\end{equation}

\begin{equation}
\label{eq:Z3app}
    \begin{split}
        Z^{(3)}_{0,1,2,3} = - 2 \big(  &V_{-0,0-3,3}A^{*(3)}_{-1-2,1,2} + V_{-3,3-0,0} A^{*(1)}_{1+2,1,2} + V_{-(0+1),1,0}A^{(1)}_{3,2,3-2} \\
        & + V_{-(0+2),2,0}A^{(1)}_{3,1,3-1} + V_{0,1,-0-1}A^{*(1)}_{2,3,2-3} + V_{0,2,-0-2}A^{*(1)}_{1,3,1-3} \big)\,.
    \end{split}
\end{equation}

\begin{equation}
\label{eq:Z4app}
    \begin{split}
        Z^{(4)}_{0,1,2,3} = - \dfrac{2}{3} \big( &V_{-(0+1),1,0} A^{(3)}_{-2-3,2,3} + V_{-(0+2),2,0} A^{(3)}_{-1-3,1,3} + V_{-(0+3),3,0} A^{(3)}_{-1-2,1,2} \\
        & + V_{0,1,-0-1}A^{*(1)}_{2+3,2,3} + V_{0,2,-0-2}A^{*(1)}_{1+3,1,3} + V_{0,3,-0-3}A^{*(1)}_{1+2,1,2} \big)\,.
    \end{split}
\end{equation}

Potentially resonant case (non-resonant subcase) of $4$-wave interactions:

\begin{equation}
\label{eq:Z2app}
    \begin{split}
        Z^{(2)}_{0,1,2,3} = 2 \big( &V_{-0,0-2,2} A^{(1)}_{3,1,3-1} + V_{-0,0-3,3} A^{(1)}_{2,1,2-1} - V_{0,1,-0-1}A^{*(3)}_{-2-3,2,3} \\
         & - V_{-2,2-0,0} A^{*(1)}_{1,3,1-3} - V_{-3,3-0,0} A^{*(1)}_{1,2,1-2} - V_{-(0+1),1,0}A^{(1)}_{2+3,2,3} \big)\,.
    \end{split}
\end{equation}

Always non-resonant case of $5$-wave interactions:

\begin{equation}\label{kernel:X1}
    \begin{split}
       -i\, X^{(1)}_{0,1,2,3,4} = &\dfrac{1}{3} \big( - V_{-0,1+2,3+4} A^{(1)}_{1+2,1,2} A^{(1)}_{3+4,3,4} - V_{-0,1+3,2+4} A^{(1)}_{1+3,1,3} A^{(1)}_{2+4,2,4} \\
        &\qquad  - V_{-0,3+2,1+4} A^{(1)}_{1+4,1,4} A^{(1)}_{2+3,3,2} +  V_{-(3+4),-1-2,0} A^{(1)}_{3+4,3,4}A^{*(3)}_{-1-2,1,2}  \\  
        &\qquad + V_{-(2+4),-1-3,0} A^{(1)}_{2+4,2,4} A^{*(3)}_{-1-3,1,3} + V_{-(2+3),-1-4,0} A^{(1)}_{2+3,2,3} A^{*(3)}_{-1-4,1,4} \\
        &\qquad + V_{-(1+4),-2-3,0} A^{(1)}_{1+4,1,4} A^{*(3)}_{-2-3,2,3} + V_{-(1+3),-2-4,0} A^{(1)}_{1+3,1,3} A^{*(3)}_{-2-4,2,4} \\
        &\qquad + V_{-(1+2),-4-3,0} A^{(1)}_{1+2,1,2} A^{*(3)}_{-3-4,3,4} + V_{0,-1-2,-3-4} A^{*(3)}_{-1-2,1,2} A^{*(3)}_{-3-4,3,4} \\
        &\qquad + V_{0,-1-3,-2-4} A^{*(3)}_{-1-3,1,3} A^{*(3)}_{-2-4,2,4} + V_{0,-1-4,-2-3} A^{*(3)}_{-1-4,1,4} A^{*(3)}_{-2-3,2,3} \big) \\
    - & \dfrac{1}{2} \big( T_{-0,2,3,1+4}A^{(1)}_{1+4,1,4}  + T_{-0,1,3,2+4}A^{(1)}_{2+4,2,4} + T_{-0,1,2,3+4}A^{(1)}_{3+4,3,4} \\ 
        &\qquad + T_{-0,3,4,1+2}A^{(1)}_{1+2,1,2}  + T_{-0,2,4,1+3}A^{(1)}_{1+3,1,3} + T_{-0,1,4,2+3}A^{(1)}_{2+3,2,3}  \\       
        &\qquad + T_{-0,-(-1-2),3,4} A^{*(3)}_{-1-2,1,2} + T_{-0,-(-1-3),2,4} A^{*(3)}_{-1-3,1,3} + T_{-0,-(-1-4),2,3} A^{*(3)}_{-1-4,1,4} \\
        &\qquad + T_{-0,-(-2-4),1,3} A^{*(3)}_{-2-4,2,4} + T_{-0,-(-3-4),1,2} A^{*(3)}_{-3-4,3,4} + T_{-0,-(-2-3),1,4} A^{*(3)}_{-2-3,2,3}  \\
        &\qquad + V_{-0,1,0-1} B^{(1)}_{0-1,2,3,4} + V_{-0,2,0-2} B^{(1)}_{0-2,1,3,4} + V_{-0,3,0-3} B^{(1)}_{0-3,1,2,4} \\
        &\qquad + V_{-0,4,0-4} B^{(1)}_{0-4,1,2,3} - V_{-1,1-0,0} B^{*(4)}_{1-0,2,3,4} - V_{-2,2-0,0} B^{*(4)}_{2-0,1,3,4} \\
        &\qquad - V_{-3,3-0,0} B^{*(4)}_{3-0,1,2,4} - V_{-4,4-0,0} B^{*(4)}_{4-0,1,2,3} \big)\,.
    \end{split}
\end{equation}

\begin{equation}\label{kernel:X4}
    \begin{split}
        -i\,X^{(4)}_{0,1,2,3,4} = &\dfrac{4}{3} \big( V_{-0,4-3,-1-2}  A^{(1)}_{4,3,4-3} A^{(3)}_{-2-1,1,2} +  V_{-0,4-2,-1-3}  A^{(1)}_{4,2,4-2} A^{(3)}_{-3-1,1,3} \\
        &\qquad +  V_{-0,4-1,-2-3}  A^{(1)}_{4,1,4-1} A^{(3)}_{-2-3,2,3} +  V_{-(4-3),1+2,0}  A^{*(1)}_{1+2,1,2} A^{*(1)}_{4,3,4-3} \\
        &\qquad  +  V_{-(4-2),1+3,0}  A^{*(1)}_{1+3,1,3} A^{*(1)}_{4,2,4-2}  +  V_{-(4-1),2+3,0}  A^{*(1)}_{2+3,2,3} A^{*(1)}_{4,1,4-1} \\
        &\qquad -  V_{-(-1-2),3-4,0} A^{*(1)}_{3,4,3-4} A^{(3)}_{-1-2,1,2}  -  V_{-(-1-3),2-4,0}  A^{*(1)}_{2,4,2-4} A^{(3)}_{-1-3,1,3} \\
       &\qquad  -   V_{-(-2-3),1-4,0}  A^{*(1)}_{1,4,1-4} A^{(3)}_{-2-3,2,3} - V_{0,1+2,3-4}A^{*(1)}_{1+2,1,2} A^{*(1)}_{3,4,3-4} \\
       & \qquad -  V_{0,1+3,2-4}A^{*(1)}_{1+3,1,3} A^{*(1)}_{2,4,2-4} - V_{0,2+3,1-4}A^{*(1)}_{2+3,2,3} A^{*(1)}_{1,4,1-4} \big) \\
    + & 2 \big( - T_{-0,-1,-2-3,4}  A^{(3)}_{-2-3,2,3} -  T_{-0,-2,-1-3,4}  A^{(3)}_{-1-3,1,3} -  T_{-0,-3,-1-2,4}  A^{(3)}_{-1-2,1,2} \\
       &\qquad  + T_{0,1+2,3,-4}A^{*(1)}_{1+2,1,2} + T_{0,1+3,2,-4}A^{*(1)}_{1+3,1,3}+ T_{0,2+3,1,-4}A^{*(1)}_{2+3,2,3} \\
       &\qquad  + T_{0,1,2,-(4-3)}A^{(1)}_{4,3,4-3} + T_{0,1,3,-(4-2)}A^{(1)}_{4,2,4-2} + T_{0,2,3,-(4-1)}A^{(1)}_{4,1,4-1}  \\
       &\qquad  - T_{0,1,2,3-4}A^{(1)}_{3,4,3-4} - T_{0,1,3,2-4}A^{(1)}_{2,4,2-4} - T_{0,2,3,1-4}A^{(1)}_{1,4,1-4}  \big) \\
     -  & \dfrac{2}{3} \big( V_{-(0+1),1,0} B^{(3)}_{0+1,2,3,4}+ V_{-(0+2),2,0} B^{(3)}_{0+2,1,3,4} + V_{-(0+3),3,0} B^{(3)}_{0+3,1,2,4} \\
       &\qquad + V_{0,1,-0-1} B^{*(2)}_{-0-1,4,2,3} + V_{0,2,-0-2} B^{*(2)}_{-0-2,4,1,3} + V_{0,3,-0-3} B^{*(2)}_{-0-3,4,1,2} \big)\\ 
     +  & 2 \big( V_{-0,0-4,4} B^{(4)}_{0-4,1,2,3} - V_{-4,4-0,0}B^{*(1)}_{4-0,1,2,3} \big) \\
     +  & \dfrac{2}{3} \big( A^{(3)}_{0,1,-0-1} \widetilde{T}^*_{-0-1,4,2,3} + A^{(3)}_{0,2,-0-2} \widetilde{T}^*_{-0-2,4,1,3} + A^{(3)}_{0,3,-0-3} \widetilde{T}^*_{-0-3,4,1,2} \big)\,.
    \end{split}
\end{equation}

\begin{equation}\label{kernel:X5}
    \begin{split}
      -i\, X^{(5)}_{0,1,2,3,4} = &\dfrac{1}{3} \big( - V_{-0,-1-2,-3-4} A^{(3)}_{-1-2,1,2} A^{(3)}_{-3-4,3,4} - V_{-0,-1-3,-2-4} A^{(3)}_{-1-3,1,3} A^{(3)}_{-2-4,2,4} \\
        &\qquad  - V_{-0,-1-4,-2-3} A^{(3)}_{-1-4,1,4} A^{(3)}_{-2-3,2,3} +  V_{0,1+2,3+4} A^{*(1)}_{1+2,1,2}A^{*(1)}_{3+4,3,4}  \\  
        &\qquad +  V_{0,1+3,2+4} A^{*(1)}_{1+3,1,3}A^{*(1)}_{2+4,2,4} + V_{0,1+4,2+3} A^{*(1)}_{1+4,1,4}A^{*(1)}_{2+3,2,3} \\
        &\qquad  + V_{-(-3-4),1+2,0}A^{*(1)}_{1+2,1,2}A^{(3)}_{-3-4,3,4}  + V_{-(-2-4),1+3,0}A^{*(1)}_{1+3,1,3}A^{(3)}_{-2-4,2,4}\\
        &\qquad  +  V_{-(-2-3),1+4,0}A^{*(1)}_{1+4,1,4}A^{(3)}_{-2-3,2,3} + V_{-(-1-2),3+4,0}A^{*(1)}_{3+4,3,4}A^{(3)}_{-1-2,1,2} \\
        &\qquad  + V_{-(-1-3),2+4,0}A^{*(1)}_{2+4,2,4}A^{(3)}_{-1-3,1,3} 
         + V_{-(-1-4),2+3,0}A^{*(1)}_{2+3,2,3}A^{(3)}_{-1-4,1,4} \big) \\
     -   & \dfrac{1}{2} \big( T_{0,1,2,-(-3-4)}A^{(3)}_{-3-4,3,4} + T_{0,1,3,-(-2-4)}A^{(3)}_{-2-4,2,4} + T_{0,1,4,-(-2-3)}A^{(3)}_{-2-3,2,3} \\
        &\qquad + T_{0,2,3,-(-1-4)}A^{(3)}_{-1-4,1,4} + T_{0,2,4,-(-1-3)}A^{(3)}_{-1-3,1,3} + T_{0,3,4,-(-1-2)}A^{(3)}_{-1-2,1,2} \\
        &\qquad + T_{0,1,2,3+4}A^{*(1)}_{3+4,3,4} + T_{0,1,3, 2+4}A^{*(1)}_{2+4,2,4} + T_{0,1,4,2+3}A^{*(1)}_{2+3,2,3} \\
         &\qquad + T_{0,2,3,1+4}A^{*(1)}_{1+4,1,4} + T_{0,2,4,1+3}A^{*(1)}_{1+3,1,3} + T_{0,3,4,1+2}A^{*(1)}_{1+2,1,2}  \\         
         &\qquad - V_{-(0+1),1,0} B^{(4)}_{0+1,2,3,4} - V_{-(0+2),2,0} B^{(4)}_{0+2,1,3,4} -  V_{-(0+3),3,0} B^{(4)}_{0+3,1,2,4} \\
         &\qquad  -  V_{-(0+4),4,0} B^{(4)}_{0+4,1,2,3} - V_{0,1,-0-1} B^{*(1)}_{-0-1,2,3,4}  - V_{0,2,-0-2} B^{*(1)}_{-0-2,1,3,4} \\
         & \qquad - V_{0,3,-0-3} B^{*(1)}_{-0-3,1,2,4} - V_{0,4,-0-4} B^{*(1)}_{-0-4,1,2,3}  \big)\,.
    \end{split}
\end{equation}

\newpage
Potentially resonant case (non-resonant subcase) of $5$-wave interactions:

\begin{equation}\label{kernel:X2}
    \begin{split}
      i\,X^{(2)}_{0,1,2,3,4} = &-\dfrac{4}{3} \big(  V_{-0,2+3,4-1} A^{(1)}_{2+3,2,3} A^{(1)}_{4,1,4-1} +  V_{-0,2+4,3-1} A^{(1)}_{2+4,2,4} A^{(1)}_{3,1,3-1} \\
        &\qquad  + V_{-0,3+4,2-1} A^{(1)}_{3+4,3,4} A^{(1)}_{2,1,2-1} -  V_{-(3+4),1-2,0} A^{(1)}_{3+4,3,4} A^{*(1)}_{1,2,1-2} \\
         &\qquad -  V_{-(2+4),1-3,0} A^{(1)}_{2+4,2,4} A^{*(1)}_{1,3,1-3}  -  V_{-(2+3),1-4,0} A^{(1)}_{2+3,2,3} A^{*(1)}_{1,4,1-4} \\
         &\qquad -  V_{-(2-1),-3-4,0} A^{(1)}_{2,1,2-1} A^{*(3)}_{-3-4,3,4} -  V_{-(3-1),-2-4,0} A^{(1)}_{3,1,3-1} A^{*(3)}_{-2-4,2,4} \\
         &\qquad -  V_{-(4-1),-2-3,0} A^{(1)}_{4,1,4-1} A^{*(3)}_{-2-3,2,3} - V_{0,1-2,-3-4} A^{*(1)}_{1,2,1-2} A^{*(3)}_{-3-4,3,4} \\
         &\qquad  - V_{0,1-3,-2-4} A^{*(1)}_{1,3,1-3} A^{*(3)}_{-2-4,2,4} - V_{0,1-4,-2-3} A^{*(1)}_{1,4,1-4} A^{*(3)}_{-2-3,2,3}   \big) \\
      -   & 2 \big( T_{-0,2,3,4-1} A^{(1)}_{4,1,4-1} + T_{-0,2,4,3-1} A^{(1)}_{3,1,3-1} + T_{-0,3,4,2-1} A^{(1)}_{2,1,2-1} \\
         &\qquad + T_{-2,-3,0,1-4}A^{*(1)}_{1,4,1-4} + T_{-2,-4,0,1-3}A^{*(1)}_{1,3,1-3} + T_{-3,-4,0,1-2}A^{*(1)}_{1,2,1-2} \\
         &\qquad + T_{-0,-1,2,3+4}  A^{(1)}_{3+4,3,4} + T_{-0,-1,3,2+4}  A^{(1)}_{2+4,2,4} + T_{-0,-1,4,2+3}  A^{(1)}_{2+3,2,3}  \\
         &\qquad +  T_{-2,1,0,-3-4}A^{*(3)}_{-3-4,3,4} + T_{-3,1,0,-2-4}A^{*(3)}_{-2-4,2,4} + T_{-4,1,0,-2-3}A^{*(3)}_{-2-3,2,3}\big) \\
      -   & \dfrac{2}{3} \big( 
         V_{-0,2,0-2} B^{(2)}_{0-2,1,3,4} +  V_{-0,3,0-3} B^{(2)}_{0-3,1,2,4}+  V_{-0,4,0-4} B^{(2)}_{0-4,1,2,3} \\
         &\qquad - V_{-2,0,2-0}B^{*(3)}_{2-0,3,4,1} - V_{-3,0,3-0}B^{*(3)}_{3-0,2,4,1} - V_{-4,0,4-0}B^{*(3)}_{4-0,2,3,1} \\
         &\qquad -3V_{-(0+1),0,1}B^{(1)}_{0+1,2,3,4} 
         - 3 V_{0,1,-0-1}B^{*(4)}_{-0-1,2,3,4} \big) \\
      +   & \dfrac{2}{3} \big( A^{(1)}_{0,2,0-2} \widetilde{T}_{3,4,1,0-2} + A^{(1)}_{0,3,0-3} \widetilde{T}_{2,4,1,0-3} + A^{(1)}_{0,4,0-4} \widetilde{T}_{2,3,1,0-4} \big)\,.
    \end{split}
\end{equation}

\begin{equation}\label{kernel:X3}
    \begin{split}
     - i\, X^{(3)}_{0,1,2,3,4} &= - 2 \big( V_{-0,3+4,-1-2} A^{(1)}_{3+4,3,4} A^{(3)}_{-1-2,1,2} + V_{-0,3-1,4-2} A^{(1)}_{3,1,3-1} A^{(1)}_{4,2,4-2} \\
       &\qquad + V_{-0,4-1,3-2} A^{(1)}_{3,2,3-2} A^{(1)}_{4,1,4-1}     
       -V_{-(3+4),1+2,0} A^{(1)}_{3+4,3,4} A^{*(1)}_{1+2,1,2} \\
       &\qquad -V_{-(-1-2),-3-4,0} A^{(3)}_{-1-2,1,2} A^{*(3)}_{-3-4,3,4} - 
       V_{-(3-1),2-4,0} A^{(1)}_{3,1,3-1} A^{*(1)}_{2,4,2-4} \\
       &\qquad   - V_{-(4-1),2-3,0} A^{(1)}_{4,1,4-1} A^{*(1)}_{2,3,2-3}
        - V_{-(3-2),1-4,0} A^{(1)}_{3,2,3-2} A^{*(1)}_{1,4,1-4} \\
        &\qquad - V_{-(4-2),1-3,0} A^{(1)}_{4,2,4-2} A^{*(1)}_{1,3,1-3}   + V_{0,1+2,-3-4}A^{*(1)}_{1+2,1,2}A^{(3)}_{-3-4,3,4}  \\
        &\qquad   -V_{0,1-4,2-3}A^{*(1)}_{1,4,1-4}A^{*(1)}_{2,3,2-3}  -V_{0,1-3,2-4}A^{*(1)}_{1,3,1-3}A^{*(1)}_{2,4,2-4}  \big)\\
       &- 3 \big( T_{-0,3,4,(-1-2)}A^{(3)}_{1,2,-1-2} + T_{-(1+2),-0,3,4}A^{*(1)}_{1+2,1,2} +  T_{-(3+4),2,1,0}A^{(1)}_{3+4,3,4}\\
       &\qquad +  T_{-0,-1,3,4-2} A^{(1)}_{4,2,4-2} +  T_{-0,-1,4,3-2} A^{(1)}_{3,2,3-2} +  T_{-0,-2,3,4-1} A^{(1)}_{4,1,4-1} \\
       &\qquad +  T_{-0,-2,4,3-1} A^{(1)}_{3,1,3-1} + T_{-3,2,0,1-4}A^{*(1)}_{1,4, 1-4} + T_{-3,1,0,2-4}A^{*(1)}_{2,4, 2-4}  
       \\
       &\qquad + T_{-4,2,0,1-3}A^{*(1)}_{1,3, 1-3} + T_{-4,1,0,2-3}A^{*(1)}_{2,3, 2-3} - T_{0,1,2,-3-4} A^{(3)}_{3,4,-3-4} \big) \\
       &-\big( V_{-0,3,0-3}B^{*(3)}_{0-3,1,2,4} + V_{-0,4,0-4}B^{*(3)}_{0-4,1,2,3} - V_{-3,0,3-0}B^{*(2)}_{3-0,4,1,2} \\
       &\qquad - V_{-4,0,4-0}B^{*(2)}_{4-0,3,1,2} -  V_{-(0+1),0,1} B^{(2)}_{0+1,2,3,4}-  V_{-(0+2),0,2} B^{(2)}_{0+2,1,3,4}\\
       &\qquad  - V_{0,1,-0-1}B^{*(3)}_{-0-1,3,4,2} - V_{0,2,-0-2}B^{*(3)}_{-0-2,3,4,1} \big) \\
       & - \big(  \widetilde{T}_{1,2,4,3-0} A^{(1)}_{3,0,3-0} +  \widetilde{T}_{1,2,3,4-0} A^{(1)}_{4,0,4-0} - \widetilde{T}_{3,4,2,0+1} A^{(1)}_{0+1,0,1} - \widetilde{T}_{3,4,1,0+2} A^{(1)}_{0+2,0,2} \big)\,.
    \end{split}
\end{equation}

\newpage

Potentially resonant case (resonant subcase) of $5$-wave interactions:

\begin{equation}
    \begin{split}
    \label{eq:P,p,Q}
        \hspace{-3cm}  -i\, p_{0,1,2,3,4}   &= \frac{1}{3} A^{(1)}_{0,2,0-2}\left(A^{(1)}_{3,1,3-1} A^{(1)}_{0-2,3-1,4} 
          + A^{(1)}_{4,1,4-1} A^{(1)}_{0-2,4-1,3}
          + A^{(1)}_{3+4,3,4} A^{(1)}_{3+4,0-2,1} 
          - A^{(3)}_{3,4,-3-4} A^{(3)}_{0-2,-3-4,1} \right)          \\
     \hspace{-3cm}                 & - \frac{2}{3}\,\frac{A^{(1)}_{2,0,2-0}}{\omega_{2-0} + \omega_3 + \omega_4 - \omega_1} \left(
                      - A^{(1)}_{1,3,1-3} V_{-(1-3),2-0,4} 
                      - A^{(1)}_{1,4,1-4} V_{-(1-4),2-0,3}
                      + A^{(1)}_{3+4,3,4} V_{-1,2-0,3+4} \right.\\
                     &\left.\qquad \qquad 
                     + A^{(3)}_{3,4,-3-4} V_{-(2-0),-3-4,1}
                      + A^{(1)}_{3,1,3-1} V_{4,2-0,3-1}
                      + A^{(1)}_{4,1,4-1} V_{3,2-0,4-1} 
                      + \frac{3}{2} T_{-1,3,4,2-0}
                      \right)\,,\\
    \hspace{-3cm}     i\, Q_{0,1,2,3,4}   
         &= A^{(1)}_{3,0,3-0} \left(A^{(1)}_{2,4,2-4} A^{(1)}_{3-0,2-4,1}
         + \frac{1}{2} A^{(1)}_{1+2,1,2} A^{(1)}_{1+2,3-0,4} 
         - \frac{1}{2} A^{(3)}_{1,2,-1-2} A^{(3)}_{4,3-0,-1-2}  \right) \\
 \hspace{-3cm}       & +A^{(1)}_{4,2,4-2} \left( A^{(1)}_{3,1,3-1}A^{(1)}_{0,4-2,3-1} 
         + A^{(3)}_{0,1,-0-1}A^{(3)}_{3,4-2,-0-1}
         - A^{(1)}_{1,3,1-3}A^{(1)}_{4-2,1-3,0}
         - A^{(1)}_{0,3,0-3}A^{(1)}_{4-2,0-3,1} \right) \\
        & + A^{(1)}_{0+2,0,2} \left( A^{(1)}_{3,1,3-1}A^{(1)}_{0+2,3-1,4} 
        + A^{(1)}_{3+4,3,4}A^{(1)}_{3+4,0+2,1}
        - A^{(1)}_{1,3,1-3}A^{(1)}_{4,0+2,1-3} \right. \\
  \hspace{-3cm}      & \left. \qquad \qquad  - A^{(1)}_{1,4,1-4}A^{(1)}_{3,0+2,1-4} 
        - A^{(3)}_{3,4,-3-4}A^{(3)}_{1,0+2,-3-4} \right) \\
  \hspace{-3cm}      & + 2 \frac{A^{(1)}_{2,4,2-4}}{\omega_{2-4} + \omega_{1} + \omega_{0} - \omega_{3}} \left( A^{(1)}_{0,3,0-3}V_{1,0-3,2-4}
        + A^{(1)}_{1,3,1-3}V_{1-3,2-4,0}
        - A^{(1)}_{3,0,3-0}V_{-(3-0),2-4,1} \right.\\
  \hspace{-3cm}      & \left. \qquad \qquad - A^{(1)}_{3,1,3-1}V_{-(3-1),2-4,0} 
        + A^{(1)}_{0+1,0,1}V_{-3,0+1,2-4} 
        + A^{(3)}_{0,1,-0-1}V_{-(2-4),-0-1,3} 
        + \frac{3}{2} T_{-3,0,1,2-4} \right) \\
  \hspace{-3cm}      & - 2 \frac{A^{(1)}_{0,3,0-3}}{\omega_{0-3} + \omega_{1} + \omega_{2} - \omega_{4}} \left(-A^{(1)}_{4,2,4-2}V_{-(4-2),0-3,1} 
        + A^{(1)}_{2,4,2-4}V_{1,0-3,2-4} 
        + (1/2)*A^{(1)}_{1+2,1,2}V_{-4,1+2,0-3} \right.\\
  \hspace{-3cm}      & \left. \qquad \qquad
        + \frac{1}{2} A^{(3)}_{1,2,-1-2}V_{-(0-3),-1-2,4}
        + \frac{3}{4} T_{-4,1,2,0-3} \right) \\
  \hspace{-3cm}      & -2 \frac{A^{(3)}_{0,2,-0-2}}{\omega_{0+2} + \omega_{3}  + \omega_{4} - \omega_{1}} \left(- A^{(1)}_{1,3,1-3}V_{-(1-3),-0-2,4}
        - A^{(1)}_{1,4,1-4}V_{-(1-4),-0-2,3}
        + A^{(1)}_{3,1,3-1}V_{4,3-1,-0-2} \right.\\
  \hspace{-3cm}      & \left. \qquad \qquad
        + A^{(1)}_{4,1,4-1}V_{3,4-1,-0-2}
        + A^{(1)}_{3+4,3,4}V_{-1,3+4,-0-2} 
        + A^{(3)}_{3,4,-3-4}V_{-(-0-2),-3-4,1}
        + \frac{3}{2}T_{-1,3,4,-0-2} \right) \\
  \hspace{-3cm}      & + 2 \frac{A^{(1)}_{3+4,3,4}}{\omega_{3+4} - \omega_{0} - \omega_{1} - \omega_{2}} \left( A^{(1)}_{0+2,0,2}V_{-(3+4),0+2,1}
        + \frac{1}{2} A^{(1)}_{1+2,1,2}V_{-(3+4),1+2,0}
        + A^{(3)}_{0,2,-0-2}V_{-1,3+4,-0-2} \right.\\
  \hspace{-3cm}      & \left. \qquad \qquad
        + \frac{1}{2} A^{(3)}_{1,2,-1-2}V_{-0,3+4,-1-2} 
        + \frac{3}{4} T_{-3+4,2,1,0} \right) \\
  \hspace{-3cm}      & + 2 \frac{A^{(3)}_{3,4,-3-4}}{\omega_{3+4} + \omega_{0}  + \omega_{1} + \omega_{2} } \left( A^{(1)}_{0+2,0,2}V_{-3-4,0+2,1} 
        + \frac{1}{2} A^{(1)}_{1+2,1,2}V_{-3-4,1+2,0}
        - A^{(3)}_{0,2,-0-2}V_{-(-0-2),-3-4,1} \right.\\
  \hspace{-3cm}      & \left. \qquad \qquad
        - \frac{1}{2} A^{(3)}_{1,2,-1-2}V_{-(-1-2),-3-4,0} 
        + \frac{3}{4} T_{0,1,2,-3-4} \right) \,.
          \end{split}
\end{equation}




\end{document}